\documentclass[twocolumn,pra]{revtex4-1}
\usepackage{verbatim}
\usepackage{graphicx} % Include figure files
\usepackage{dcolumn}  % Align table columns on decimal point
\usepackage{bm}       % bold math
\usepackage{amssymb}

\usepackage[english]{babel}
\usepackage[normalem]{ulem}
\usepackage{color}
\usepackage[hypcap]{caption}
\usepackage{amsmath}
\usepackage[mathscr]{eucal}

\usepackage[]{hyperref}

\hypersetup{
  pdftitle={Your title here},
  pdfauthor={Your name here},
  pdfsubject={Your subject here},
  pdfkeywords={keyword1, keyword2},
  bookmarksnumbered=true,     
  bookmarksopen=true,         
  bookmarksopenlevel=1,       
  colorlinks=true,            
  pdfstartview=Fit,           
  pdfpagemode=UseOutlines,
  pdfpagelayout=TwoPageRight
}

\usepackage{ifthen}
  \newboolean{pt-br} 
  \setboolean{pt-br}{false}
  \newboolean{fig} 
  \setboolean{fig}{false}

\newcommand*\ptbr{\textbf}%{\textcolor[rgb]{0.75,0.75,0.75}}

\usepackage{lineno}
\usepackage{setspace}
\begin{document}

\preprint{APS/123-QED}

\title{The Clauser-Horne-Shimony-Holt inequality  in the context of a broad random variable }

\author{Felipe Andrade Velozo}
  \email{felipe.andrade.velozo@gmail.com}
  \affiliation{Instituto de Ci\^encias Sociais Aplicadas (ICSA). Universidade Federal de Alfenas (UNIFAL), Campus Avan\c{c}ado de Varginha-MG, CEP 37048-395, Brazil}

\author{Jos\'e A. C. Nogales}%
  \email{jnogales@dfi.ufla.br}
  \affiliation{Departamento de F\'isica (DFI) and Museu de Historia Natural (MHN), Universidade Federal de Lavras (UFLA), Lavras-MG, Caixa postal 3037, CEP 37000-000, Brazil.}%

\author{Gustavo Figueiredo Ara\'ujo}
  \email{kustavo@gmail.com}
  \affiliation{Departamento de Ci\^encia da Computa\c{c}\~ao (DCC), Universidade Federal de Lavras (UFLA), Lavras-MG, CEP 37000-000, Brazil.}%

\date{\today}

\begin{abstract}
In this work we  aim to analyze the Clauser-Horne-Shimony-Holt CHSH inequality strictly in the context of probability theory. 
In the course of assembling inequality we have to take care not to produce assumptions a priori, that is, physically or intuitively accepted suppositions. Of course, this does not mean that after these considerations, we put the adequate physical conditions suitable and generally accepted in these contexts. This allows us to clearly visualize the possibility of finding a greater inequality than that of CHSH in which it is included. This inequality does not contradict the CHSH inequality. This result is suported by using a robust computational simulation, showing the possibility for obtaining an inequality for quantum mechanics that is not violated and allows random hidden variables without any conflict with Bell's inequality.

\end{abstract}


\maketitle

\section{Introduction}

\ifthenelse{\boolean{pt-br}}
  {\ptbr
  {Em 1935, A. Einstein, juntamente com B. Podolsky e N. Rosen publicam um artigo \cite{einstein1935b}
  sobre Mec\^anica Qu\^antica, cujo t\'itulo traduzido \'e: ``Pode a descri\c{c}\~ao da Mec\^anica Qu\^antica
  sobre a realidade f\'isica ser considerada completa?''. Argumentando a possibilidade de haver ``vari\'aveis
  escondidas'' (vari\'aveis aleat\'orias) que, se seus valores fossem conhecidos, um experimento da
  Mec\^anica Qu\^antica deixaria de ser aleat\'orio e passaria a ser determin\'istico, ou seja, a aleatoriedade
  do experimento qu\^antico vem da falta de informa\c{c}\~ao de tais vari\'aveis.}}

In 1935, A. Einstein, together with B. Podolsky and N. Rosen published an article \cite{einstein1935b} on quantum mechanics, whose translated title can be described as ``can quantum mechanics on the physical reality be considered complete?'' Arguing the possibility of "hidden variances" (random variables) that, if their values were known, a quantum mechanics experiment would no longer be random and would become deterministic, that is, the randomness of the quantum experiment comes from the lack of information of such variables.

\ifthenelse{\boolean{pt-br}}{
  \ptbr{
  Em 1964, John S. Bell (em resposta ao paradoxo de Einstein, Podolsky e Rosen) publica um artigo \cite{Bell1964a}
  em que desenvolve uma desigualdade envolvendo correla\c{c}\~ao estat\'istica e, a partir da suposi\c{c}\~ao
  de que a Mec\^anica Qu\^antica seria uma teoria estat\'istica, dever-se-ia possuir uma vari\'avel
  aleat\'oria envolvida com as observa\c{c}\~oes. Assim, uma vari\'avel em que, se houvesse a possibilidade
  de conhecer seu valor, o resultado do experimento seria completamente previs\'ivel. Portanto, a falta
  de previsibilidade do experimento seria devida � ignor\^ancia sobre o valor que tal vari\'avel assume
  na realiza\c{c}\~ao do experimento. Por\'em, John S. Bell ao usar a f\'ormula obtida pelo c\'alculo
  de probabilidades no experimento de Mec\^anica Qu\^antica, encontra um conjunto de valores em que
  a desigualdade \'e violada, e logo conclui que os axiomas de probabilidade de Kolmogorov n\~ao s\~ao
  suficientes para descrever fen\^omenos qu\^anticos.
  }}

In 1964, John S. Bell (in response to the paradox of Einstein, Podolsky and Rosen) published an article \cite{Bell1964a} in which he developed an inequality involving statistical correlation and, from of the assumption that quantum mechanics would be a statistical theory, one should have a random variable involved with observations. Thus, a variable in which, if it were possible to know its value, the result of the experiment would be completely predictable. Therefore, the lack of predictability of the experiment would be due to ignorance about the value that such variable assumes in the experiment's performance. By John S. Bell, using the formula obtained by calculating probabilities in the Quantum Mechanics experiment, finds a set of values in which the inequality is violated, and hence concludes the Kolmogorov's axioms of probability are not sufficient to describe quantum phenomena.

\ifthenelse{\boolean{pt-br}}{
  \ptbr{
  Em 1969 J. F. Clauser, M. A. Horne, A. Shimony e R. A. Holt \cite{PhysRevLett_23_880} adaptam a desigualdade
  de Bell para um experimento vi\'avel. Em 1982, Aspect, Dalibard e Roger \cite{citeulike_679960} realizaram
  um experimento para observar a viola\c{c}\~ao da desigualdade de Clauser-Horne-Shimony-Holt na pr\'atica.
  Ap\'os realizar o experimento, usaram os dados na desigualdade e conclu\'iram que a desigualdade,
  obtida por meio de argumentos probabil\'isticos, era violada. Confirma-se, portanto, as conclus\~oes
  obtidas por John S. Bell de que n\~ao era poss\'ivel uma teoria de vari\'aveis ocultas nas condi\c{c}\~oes
  propostas por Einstein, Podolsky e Rosen.
  }}

In 1969 J. F. Clauser, M. A. Horne, A. Shimony and R. A. Holt \cite{PhysRevLett_23_880} fit the Bell inequality for a viable experiment. In 1982, Aspect, Dalibard and Roger \cite{citeulike_679960} conducted an experiment to observe the violation of Clauser-Horne-Shimony-Holt inequality in practice. After performing the experiment, they used the data in the inequality and concluded that this inequality, obtained by means of probabilistic arguments, was violated. It is confirmed, therefore, that the conclusions obtained by John S. Bell about the theory of hidden variables was not possible in the conditions proposed by Einstein, Podolsky and Rosen.

\ifthenelse{\boolean{pt-br}}{
  \ptbr{
  Desde ent\~ao, pode-se encontrar trabalhos que procuram estabelecer uma probabilidade qu\^antica ou
  o uso de outros sistemas de axiomas de probabilidades \cite{citeulike_3633707}.
  }}

Since then, works that seek to establish a quantum probability or the use of other systems of probability axioms \cite{citeulike_3633707} can be found.

The analysis of the probabilistic assumptions of Bell's arguments is extremely important for modern
quantum physics and the consequences of the modern interpretation of the violation of Bell's inequality
for the foundations of quantum mechanics are really relevant from a conceptual and practical reason.
Hence, the conditions for deriving this inequality should be carefully checked. 

\ifthenelse{\boolean{pt-br}}
  {\ptbr
  {Aqui, o foco de nossas consider\c{c}\~oes \'e analisar rigorosamente as condi\c{c}\~oes
  probabil\'isticas que foram assumidas para a demonstra\c{c}\~ao da desigualdade de CHSH. Uma vez realizado
  o estudo te\'orico dos pressupostos b\'asicos para a desigualdade de CHSH, verifica-se, por meio de
  simula\c{c}\~oes, a maneira como os dados das amostras devem ser utilizados para que tais pressupostos
  sejam obedecidos. Dessa forma, tanto os aspectos populacionais quanto os amostrais estar\~ao demonstrados.}
  }

Here the focus of our considerations is to strictly analyze the probabilistic conditions that have been assumed for the demonstration of CHSH inequality. Once the theoretical study of the basic assumptions for the CHSH inequality has been made, it is verified, through simulations, the manner in which the data of the samples should be used for such assumptions to be obeyed. In this way, both population and sample aspects shall be demonstrated.

\ifthenelse{\boolean{pt-br}}
  {\ptbr
  {Come\c{c}aremos, na se\c{c}\~ao \ref{sec:CHSH}, por apresentar o experimento e o conjunto de fun\c{c}\~oes
  de probabilidade associadas a ele, juntamente com a aplica\c{c}\~ao da desigualdade de CHSH, encontrada
  geralmente na literatura. A se\c{c}\~ao \ref{sec:CHSH} e a primeira parte da se\c{c}\~ao \ref{sec:Sim_Resultados}
  \'e apresentado o que se encontra na literatura, nas demais se\c{c}\~oes e a segunda parte da se\c{c}\~ao
  \ref{sec:Sim_Resultados} s\~ao resultados exclusivos deste trabalho.}
  }

We will start, in section \ref{sec:CHSH}, by presenting the experiment and the set of Probability functions associated with it, together with the application of the CHSH inequality, generally found in the literature. Section \ref{sec:CHSH} and the first part of section \ref{sec:Sim_Resultados} presents what is found in the literature, in the Other sections and the second part of section \ref{sec:Sim_Resultados} are the exclusive results of this work.

\ifthenelse{\boolean{pt-br}}
  {\ptbr
  {Ap\'os essa apresenta\c{c}\~ao do experimento, apresentamos a proposta deste trabalho de analisar
  a desigualdade, come\c{c}ando por analisar a quest\~ao da vari\'avel oculta $\lambda$, na se\c{c}\~ao
  \ref{sec:variavel_oculta}. As conclus\~oes a que chegamos nesta se\c{c}\~ao nos servem como justificativa
  para nos concentrarmos apenas nas vari\'aveis alet\'orias $Z_j$.}
  }

After this presentation of the experiment, we present that the proposal of this work is to analyze the inequality, starting by analyzing the question of the hidden variable $\lambda$, in the section \ref{sec:variavel_oculta}. The conclusions we reach in this section serve as a justification for concentrating only on the random variables $Z_j$.

\ifthenelse{\boolean{pt-br}}
  {\ptbr
  {Na se\c{c}\~ao \ref{sec:CHSH-Kolmogorov} e na se\c{c}\~ao \ref{sec:Sistema} demonstramos que o conjunto
  de fun\c{c}\~oes de probabilidade, que s\~ao usadas na literatura, s\~ao consistentes (ou seja, \'e
  poss\'ivel de se atribuir valores \`as fun\c{c}\~oes de probabilidades de $(Z_1,Z_2,Z_3,Z_4)$ de tal
  forma que se pode encontrar probabilidades marginais de $(Z_j,Z_k)$, com $j\in\{1,2\}$ e $k\in\{3,4\}$).
  Contudo, esse conjunto de fun\c{c}\~oes de probabilidades (encontradas na literatura), para determinados
  valores dos par\^ametros, levam \`a viola\c{c}\`ao da desigualdade de CHSH. Demonstramos que essa
  viola\c{c}\`ao s\'o se d\'a quando h\'a a viola\c{c}\`ao dos axiomas de Kolmogorov (ou seja, ou os
  valores atribuidoa \`as fun\c{c}\~oes de probabilidade de $(Z_1,Z_2,Z_3,Z_4)$ viola o axioma que diz
  que as probabilidades  devem ser maiores ou iguais a zero, ou viola o axioma que diz que a soma das
  probabilidades deve ser 1.}
  }

In the section \ref{sec:CHSH-Kolmogorov} and in the section \ref{sec:Sistema} it is shown that the set of probability functions, which are used in the literature, are consistent (ie it is possible to assign values to the functions of probabilities of $(Z_1,Z_2,Z_3,Z_4)$ such that one can find marginal probabilities of $(Z_j,Z_k)$, with $j\in\{1,2\}$ and $k\in\{3,4\}$). However, this set of probability functions (found in the literature), for certain values of the parameters, leads to a violation of CHSH inequality. We show that this violation only occurs when there is a violation of Kolmogorov's axioms (that is, the values attributed to the probability functions of $(Z_1,Z_2,Z_3,Z_4)$ violate the axiom that says the probabilities must be greater than or equal to zero, or violates the axiom which says that the sum of the probabilities must be 1).

\ifthenelse{\boolean{pt-br}}
  {\ptbr
  {Na se\c{c}\~ao \ref{sec:Desig_Basica} desenvolvemos uma desigualdade b\'asica, da qual se pode demonstrar
  tanto a desigualdade de Bell quanto a desigualdade de CHSH. Tamb\'em demonstramos, na se\c{c}\~ao
  \ref{sec:Bell_CHSH}, que da desigualdade de Bell pode-se chegar na desigualdade de CHSH, e mostramos
  a rela\c{c}\~ao entre as regi\~oes onde h\'a a viola\c{c}\~ao para as desigualdades de Bell e a de
  CHSH.}
  }

In the section \ref{sec:Desig_Basica} we have developed a basic inequality, from which both the Bell inequality and the CHSH inequality can be demonstrated. We also show, in the section \ref{sec:Bell_CHSH}, that from the Bell inequality we can arrive at the CHSH inequality, and the relation between the regions where there is the violation for the Bell and CHSH inequalities.

\ifthenelse{\boolean{pt-br}}
  {\ptbr
  {Na se\c{c}\~ao \ref{sec:Prob_Cond} propomos o uso de probabilidade condicional na desigualdade de
  CHSH, e justificamos tal proposta atrav\'es do esquema experimental encontrado na literatura. Provamos
  que, calculando os valores esperados, com base no uso das probabilidades condicionais, qualquer possibilidade
  de viola\c{c}\~ao da desigualdade de CHSH desaparece.}
  }

In section \ref{sec:Prob_Cond} we proposed the use of the  conditional probability in the CHSH inequality., we have justified such proposal through the experimental scheme that was found in the literature. We proved that, calculating the expected values, based on the use of conditional probabilities, any possibility of violation from the CHSH inequality  disappears.

\ifthenelse{\boolean{pt-br}}
  {\ptbr
  {Na se\c{c}\~ao \ref{sec:Sim} \'e descrito o algoritmo usado nas simula\c{c}\~oes. Na se\c{c}\~ao
  \ref{sec:Sim_Resultados} primeiramente apresentamos como as amostras geradas s\~ao usadas na literatura,
  surgindo a viola\c{c}\~ao da desigualdae de CHSH, para logo depois apresentarmos a proposta deste
  trabalho de como usar as amostras de modo que desapare\c{c}a qualquer viola\c{c}\~ao da desigualdade
  de CHSH a apresentamos justificativas que se adotar tal uso.}
  }

In section \ref{sec:Sim} it is described  the algorithm used in the simulations. In section \ref{sec:Sim_Resultados}, first we have presented how the generated samples are used in the literature, emerging the violation of the CHSH inequality, so we after have presented the proposal of this work of how to use the samples so that any violation of the CHSH inequality disapears and we presented justifications for such use.

\ifthenelse{\boolean{pt-br}}
  {\ptbr
  {Finalmente, na se\c{c}\~ao \ref{sec:Conclusao} apresentamos nossas conclus\~oes acerca da modelagem
  do experimento, assim como os aspectos te\'oricos relacionados com o problema.}
  }

Finally, in the section \ref{sec:Conclusao} we present our conclusions about the modeling of the experiment, as well as the theoretical aspects related to the problem.

\section{The experiment}
\label{sec:CHSH}

In this section we describe and analyze the experiment proposed by Aspect et. al. 1982 \cite{citeulike_679960}. The experiment consists of a source that emits entangled pairs of photons with correlated polarizations, which are emitted in opposite directions to two polarizers by the source. Each polarizer is implemented in a way that it acts according to the orientation angle ${\theta}_{1}$ and ${\theta}_{2}$, respectively. The angle ${\theta}_k$ provides the vector ${\vec{r}}_k=\rm{cos}({\theta}_k)\cdot\vec{i}+\rm{sin}({\theta}_k)\cdot\vec{j}$, that represent the orientation of the polarization . Each polarizer record it , if the photon will cross it ($Z_k(w)=+1$), or not ($Z_k(w)=-1$). We represent $k=1$ for the 1st. photon and $k=2$ for the 2nd photon. The probability  of each photon to cross or not the polarizer \cite{citeulike_9323624}, it is given by 
\begin{widetext}

  \begin{eqnarray}
  \label{eq:ProbZjZk}
    {{\mathcal P}}_{\dot Z_{1,2}}(Z_{1}(w),Z_{2}(w);{\bar\theta}_{1,2})
    =\left\{
    \begin{array}{c}
      \frac{{\rm{cos}}^{2}({\bar\theta}_{1,2})}{2}\Leftarrow (Z_{1}(w),Z_{2}(w))\in C 
      \\ 
      \frac{{\rm{sin}}^{2}({\bar\theta}_{1,2})}{2}\Leftarrow (Z_{1}(w),Z_{2}(w))\in \bar C
    \end{array}\right.,
    \qquad
    \left\{
    \begin{array}{l}
      C:=\{(-1,-1),(+1,+1)\}
      \\
      \bar C:=\{(-1,+1),(+1,-1)\}
      \\
      \bar\theta_{j,k}:=\theta_{k}-\theta_{j}
      \\
      \dot Z_{j,k}:=(Z_j,Z_k)
    \end{array}\right.\ 
  \end{eqnarray}

\end{widetext}

\noindent
where ${\bar\theta}_{1,2}={\theta}_{2}-{\theta}_{1}$ (is the angles' difference from the orientations
of the polarizers) and the random variables $Z_{1}$ and $Z_{2}$ are the passage or not for their respective
polarizers. In search of determining the $w$ associate to the experiment of measuring the crossing or
not of the photons that have been emitted with the polarization property correlated, it will be admitted
more two polarizer's orientations (totalling 4 orientations: $a_{1}$, $a_{2}$, $b_{1}$ and $b_{2}$)
and more two measures (totalling 4 random variables: $Z_{1}(w),Z_{2}(w),Z_{3}(w),Z_{4}(w)\in \{-1;+1\}$.
Therefore, the experiment was conducted for the same number of times for the polarizers, with the following
directions: a) Direction ${\theta}_{1}$ in the polarizer $I$ and ${\theta}_{3}$ in the $III$; b) Direction
${\theta}_{1}$ in the polarizer $I$ and ${\theta}_{4}$ in the $IV$; c) Direction ${\theta}_{2}$ in the
polarizer $II$ and ${\theta}_{2}$ in the $III$; d) Direction ${\theta}_{2}$ in the polarizer $II$ and
${\theta}_{4}$ in the $IV$.  The demonstration of the inequality \cite{citeulike_8683605,citeulike_9323624}
is based on Statistical arguments. We compared the predictions from the statistics with the results
obtained by the Quantum Mechanics. Thus, we omitted the parameters ${\theta}_j$ and $w$ (example: 
$\mathcal{P}_{Z_j,Z_k}(Z_j(w),Z_k(w);\bar\theta_{j,k})\equiv\mathcal{P}_{Z_j,Z_k}(Z_j,Z_k)$ or 
$\mathcal{P}_{Z_j,Z_k}(a,b;\bar\theta_{j,k})\equiv\mathcal{P}_{Z_j,Z_k}(a,b)$).

Calculating the expected value of $Z_{1}\cdot Z_{2}-Z_{1}\cdot Z_{4}+Z_{2}\cdot Z_{3}+Z_{2}\cdot Z_{4}$
and observing that $-2\le Z_{1}\cdot\left(Z_{3}-Z_{4}\right)+Z_{2}\cdot\left(Z_{3}+Z_{4}\right)\le 2$
is enough to substitute the values, convincing the validity of that inequality, then

\begin{equation}\nonumber
  |Z_{1}\cdot Z_{2}-Z_{1}\cdot Z_{4}+Z_{2}\cdot Z_{3}+Z_{2}\cdot Z_{4}|\leq 2
\end{equation}

\ifthenelse{\boolean{pt-br}}{
  \ptbr{
  Se multiplicamos ambos os membros da desigualdade por uma quantidade que n\~ao seja negativa, a desigualdade
  se mant\'em. Supondo a exist\^encia da fun\c{c}\~ao de probabilidade $\mathcal{P}_{\ddot Z}(Z_1,Z_2,Z_3,Z_4)$
  (que, de acordo com os axiomas de Kolmogorov \cite{Mood_IntroTheoryStats,magalhaes2006probabilidade},
  deve ser maior ou igual a zero, para quaisquer valores de $(Z_1,Z_2,Z_3,Z_4)$, ou seja, $|\mathcal{P}_{\ddot
  Z}(Z_1,Z_2,Z_3,Z_4)|=\mathcal{P}_{\ddot Z}(Z_1,Z_2,Z_3,Z_4)$), temos}}

If we multiply both members of the inequality by an amount that is not negative, the inequality remains. Assuming the existence of probability function $\mathcal{P}_{\ddot Z}(Z_1,Z_2,Z_3,Z_4)$
(which, according to the axioms of Kolmogorov \cite{Mood_IntroTheoryStats,magalhaes2006probabilidade}, must be greater than or equal to zero for any values $(Z_1,Z_2,Z_3,Z_4)$, that is, $|\mathcal{P}_{\ddot Z}(Z_1,Z_2,Z_3,Z_4)|=\mathcal{P}_{\ddot Z}(Z_1,Z_2,Z_3,Z_4)$), we have

\begin{widetext}
  
  \begin{equation}
  \label{eq:CHSHdem}
    \forall_{(Z_1,Z_2,Z_3,Z_4)\in\{-1,+1\}^4}
    (|Z_{1}\cdot Z_{2}-Z_{1}\cdot Z_{4}+Z_{2}\cdot Z_{3}+Z_{2}\cdot Z_{4}|\cdot
    \underbrace{\mathcal{P}_{\ddot Z}(Z_1,Z_2,Z_3,Z_4)}_{=|\mathcal{P}_{\ddot Z}(Z_1,Z_2,Z_3,Z_4)|}
    \leq 2\cdot\mathcal{P}_{\ddot Z}(Z_1,Z_2,Z_3,Z_4))
  \end{equation}
  
  \ifthenelse{\boolean{pt-br}}{
  \ptbr{
  sendo que $\ddot Z:=Z_{1},Z_{2},Z_{3},Z_{4}$. Portanto, temos que
  }}
  
  \noindent
  being that $\ddot Z:=Z_{1},Z_{2},Z_{3},Z_{4}$. Therefore
  
  \begin{equation}\nonumber
    |(Z_{1}\cdot Z_{2}-Z_{1}\cdot Z_{4}+Z_{2}\cdot Z_{3}+Z_{2}\cdot Z_{4})\cdot\mathcal{P}_{\ddot Z}(Z_1,Z_2,Z_3,Z_4)|
    \leq 2\cdot\mathcal{P}_{\ddot Z}(Z_1,Z_2,Z_3,Z_4)
  \end{equation}
  
  \begin{equation}\nonumber
    -2\cdot\mathcal{P}_{\ddot Z}(Z_1,Z_2,Z_3,Z_4)\leq
    (Z_{1}\cdot Z_{2}-Z_{1}\cdot Z_{4}+Z_{2}\cdot Z_{3}+Z_{2}\cdot Z_{4})\cdot\mathcal{P}_{\ddot Z}(Z_1,Z_2,Z_3,Z_4)
    \leq 2\cdot\mathcal{P}_{\ddot Z}(Z_1,Z_2,Z_3,Z_4)
  \end{equation}

  \ifthenelse{\boolean{pt-br}}{
  \ptbr{
  Uma vez que tal desigualdade \'e v\'alida para quaisquer valores de $(Z_1,Z_2,Z_3,Z_4)$, ent\~ao o
  somat\'orio de todos os valores poss\'iveis de $(Z_1,Z_2,Z_3,Z_4)$ para o membro esquerdo da desigualdade,
  continuar\'a sendo menor que o mesmo somat\'orio para o membro do meio, e este ser\'a menor que o
  somat\'orio do membro direito
  }}

  Since such an inequality is valid for any $(Z_1,Z_2,Z_3,Z_4)$, then the sum of all possible values of $(Z_1,Z_2,Z_3,Z_4)$ for the left side of the inequality, will remain smaller than the same sum for the middle side, and this will be smaller than the sum on the right side
  
  \begin{eqnarray}
  \label{eq:CHSHsoma}
    \sum\limits_{\ddot Z\in\{-1,+1\}^4}(
    -2\cdot\mathcal{P}_{\ddot Z}(\ddot Z))
    \leq
    \sum\limits_{\ddot Z\in\{-1,+1\}^4}(
    (Z_{1}\cdot Z_{2}-Z_{1}\cdot Z_{4}+Z_{2}\cdot Z_{3}+Z_{2}\cdot Z_{4})\cdot\mathcal{P}_{\ddot Z}(\ddot
    Z))
    \leq\sum\limits_{\ddot Z\in\{-1,+1\}^4}(2\cdot\mathcal{P}_{\ddot Z}(\ddot Z))
  \end{eqnarray}
  
  \ifthenelse{\boolean{pt-br}}{
  \ptbr{
  Tais somat\'orios s\~ao os valores esperados, resultando em
  }}

  Such sums are the expected values, resulting in
  
  \begin{equation}
  \label{eq:CHSH-Abs}
    -2\le {{\mathcal E}}_{\ddot Z}\left(Z_{1}\cdot Z_{3}\right)-{{\mathcal E}}_{\ddot Z}\left(Z_{1}\cdot     Z_{4}\right)+{{\mathcal E}}_{\ddot Z}\left(Z_{2}\cdot Z_{3}\right)+{{\mathcal E}}_{\ddot Z}\left(Z_{2}\cdot     Z_{4}\right)\le 2,
  \end{equation}
  
  \noindent
  and therefore
  
  \begin{equation}
  \label{eq:CHSH}
    \left|{{\mathcal E}}_{Z_{1},Z_{3}}\left(Z_{1}\cdot Z_{3}\right)
    -{{\mathcal E}}_{Z_{1},Z_{4}}\left(Z_{1}\cdot Z_{4}\right)
    +{{\mathcal E}}_{Z_{2},Z_{3}}\left(Z_{2}\cdot Z_{3}\right)
    +{{\mathcal E}}_{Z_{2},Z_{4}}\left(Z_{2}\cdot Z_{4}\right)\right|
    \le 2 
  \end{equation}
\end{widetext}

\ifthenelse{\boolean{pt-br}}{
  \ptbr{
  Observe que a desigualdade pode ser obtida atrav\'es da suposi\c{c}\~ao da exist\^encia de $\mathcal{P}_{\ddot
  Z}$ e que $|\mathcal{P}_{\ddot Z}(Z_1,Z_2,Z_3,Z_4)|=\mathcal{P}_{\ddot Z}(Z_1,Z_2,Z_3,Z_4)$. Falaremos
  mais da import\^ancia disso na se\c{c}\~ao \ref{sec:Sistema}.
  }}

It is noticeable that this inequality can be obtained through the assumption of the existence of $\mathcal{P}_{\ddot
Z}$ and $|\mathcal{P}_{\ddot Z}(Z_1,Z_2,Z_3,Z_4)|=\mathcal{P}_{\ddot Z}(Z_1,Z_2,Z_3,Z_4)$. We'll talk more about the importance of this in the section \ref{sec:Sistema}.

The covariance is given by 

\begin{equation}\nonumber
  \mathrm{Cov}(Z_j,Z_k)
  ={{\mathcal E}}_{Z_j,Z_k}(Z_j\cdot Z_k)-{{\mathcal E}}_{Z_j}(Z_j)\cdot {{\mathcal E}}_{Z_k}(Z_k)
\end{equation}

\noindent
however, it is observed that ${{\mathcal E}}_{Z_j}(Z_j)=0$ for any $j$. Making calculations starting
from the functions of supplied probability,

\begin{eqnarray}\nonumber
  &&{{\mathcal E}}_{Z_j,Z_k}(Z_j\cdot Z_k)
  =\rm{cos}(2({\theta}_j-{\theta}_k))
  =\rm{cos}(2{\bar\theta}_{j,k}),
  \\\nonumber
  &&{\bar\theta}_{j,k}:={\theta}_j-{\theta}_k
\end{eqnarray}

Substituting in the inequality, 

\begin{eqnarray}\nonumber
  \left|\rm{cos}\left(2 {\bar\theta}_{1,3}\right)
  -\rm{cos}\left(2 {\bar\theta}_{1,4}\right)
  +\rm{cos}\left(2 {\bar\theta}_{2,3}\right)
  +\rm{cos}\left(2 {\bar\theta}_{2,4}\right)\right|
  \le 2
\end{eqnarray} 

By choosing ${\theta}_{1}=-\frac{\pi}{3}$, ${\theta}_{2}=0$, ${\theta}_{3}=\frac{\pi}{3}$ e ${\theta}_{4}=\frac{2\cdot\pi}{3}$,
we had ${\bar\theta}_{1,3}=\frac{2\cdot\pi}{3}$, ${\bar\theta}_{1,4}=\pi $, ${\bar\theta}_{2,3}=\frac{\pi}{3}$,
${\bar\theta}_{2,4}=\frac{2\cdot\pi}{3}$, therefore

\begin{equation}
\label{eq:CHSHex}
  \overbrace{|\underbrace{\rm{cos}(2{\bar\theta}_{1,3})}_{=-1/2}
  -\underbrace{\rm{cos}(2{\bar\theta}_{1,4})}_{=1}
  +\underbrace{\rm{cos}(2{\bar\theta}_{2,3})}_{=-1/2}
  +\underbrace{\rm{cos}(2{\bar\theta}_{2,4})}_{=-1/2}|
  }^{=|-5/2|=5/2}
  \le 2
\end{equation}

But $\frac{5}{2}=2.5$ , which is greater than 2 and as a result, the inequality is not obeyed in the Quantum Mechanics.

\section{New Perspective of Bell inequality}
\label{sec:variavel_oculta}

In this section we start the argument by clarifying the relationship between the hidden variable $\lambda $ and free parameter $\theta $ with a random variable $Z$, which, experimentally, it is the result of the photons after crossing the polarizer. Thereby, the dependence of the continuous random variable $\lambda $ and a parameter $\theta $ is given by 

\begin{equation}\nonumber
  Z_j=Z_j\left(\lambda ,\theta\right),
  \qquad\lambda\in\Lambda,
  \qquad\theta\in\Theta
\end{equation}

\noindent
and the expected value of $Z_j\cdot Z_k$ is given by 

\begin{eqnarray}
  &&{{\mathcal E}}_{\lambda}(Z_j\left(\lambda,\theta\right)\cdot Z_k\left(\lambda ,\theta\right))
  =\nonumber
  \\
  &&=\int\limits_{\lambda\in\Lambda}
    {Z_j\left(\lambda,\theta\right)\cdot Z_k\left(\lambda,\theta\right)
    \cdot\rho\left(\lambda;\theta\right)\cdot d\lambda}\nonumber 
\end{eqnarray}

\noindent
where $\theta $ is a controlled parameter that is fixed, representing a constant. The random variable
$\lambda $ is not fixed, and it assumes several values in $\Lambda $ in the calculation of the integral.
The function $Z_j\cdot Z_k$ does not depend explicitly of $\lambda $. We evaluated the integral based
on the values assumed by the variables $Z$ 

\begin{equation}\nonumber
  {\Lambda}_{a,b}
  =\left\{\lambda\in\Lambda:
  \left(Z_j\left(\lambda ,\theta\right)=a\right)\wedge\left(Z_k\left(\lambda ,\theta\right)=b\right)\right\}
\end{equation}

So, the integral of the product $Z_j\cdot Z_k$ for values of $\lambda\in {\Lambda}_{a,b}$ is given by

\begin{eqnarray}\nonumber
  \int\limits_{\lambda\in{\Lambda}_{a,b}}
    {Z_j\left(\lambda,\theta\right)\cdot Z_k\left(\lambda,\theta\right)
    \cdot\rho\left(\lambda;\theta\right)\cdot d\lambda}
  &=&\int\limits_{\lambda\in{\Lambda}_{a,b}}
    {a\cdot b\cdot\rho\left(\lambda;\theta\right)\cdot d\lambda}
  \\\nonumber
  &=&a\cdot b\cdot\int\limits_{\lambda\in {\Lambda}_{a,b}}{\rho\left(\lambda;\theta\right)\cdot d\lambda}
\end{eqnarray} 

Note that the parameter $\theta $ is presented only in the function of density of probability $\rho
$; in other words, if the random variable $Z_j$ does not depend on $\theta $, the conclusion would be
the same if the probability carries the information of the parameter $\theta $. The integral $\int_{\lambda\in
{\Lambda}_{a,b}}{\rho\left(\lambda;\theta\right)\cdot d\lambda}$ is nothing more than the probability
of the event ${\Lambda}_{a,b}$ 

\[
  {{\mathcal P}}_{\lambda}\left({\Lambda}_{a,b}\right)
  =\int\limits_{\lambda\in {\Lambda}_{a,b}}
    {\rho\left(\lambda;\theta\right)\cdot d\lambda}
\]

As $\lambda\in{\Lambda}_{a,b}$ the probability of the value ${\dot{Z}}_{j,k}:=(Z_j,Z_k)\in\{(a,b)\}$
corresponds to 

\[
  {{\mathcal P}}_{\lambda}\left({\Lambda}_{a,b}\right)
  ={{\mathcal P}}_{{\dot{Z}}_{j,k}}\left(a,b\right)
\]

\noindent
and

\begin{eqnarray}\nonumber
  {{\mathcal P}}_{\lambda}\left({\Lambda}_{a,-1}\cup{\Lambda}_{a,+1}\right)
  &=&{{\mathcal P}}_{{\dot{Z}}_{j,k}}\left(a,-1\right)+{{\mathcal P}}_{{\dot{Z}}_{j,k}}\left(a,+1\right)=
  \\
  \nonumber
  &=&{{\mathcal P}}_{Z_j}\left(a\right)
\end{eqnarray}

\noindent
so the expected value is 

\begin{widetext}

  \begin{eqnarray}\nonumber
    {{\mathcal E}}_{\lambda}(Z_j\left(\lambda ,\theta\right))
    &=&\sum\limits_{\substack{a\in\left\{-1,+1\right\}\\ b\in\{-1,+1\}}}
      {\Bigg(\int\limits_{\lambda\in {\Lambda}_{a,b}}
        {Z_j(\lambda ,\theta )\cdot\rho (\lambda;\theta )\cdot d\lambda}\Bigg)}
    =\sum\limits_{\substack{a\in\left\{-1,+1\right\}\\ b\in\{-1,+1\}}}
      {\Bigg(a\cdot\int\limits_{\lambda\in\Lambda_{a,b}}{\rho\left(\lambda;\theta\right)\cdot d\lambda}\Bigg)}=
    \\
    \nonumber
    &=&\sum\limits_{a\in\left\{-1,+1\right\}}
      {\Bigg(a\cdot\sum\limits_{b\in\left\{-1,+1\right\}}
        {({{\mathcal P}}_{{\dot{Z}}_{j,k}}\left(a,b\right))}\Bigg)}
    =\sum\limits_{a\in\left\{-1,+1\right\}}
      {(a\cdot {{\mathcal P}}_{Z_j}\left(a\right))}=
    {{\mathcal E}}_{Z_j}(Z_j)
  \end{eqnarray}
  
  \noindent
  and the expected value of the product is
  
  \begin{eqnarray}\nonumber
    {{\mathcal E}}_{\lambda}(Z_j\left(\lambda ,\theta\right)\cdot Z_k\left(\lambda ,\theta\right))
    &&=\sum\limits_{a\in\left\{-1,+1\right\}\atop b\in\{-1,+1\}}
      {\Bigg(\int\limits_{\lambda\in {\Lambda}_{a,b}}
        {Z_j(\lambda,\theta)\cdot Z_k(\lambda,\theta)\cdot\rho(\lambda;\theta)\cdot d\lambda}\Bigg)}=
    \\
    \nonumber
    &&=\sum\limits_{a\in\left\{-1,+1\right\}\atop b\in\{-1,+1\}}
      {\Bigg(a\cdot b\cdot
      \int\limits_{\lambda\in {\Lambda}_{a,b}}
        {\rho (\lambda;\theta )\cdot d\lambda}\Bigg)}
    =\sum\limits_{a\in\left\{-1,+1\right\}\atop b\in\{-1,+1\}}
      {\Bigg(a\cdot b\cdot {{\mathcal P}}_{{\dot{Z}}_{j,k}}\left(a,b\right)\Bigg)}=
    \\\nonumber
    &&={{\mathcal E}}_{{\dot{Z}}_{j,k}}(Z_j\cdot Z_k)
  \end{eqnarray}
\end{widetext}

Therefore, the expected value in the continuous variable $\lambda $ becomes an expected value in the
discret variable $Z$. Therefore, the integral becomes the sum of $a$ value. Observe that the parameter
is presented in the function of probability 

\ifthenelse{\boolean{pt-br}}
  {\ptbr
  {Rela\c{c}\~ao entre a desigualdade de CHSH e os axiomas de Kolmogorov}}

\section{Relationship between CHSH inequality and Kolmogorov axioms}
\label{sec:CHSH-Kolmogorov}

\ifthenelse{\boolean{pt-br}}
  {\ptbr
  {Nesta se\c{c}\~ao, veremos como a viola\c{c}\~ao da desigualdade de CHSH est\'a relacionada com a
  viola\c{c}\~ao dos axiomas de Kolmogorov \cite{Mood_IntroTheoryStats}, mais especificamente, a viola\c{c}\~ao
  do axioma que afirma que para qualquer conjunto $A$, pertencente ao dom\'inio da fun\c{c}\~ao de probabilidade,
  $\mathcal{P}(A)$ ser\'a maior ou igual a 0, e o axioma que afirma que $\mathcal{P}(\Omega)=1$ (ou
  seja, a probabilidade de ocorrer qualquer resultado, representado pelo conjunto $\Omega$, ser\'a igual
  a 1).}
  }

In this section, we shall see how the violation of the CHSH inequality is related to the violation of Kolmogorov's axioms, more specifically, the violation of the axiom that states that for any set $A$, belonging to the domain of the probability function, $\mathcal{P}(A)$ will be greater than or equal to 0, and the axiom that states that $\mathcal{P}(\Omega)=1$ (that is, the probability of any result occurring, represented by the set $\Omega$, will be equal to 1 ).

\ifthenelse{\boolean{pt-br}}
  {\ptbr
  {Para facilitar a escrita, vamos adotar a seguinte nota\c{c}\~ao}}

In order to make it easier to read, we will adopt the following notation:

\begin{eqnarray}\nonumber
  \mathcal{P}_{\dot{Z}_{j,k}}(-1,-1)=P_{j,k}^{--},
  \quad
  &\mathcal{P}_{\dot{Z}_{j,k}}(-1,+1)=P_{j,k}^{-+}
  \\
  \nonumber
  \mathcal{P}_{\dot{Z}_{j,k}}(+1,-1)=P_{j,k}^{+-},
  \quad
  &\mathcal{P}_{\dot{Z}_{j,k}}(+1,+1)=P_{j,k}^{++}
  \\
  \nonumber
  \mathcal{P}_{\dot Z_{j,k}}(C)=P_{j,k}^{\pm\pm},
  \quad
  &\mathcal{P}_{\dot Z_{j,k}}(\bar C)=P_{j,k}^{\pm\mp},
  \\
  \nonumber
  \mathcal{P}_{\dot{Z}_{j,k,l}}(-1,-1,-1)=P_{j,k,l}^{---},
  \quad
  &\mathcal{P}_{\dot{Z}_{j,k,l}}(-1,-1,+1)=P_{j,k,l}^{--+}
  \\
  \nonumber
  \mathcal{P}_{\dot{Z}_{j,k,l}}(-1,+1,-1)=P_{j,k,l}^{-+-},
  \quad
  &\mathcal{P}_{\dot{Z}_{j,k,l}}(-1,+1,+1)=P_{j,k,l}^{-++}
  \\
  \nonumber
  \mathcal{P}_{\dot{Z}_{j,k,l}}(+1,-1,-1)=P_{j,k,l}^{+--},
  \quad
  &\mathcal{P}_{\dot{Z}_{j,k,l}}(+1,-1,+1)=P_{j,k,l}^{+-+}
  \\
  \nonumber
  \mathcal{P}_{\dot{Z}_{j,k,l}}(+1,+1,-1)=P_{j,k,l}^{++-},
  \quad
  &\mathcal{P}_{\dot{Z}_{j,k,l}}(+1,+1,+1)=P_{j,k,l}^{+++}
\end{eqnarray}

\ifthenelse{\boolean{pt-br}}
  {\ptbr
  {Observa-se que a desigualdade de CHSH \'e dada em termos de valores esperados $\mathcal{E}_{\dot{Z}_{j,k}}(Z_j\cdot
  Z_k)$, contudo, os valores esperados podem ser expressos em termos das probabilidades $\mathcal{P}_{\dot{Z}_{j,k}}(z_j,z_k)$,
  da seguinte forma}
  }

It is noted that the CHSH inequality is given in terms of expected values $\mathcal{E}_{\dot{Z}_{j,k}}(Z_j\cdot
Z_k)$. However, the expected values can be expressed in terms of the probabilities $\mathcal{P}_{\dot{Z}_{j,k}}(z_j,z_k)$,
as follow

\begin{eqnarray}
\label{eq:Correlacao}
  \nonumber
  \mathcal{E}_{\dot{Z}_{j,k}}(Z_j\!\cdot\!Z_k)
  &=&\sum\limits_{z_j\in\{-1,+1\}\atop z_k\in\{-1,+1\}}(z_j\cdot z_k\cdot\mathcal{P}_{\dot{Z}_{j,k}}(z_j,z_k))
  \\
  &=&\underbrace{P_{j,k}^{--}+P_{j,k}^{++}}_{P_{j,k}^{\pm\pm}=1-P_{j,k}^{\pm\mp}}
    -\underbrace{(P_{j,k}^{-+}+P_{j,k}^{+-})}_{P_{j,k}^{\pm\mp}=1-P_{j,k}^{\pm\pm}}
\end{eqnarray}

\begin{equation}\nonumber
  \mathcal{E}_{\dot Z_{j,k}}(Z_j\cdot Z_k)
  =\left\{
    \begin{array}{l}
      2\cdot\mathcal{P}_{\dot{Z}_{j,k}}(C)-1
      \\
      1-2\cdot\mathcal{P}_{\dot{Z}_{j,k}}(\bar{C})
    \end{array}\right.\ 
\end{equation}

\ifthenelse{\boolean{pt-br}}
  {\ptbr
  {ao expressar a desiguladade de CHSH em termos das probabilidades, damos o primeiro passo para relacionar
  a viola\c{c}\~ao da desigualdade com a viola\c{c}\~ao dos axiomas de Kolmogorov, uma vez que esses
  axiomas est\~ao relacionados com as fun\c{c}\~oes de probabilidade.}
  }

Expressing the CHSH inequality in terms of probabilities, we take the first step relating the violation
of inequality to the violation of Kolmogorov's axioms, since these axioms are related to the probability
functions.

\ifthenelse{\boolean{pt-br}}
  {\ptbr
  {Agora, com a express\~ao do valor esperado e a f\'ormula da desigualdade de CHSH}
  }

Now, with this expression, the expected value and the CHSH inequality formula we obtain

\begin{widetext}

\begin{equation}\nonumber
  \left\{
  \begin{array}{l}
    \mathcal{E}_{\dot{Z}_{j,k}}
    =2\mathcal{P}_{\dot{Z}_{j,k}}(C)-1
    \\
    -2\leq\mathcal{E}_{\dot{Z}_{1,3}}(Z_1\cdot Z_3)-\mathcal{E}_{\dot{Z}_{1,4}}(Z_1\cdot Z_4)
      +\mathcal{E}_{\dot{Z}_{2,3}}(Z_2\cdot Z_3)+\mathcal{E}_{\dot{Z}_{2,4}}(Z_2\cdot Z_4)
      \leq 2
  \end{array}\right.
\end{equation}
\end{widetext}

\ifthenelse{\boolean{pt-br}}
  {\ptbr
  {realizando as substitui\c{c}\~oes, temos a desigualdade de CHSH em termos das probabilidades}
  }

\noindent
making the appropriate substitutions, we have the CHSH inequality in terms of the probabilities

\begin{eqnarray}\nonumber
  -2&\leq&(2P_{1,3}^{\pm\pm}-1)-(2P_{1,4}^{\pm\pm}-1)+
    \\\nonumber
    &&+(2P_{2,3}^{\pm\pm}-1)+(2P_{2,4}^{\pm\pm}-1)\leq2
\end{eqnarray}

\ifthenelse{\boolean{pt-br}}
  {\ptbr
  {que, simplificando resulta em}
  }

\noindent
after some simplifications we get

\begin{eqnarray}\nonumber
  -2&\leq&2P_{1,3}^{\pm\pm}-2P_{1,4}^{\pm\pm}
  +2P_{2,3}^{\pm\pm}+2P_{2,4}^{\pm\pm}-2\leq2
\end{eqnarray}

\ifthenelse{\boolean{pt-br}}
  {\ptbr
  {Dividindo todos os membros da desigualdade por 2, tem-se}
  }

Dividing all the members of the inequality by 2, we have

\begin{equation}\nonumber
  -1\leq P_{1,3}^{\pm\pm}-P_{1,4}^{\pm\pm}
    +P_{2,3}^{\pm\pm}+P_{2,4}^{\pm\pm}-1\leq1
\end{equation}

\ifthenelse{\boolean{pt-br}}
  {\ptbr
  {e somando 1 em todos os membros, tem-se}
  }

\noindent
and adding 1 in all members, we gather

\begin{equation}\nonumber
  0\leq P_{1,3}^{\pm\pm}-P_{1,4}^{\pm\pm}
    +P_{2,3}^{\pm\pm}+P_{2,4}^{\pm\pm}\leq2
\end{equation}

\ifthenelse{\boolean{pt-br}}
  {\ptbr
  {Observamos que a desigualdade ser\'a violada se a express\~ao (com as fun\c{c}\~oes de probabilidade)
  for menor que 0 (ou seja, se for negativa) ou se for maior que 2.}
  }

We observe that the inequality will be violated if the expression (with the probability functions) is
less than 0 (that is, if it is negative) or if it is greater than 2.

\ifthenelse{\boolean{pt-br}}
  {\ptbr
  {No segundo membro da desigualdade, suma-se e subtrai-se 1, e faz-se uso da propriedade das fun\c{c}\~oes
  de probabilidade em que $(\mathcal{P}(A)+\mathcal{P}(\bar{A})=1)$, sendo $\bar{A}$ o evento complementar
  de $A$, resultando em}
  }

In the second member of the inequality, 1 is added and subtracted, and use is made of the probability
functions where$(\mathcal{P}(A)+\mathcal{P}(\bar{A})=1)$, wherence  $\bar{A}$ is the complementary event
of $A$, resulting in

\begin{equation}\nonumber
  0\leq P_{1,3}^{\pm\pm}\underbrace{-P_{1,4}^{\pm\pm}+1}_{=P_{1,4}^{\pm\mp}}
    \underbrace{-1+P_{2,3}^{\pm\pm}}_{=-P_{2,3}^{\pm\mp}}+P_{2,4}^{\pm\pm}\leq2
\end{equation}

\ifthenelse{\boolean{pt-br}}
  {\ptbr
  {portanto, temos a desigualdade de CHSH expressa da seguinte forma}
  }

\noindent
therefore, we have the inequality of CHSH expressed as follows

\begin{equation}
\label{eq:CHSHProb}
  0\leq P_{1,3}^{\pm\pm}+P_{1,4}^{\pm\mp}-P_{3,4}^{\pm\pm}
    +P_{3,4}^{\pm\pm}-P_{2,3}^{\pm\mp}+P_{2,4}^{\pm\pm}\leq2
\end{equation}

\ifthenelse{\boolean{pt-br}}
  {\ptbr
  {que nada mais \'e do que a soma de duas desigualdades de Wigner \cite{Koc1993}.}
  }

\noindent
which is nothing more than the sum of two inequalities of Wigner.

\ifthenelse{\boolean{pt-br}}
  {\ptbr
  {At\'e o momento, apenas foi trabalhado com fun\c{c}\~oes de probabilidade de duas vari\'aveis aleat\'orias,
  contudo a desigualdade de CHSH se utiliza de quatro vari\'aveis aleat\'orias. Uma vez qua a deaigualdade
  de CHSH nada mais \'e que uma desigualdade demonstrada diretamente de uma argumenta\c{c}\~ao puramente
  estat\'istica, fica subentendido a exist\^encia de uma fun\c{c}\~ao de probabilidade de quatro vari\'aveis
  que modele o experimento como um todo. Nesse contexto, pode-se afirmar que para se obter a fun\c{c}\~ao
  de probabilidade de dus vari\'aveis, basta somar todas as possibilidades relacionadas com as outras
  vari\'aveis que sobram, portanto}
  }

So far, it has only been worked with probability functions of two random variables. However the CHSH
inequality is used in four random variables. Since the CHSH inequality is nothing more than an inequality
directly demonstrated by purely statistical argumentation, the existence of a four-variable probability
function is understood as the model of the experiment as a whole. In this context, it can be affirmed
that in order to obtain the probability function of two variables, it is enough to sum up all the possibilities
related to the other variables that remain, therefore 

\begin{eqnarray}\nonumber
  \mathcal{P}_{\dot{Z}_{j,k}}(z_j,z_k)
  &=&\sum\limits_{z_l\in\{-1,+1\}}(\mathcal{P}_{\dot{Z}_{j,k,l}}(z_j,z_k,z_l))
  \\\nonumber
  &=&\sum\limits_{z_l\in\{-1,+1\} \atop z_m\in\{-1,+1\}}(\mathcal{P}_{\dot{Z}_{j,k,l,m}}(z_j,z_k,z_l,z_m))
\end{eqnarray}

\ifthenelse{\boolean{pt-br}}
  {\ptbr
  {Agora, vamos escrever a probabilidade de duas vari\'aveis aleat\'orias serem iguais ($Z_j=Z_k=-1$
  ou $Z_j=Z_k=+1$) em termos de probabilidades de tr\^es vari\'aveis ($\mathcal{P}_{\dot{Z}_{j,k,l}}(z_j,z_k,z_l)$),
  da seguinte forma}
  }

Now let's write the probability of two random variables being equal ($Z_j=Z_k=-1$ or
$Z_j=Z_k=+1$) in terms of probabilities of three variables ($\mathcal{P}_{\dot{Z}_{j,k,l}}(z_j,z_k,z_l)$),
as follow

\begin{eqnarray}\nonumber
  \mathcal{P}_{\dot{Z}_{j,k}}(C)
  &=&\mathcal{P}_{\dot{Z}_{j,k}}(\{(-1,-1),(+1,+1)\})
  \\\nonumber
  &=&P_{j,k}^{--}+P_{j,k}^{++}
  \\\nonumber
  &=&P_{j,k,l}^{---}+P_{j,k,l}^{--+}+P_{j,k,l}^{++-}+P_{j,k,l}^{+++}
\end{eqnarray}

\ifthenelse{\boolean{pt-br}}
  {\ptbr
  {De forma semelhante, escreveremos para o caso em que duas vari\'aveis aleat\'orias diferem entre
  si (ou seja, $Z_j=-Z_k=-1$ ou $Z_j=-Z_k=+1$), resultando em}
  }

\noindent
Similarly, we will write for the case where two random variables differ from each other (that is, $Z_j=-Z_k=-1$
or $Z_j=-Z_k=+1$), resulting in

\begin{eqnarray}\nonumber
  \mathcal{P}_{\dot{Z}_{j,k}}(\bar{C})
  &=&\mathcal{P}_{\dot{Z}_{j,k}}(\{(-1,+1),(+1,-1)\})
  \\\nonumber
  &=&P_{j,k}^{-+}+P_{j,k}^{+-}
  \\\nonumber
  &=&P_{j,k,l}^{-+-}+P_{j,k,l}^{-++}+P_{j,k,l}^{+--}+P_{j,k,l}^{+-+}
\end{eqnarray}

\ifthenelse{\boolean{pt-br}}
  {\ptbr
  {Agora podemos reescrever as probabilidades encontradas na desigualdade de CHSH. Focando nas tr\^es
  primeiras probabilidades da f\'ormula (\ref{eq:CHSHProb}), temos}
  }

We can now rewrite the probabilities found in the CHSH inequality. Focusing on the first three probabilities of the formula (\ref{eq:CHSHProb}), we have

\begin{equation}\nonumber
  \mathcal{P}_{\dot{Z}_{1,3}}(C)
  =\underbrace{ P_{1,3,4}^{---}+ P_{1,3,4}^{--+}}
    _{=\mathcal{P}_{\dot{Z}_{1,3}}(-1,-1)}
    +\underbrace{ P_{1,3,4}^{++-}+ P_{1,3,4}^{+++}}
    _{=\mathcal{P}_{\dot{Z}_{1,3}}(+1,+1)}
\end{equation}

\begin{equation}\nonumber
  \mathcal{P}_{\dot{Z}_{1,4}}(\bar{C})
  =\underbrace{ P_{1,3,4}^{--+}+ P_{1,3,4}^{-++}}
    _{=\mathcal{P}_{\dot{Z}_{1,4}}(-1,+1)}
    +\underbrace{ P_{1,3,4}^{+--}+ P_{1,3,4}^{++-}}
    _{=\mathcal{P}_{\dot{Z}_{1,4}}(+1,-1)}
\end{equation}

\begin{equation}\nonumber
  \mathcal{P}_{\dot{Z}_{3,4}}(C)
  =\underbrace{ P_{1,3,4}^{---}+ P_{1,3,4}^{+--}}
    _{=\mathcal{P}_{\dot{Z}_{3,4}}(-1,-1)}
    +\underbrace{ P_{1,3,4}^{-++}+ P_{1,3,4}^{+++}}
    _{=\mathcal{P}_{\dot{Z}_{3,4}}(+1,+1)}
\end{equation}

\ifthenelse{\boolean{pt-br}}
  {\ptbr
  {Substituindo, encontramos}
  }

\noindent
Replacing, we found

\begin{eqnarray}\nonumber
  &&\mathcal{P}_{\dot{Z}_{1,3}}(C)+\mathcal{P}_{\dot{Z}_{1,4}}(\bar{C})-\mathcal{P}_{\dot{Z}_{3,4}}(C)=
  \\
  \nonumber
  &&= P_{1,3,4}^{--+}+ P_{1,3,4}^{--+}
    + P_{1,3,4}^{++-}+ P_{1,3,4}^{++-}
  \\
  &&=2\cdot( P_{1,3,4}^{--+}
    + P_{1,3,4}^{++-})
\label{eq:Wigner1}
\end{eqnarray}

\ifthenelse{\boolean{pt-br}}
  {\ptbr
  {Nessa f\'ormula, fica claro que, apesar de haver a subtra\c{c}\~ao de uma probabilidade ($-\mathcal{P}_{\dot{Z}_{3,4}}(C)$),
  a soma das outras duas probabilidades acabam por compensar a express\~ao que \'e subtra\'ida, resultando
  na soma de duas fun\c{c}\~oes de probabildiade. Disso conclui-se que o resultado (\ref{eq:Wigner1}),
  ou melhor, a metade dele, uma vez que est\'a sendo multiplicado por 2, deve estar entre 0
  e 1, devido aos axiomas de Kolmogorov.}
  }

\noindent
In this formula, it is clear that, although there is a subtraction of a probability ($-\mathcal{P}_{\dot{Z}_{3,4}}(C)$),
the sum of the other two probabilities ends by compensating the expression that is subtracted, resulting
in the sum of two probability functions. From this, it is concluded that the result (\ref{eq:Wigner1}),
or rather, half of it, since it is being multiplied by 2, must be between 0 and 1, due to Kolmogorov's
axioms.

\ifthenelse{\boolean{pt-br}}
  {\ptbr
  {Partindo para a reescrita das outras tr\^es parcelas restantes, temos que}
  }

Starting by rewriting the other three remaining portions, we have

\begin{equation}\nonumber
  \mathcal{P}_{\dot{Z}_{2,4}}(C)
  =\underbrace{ P_{2,3,4}^{---}+ P_{2,3,4}^{-+-}}
    _{=\mathcal{P}_{\dot{Z}_{2,4}}(-1,-1)}
    +\underbrace{ P_{2,3,4}^{+-+}+ P_{2,3,4}^{+++}}
    _{=\mathcal{P}_{\dot{Z}_{2,4}}(+1,+1)}
\end{equation}

\begin{equation}\nonumber
  \mathcal{P}_{\dot{Z}_{3,4}}(C)
  =\underbrace{ P_{2,3,4}^{---}+ P_{2,3,4}^{+--}}
    _{=\mathcal{P}_{\dot{Z}_{3,4}}(-1,-1)}
    +\underbrace{ P_{2,3,4}^{-++}+ P_{2,3,4}^{+++}}
    _{=\mathcal{P}_{\dot{Z}_{3,4}}(+1,+1)}
\end{equation}

\begin{equation}\nonumber
  \mathcal{P}_{\dot{Z}_{2,3}}(\bar{C})
  =\underbrace{ P_{2,3,4}^{-+-}+ P_{2,3,4}^{-++}}
    _{=\mathcal{P}_{\dot{Z}_{2,3}}(-1,+1)}
    +\underbrace{ P_{2,3,4}^{+--}+ P_{2,3,4}^{+-+}}
    _{=\mathcal{P}_{\dot{Z}_{2,3}}(+1,-1)}
\end{equation}

\ifthenelse{\boolean{pt-br}}
  {\ptbr
  {Substituindo, encontramos}
  }

Replacing, we found

\begin{eqnarray}\nonumber
  &&\mathcal{P}_{\dot{Z}_{3,4}}(C)-\mathcal{P}_{\dot{Z}_{2,3}}(\bar{C})+\mathcal{P}_{\dot{Z}_{2,4}}(C)
  \\\nonumber
  &&= P_{2,3,4}^{---}+ P_{2,3,4}^{+++}
    + P_{2,3,4}^{---}+ P_{2,3,4}^{+++}
  \\
  &&=2\cdot( P_{2,3,4}^{---}+ P_{2,3,4}^{+++})
\label{eq:Wigner2}
\end{eqnarray}

\ifthenelse{\boolean{pt-br}}
  {\ptbr
  {Novamente, conclui-se que a metade do resultado de (\ref{eq:Wigner2}) tamb\'em est\'a entre 0 e 1
  devido aos axiomas de Kolmogorov.}
  }

Again, one concludes that half of the result of (\ref{eq:Wigner2}) is also between 0 and 1 due to Kolmogorov's
axioms.

\ifthenelse{\boolean{pt-br}}
  {\ptbr
  {Agora vamos substiruim os resultados encontrados em (\ref{eq:Wigner1}) e (\ref{eq:Wigner2}) na f\'ormula
  (\ref{eq:CHSHProb}), de tal forma que tenhamos uma express\~ao mais uniforma, apresentado apenas fun\c{c}\~oes
  de probabilidade que dependa das mesmas vari\'aveis, ou seja, uma express\~ao contendo as probabildiades
  $\mathcal{P}_{\dot{Z}_{1,2,3,4}}(z_1,z_2,z_3,z_4)$, resultando em}
  }

Now let's simplify the results found in (\ref{eq:Wigner1}) and (\ref{eq:Wigner2}) in the formula (\ref{eq:CHSHProb}),
so that we have a more uniform expression, presenting only Functions of probability that depend on the
same variables, that is, an expression containing the probabilities, $\mathcal{P}_{\dot{Z}_{1,2,3,4}}(z_1,z_2,z_3,z_4)$,
resulting in 

\begin{eqnarray}\nonumber
  0&\leq&\underbrace{2\cdot( P_{1,3,4}^{--+}+ P_{1,3,4}^{++-})}
    _{=P_{1,3}^{\pm\pm}+P_{1,4}^{\pm\mp}-P_{3,4}^{\pm\pm}}
    +\underbrace{2\cdot( P_{2,3,4}^{---}+ P_{2,3,4}^{+++})}
    _{=P_{3,4}^{\pm\pm}-P_{2,3}^{\pm\mp}+P_{2,4}^{\pm\pm}}
  \\\nonumber
  &=&2\cdot
    (P_{1,2,3,4}^{---+}+P_{1,2,3,4}^{-+-+}+P_{1,2,3,4}^{+-+-}+P_{1,2,3,4}^{+++-}+
  \\\nonumber
  &&+P_{1,2,3,4}^{----}+P_{1,2,3,4}^{+---}
    +P_{1,2,3,4}^{-+++}+P_{1,2,3,4}^{++++})\leq
  \\\nonumber
  &\leq&2\cdot\bigg(\sum\limits_{(z_1,z_2,z_3,z_4)\in\{-1,+1\}^4}(\mathcal{P}_{\dot{Z}_{1,2,3,4}}(z_1,z_2,z_3,z_4))\bigg)
  \\\nonumber
  &=&2\cdot\mathcal{P}_{\dot{Z}_{1,2,3,4}}\bigg(\bigcup\limits_{(z_1,z_2,z_3,z_4)\in\{-1,+1\}^4}\{(z_1,z_2,z_3,z_4)\}\bigg)
  \\\nonumber
  &=&2\cdot 1=2
\end{eqnarray}

\ifthenelse{\boolean{pt-br}}
  {\ptbr
  {ou seja}
  }

\noindent
that is

\begin{equation}\nonumber
  0\leq\underbrace{2( P_{1,3,4}^{--+}+ P_{1,3,4}^{++-})}
    _{=P_{1,3}^{\pm\pm}+P_{1,4}^{\pm\mp}-P_{3,4}^{\pm\pm}}
  +\underbrace{2( P_{2,3,4}^{---}+ P_{2,3,4}^{+++})}
    _{=P_{3,4}^{\pm\pm}-P_{2,3}^{\pm\mp}+P_{2,4}^{\pm\pm}}
    \leq2
\end{equation}

\ifthenelse{\boolean{pt-br}}
  {\ptbr
  {Portanto, verificamos que os limites obtidos pela desigualdade de CHSH s\~ao novamente encontrados,
  mas agora atrav\'es dos axiomas de Kolmogorov.}
  }

Therefore, we find that the limits obtained by the CHSH inequality are again found, but now by Kolmogorov's axioms.

\ifthenelse{\boolean{pt-br}}
  {\ptbr
  {Assim, considerando o primeiro e o segundo membros da desigualdade, em que se tem que a express\~ao
  deve ser maior ou igual a 0, temos que tal desigualdade s\'o pode ser ferida caso exista ao menos
  uma das fun\c{c}\~oes de probabilidade com valor suficientemente negativo para que a express\~ao se
  torne negativa. Como sabemos, os axiomas de Kolmogorov afirmam que a fun\c{c}\~ao de probabilidade
  \'e sempre maior ou igual a 0, portanto ferir a desigualdade formada pelo primeiro e o segundo membro
  equivale a ferir o axioma que afirma que $\mathcal{P}_{\dot{Z}_{j,k,l,m}}(z_j,z_k,z_l,z_m)\geq0$.}
  }

Thus, considering the first and second members of the inequality, where we have that the expression
must be greater than or equal to 0, we have that such inequality can only be brokendown if there is
at least one of the probability functions with sufficiently negative value so that expression becomes
negative. As we know, Kolmogorov's axioms assert that the probability function is always greater than
or equal to 0, so to collapse the inequality formed by the first and second terms is to drop the axiom
which asserts that $\mathcal{P}_{\dot{Z}_{j,k,l,m}}(z_j,z_k,z_l,z_m)\geq0$.

\ifthenelse{\boolean{pt-br}}
  {\ptbr
  {J\'a na considera\c{c}\~ao do segundo e terceiro membros da desigualdade, em que a express\~ao deve
  ser menor ou igual a 2, s\'o ser\'a violada se houver um ou mais dos eventos $(z_1,z_2,z_3,z_4)$ em
  que a probabilidade $\mathcal{P}_{\dot{Z}_{1,2,3,4}}(z_1,z_2,z_3,z_4)$ seja suficientemente grande
  para que a express\~ao seja maior que 2. Tal viola\c{c}\~ao s\'o pode acontecer, se a probabilidade
  de ocorrer qualquer resultado (que nada mais \'e que a soma das probabilidades de todos os resutlados)
  ter\'a que ser maior que 1, o que viola outro axioma de Kolmogorov.}
  }

Already considering of the second and third members of the inequality, where the expression must be
less than or equal to 2, it will only be violated if there is one or more of the events, $(z_1,z_2,z_3,z_4)$,
where the probability $\mathcal{P}_{\dot{Z}_{1,2,3,4}}(z_1,z_2,z_3,z_4)$  is large enough so that the
expression is greater than 2. Such violation alone can happen if the probability of any result occurs
(which is nothing more than the sum of the probabilities of all answers) must be greater than 1, which
violates another Kolmogorov axiom.

\ifthenelse{\boolean{pt-br}}
  {\ptbr
  {Assim, se houver viola\c{c}\~ao da desigualdade de CHSH, ent\~ao deve ter tido a viola\c{c}\~ao de
  um dos axiomas de Kolmogorov.}
  }

Thus, if there is a violation of the CHSH inequality, then it must have had the violation of one of
Kolmogorov's axioms.

\begin{widetext}

\begin{eqnarray}\nonumber
  &&\Big|\mathcal{E}_{\dot{Z}_{1,3}}\!(Z_1\!\cdot\!Z_3)\!-\!\mathcal{E}_{\dot{Z}_{1,4}}\!(Z_1\!\cdot\!Z_4)
  +\mathcal{E}_{\dot{Z}_{2,3}}\!(Z_2\!\cdot\!Z_3)\!+\!\mathcal{E}_{\dot{Z}_{2,4}}\!(Z_2\!\cdot\!Z_4)\Big|
    > 2\Rightarrow
  \\%\nonumber
  &&\Rightarrow\Bigg(\exists_{(z_1,z_2,z_3,z_4)}(\mathcal{P}_{\dot Z_{1,2,3,4}}(z_1,z_2,z_3,z_4)<0)
  \vee
  \bigg(\sum\limits_{(z_1,z_2,z_3,z_4)\in\{-1,+1\}^4}(\mathcal{P}_{\dot Z_{1,2,3,4}}(z_1,z_2,z_3,z_4))>1\bigg)\Bigg)
\end{eqnarray}

\end{widetext}

\ifthenelse{\boolean{pt-br}}
  {\ptbr
  {Nesta se\c{c}\~ao, conluimos que os limites dados pela desigualdade de CHSH coincidem com os limites
  obtidos pelos axiomas de Kolmogorov. Assim, s� se pode ferir a desigualdade de CHSH s� os axiomas
  de Kolmogorv forem feridos.}
  }

In this section, we conclude that the boundaries given by the CHSH inequality coincide with the limits
obtained by the Kolmogorov axioms. Thus, only the inequality of CHSH can be brokendown only if Kolmogorov's
axioms are broken.

\ifthenelse{\boolean{pt-br}}
  {\ptbr
  {Sistema linear de equa\c{c}\~oes}
  }

\section{Linear system of equations}
\label{sec:Sistema}

\ifthenelse{\boolean{pt-br}}
  {\ptbr
  {Nesta se\c{c}\~ao, abordaremos a quest\~ao relacionada \`a determina\c{c}\~ao dos poss\'iveis valores
  das probabilidades $\mathcal{P}_{\dot{Z}_{1,2,3,4}}(z_1,z_2,z_3,z_4)$. Estudaremos se o sistema de
  equa\c{c}\~oes fornecidas pela teoria trata-se de um sistema resolv\'ivel ou n\~ao. Se n\~ao for poss\'ivel
  de resolver, ent\~ao n\�o far\'a sentido supor a exist\^encia das probabilidades 
  $\mathcal{P}_{\dot{Z}_{1,2,3,4}}(z_1,z_2,z_3,z_4)$, assim teremos uma inconsist\^encia, impossibilitando
  a modelagem probabil\'istica do problema.}
  }

In this section, we will address the question related to the determination of the possible values of
probabilities $\mathcal{P}_{\dot{Z}_{1,2,3,4}}(z_1,z_2,z_3,z_4)$. We will study if the system of equations
provided by the theory is a system that can be solved or not. If it is not possible to solve, then it
will not make sense to suppose the existence of the probabilities $\mathcal{P}_{\dot{Z}_{1,2,3,4}}(z_1,z_2,z_3,z_4)$,
So we will have an inconsistency, making it impossible to model the probabilistic problem.

\ifthenelse{\boolean{pt-br}}
  {\ptbr
  {Da teoria, obtemos o seguinte sistema}
  }

From theory, we obtain the following system:

\begin{widetext}

\begin{equation}
\label{eq:CHSHmatriz}
  \underbrace{\left[
  \begin{smallmatrix}%{cccccccccccccccc}
    1 & 1 & 0 & 0 & 1 & 1 & 0 & 0 & 0 & 0 & 0 & 0 & 0 & 0 & 0 & 0 \\
    0 & 0 & 1 & 1 & 0 & 0 & 1 & 1 & 0 & 0 & 0 & 0 & 0 & 0 & 0 & 0 \\
    0 & 0 & 0 & 0 & 0 & 0 & 0 & 0 & 1 & 1 & 0 & 0 & 1 & 1 & 0 & 0 \\
    0 & 0 & 0 & 0 & 0 & 0 & 0 & 0 & 0 & 0 & 1 & 1 & 0 & 0 & 1 & 1 \\
    1 & 0 & 1 & 0 & 1 & 0 & 1 & 0 & 0 & 0 & 0 & 0 & 0 & 0 & 0 & 0 \\
    0 & 1 & 0 & 1 & 0 & 1 & 0 & 1 & 0 & 0 & 0 & 0 & 0 & 0 & 0 & 0 \\
    0 & 0 & 0 & 0 & 0 & 0 & 0 & 0 & 1 & 0 & 1 & 0 & 1 & 0 & 1 & 0 \\
    0 & 0 & 0 & 0 & 0 & 0 & 0 & 0 & 0 & 1 & 0 & 1 & 0 & 1 & 0 & 1 \\
    1 & 1 & 0 & 0 & 0 & 0 & 0 & 0 & 1 & 1 & 0 & 0 & 0 & 0 & 0 & 0 \\
    0 & 0 & 1 & 1 & 0 & 0 & 0 & 0 & 0 & 0 & 1 & 1 & 0 & 0 & 0 & 0 \\
    0 & 0 & 0 & 0 & 1 & 1 & 0 & 0 & 0 & 0 & 0 & 0 & 1 & 1 & 0 & 0 \\
    0 & 0 & 0 & 0 & 0 & 0 & 1 & 1 & 0 & 0 & 0 & 0 & 0 & 0 & 1 & 1 \\
    1 & 0 & 1 & 0 & 0 & 0 & 0 & 0 & 1 & 0 & 1 & 0 & 0 & 0 & 0 & 0 \\
    0 & 1 & 0 & 1 & 0 & 0 & 0 & 0 & 0 & 1 & 0 & 1 & 0 & 0 & 0 & 0 \\
    0 & 0 & 0 & 0 & 1 & 0 & 1 & 0 & 0 & 0 & 0 & 0 & 1 & 0 & 1 & 0 \\
    0 & 0 & 0 & 0 & 0 & 1 & 0 & 1 & 0 & 0 & 0 & 0 & 0 & 1 & 0 & 1 \\
  \end{smallmatrix}
  \right]}_{\ddot\Sigma:=}
  \cdot
  \underbrace{
  \left[
  \begin{smallmatrix}
    P_{1,2,3,4}^{----}\\
    P_{1,2,3,4}^{---+}\\
    P_{1,2,3,4}^{--+-}\\
    P_{1,2,3,4}^{--++}\\
    P_{1,2,3,4}^{-+--}\\
    P_{1,2,3,4}^{-+-+}\\
    P_{1,2,3,4}^{-++-}\\
    P_{1,2,3,4}^{-+++}\\
    P_{1,2,3,4}^{+---}\\
    P_{1,2,3,4}^{+--+}\\
    P_{1,2,3,4}^{+-+-}\\
    P_{1,2,3,4}^{+-++}\\
    P_{1,2,3,4}^{++--}\\
    P_{1,2,3,4}^{++-+}\\
    P_{1,2,3,4}^{+++-}\\
    P_{1,2,3,4}^{++++}\\
  \end{smallmatrix}
  \right]}_{\ddot P:=}
  =
  \left[
  \begin{smallmatrix}
    P_{1,3}^{--}\\
    P_{1,3}^{-+}\\
    P_{1,3}^{+-}\\
    P_{1,3}^{++}\\
    P_{1,4}^{--}\\
    P_{1,4}^{-+}\\
    P_{1,4}^{+-}\\
    P_{1,4}^{++}\\
    P_{2,3}^{--}\\
    P_{2,3}^{-+}\\
    P_{2,3}^{+-}\\
    P_{2,3}^{++}\\
    P_{2,4}^{--}\\
    P_{2,4}^{-+}\\
    P_{2,4}^{+-}\\
    P_{2,4}^{++}\\
  \end{smallmatrix}
  \right]
  =
  \underbrace{
  \left[
  \begin{smallmatrix}%{c}
    \frac{1}{2} \cdot \sin ^2( \bar{\theta} _{1,3} ) \\
    \frac{1}{2} \cdot \cos ^2( \bar{\theta} _{1,3} ) \\
    \frac{1}{2} \cdot \cos ^2( \bar{\theta} _{1,3} ) \\
    \frac{1}{2} \cdot \sin ^2( \bar{\theta} _{1,3} ) \\
    \frac{1}{2} \cdot \sin ^2( \bar{\theta} _{1,4} ) \\
    \frac{1}{2} \cdot \cos ^2( \bar{\theta} _{1,4} ) \\
    \frac{1}{2} \cdot \cos ^2( \bar{\theta} _{1,4} ) \\
    \frac{1}{2} \cdot \sin ^2( \bar{\theta} _{1,4} ) \\
    \frac{1}{2} \cdot \sin ^2( \bar{\theta} _{2,3} ) \\
    \frac{1}{2} \cdot \cos ^2( \bar{\theta} _{2,3} ) \\
    \frac{1}{2} \cdot \cos ^2( \bar{\theta} _{2,3} ) \\
    \frac{1}{2} \cdot \sin ^2( \bar{\theta} _{2,3} ) \\
    \frac{1}{2} \cdot \sin ^2( \bar{\theta} _{2,4} ) \\
    \frac{1}{2} \cdot \cos ^2( \bar{\theta} _{2,4} ) \\
    \frac{1}{2} \cdot \cos ^2( \bar{\theta} _{2,4} ) \\
    \frac{1}{2} \cdot \sin ^2( \bar{\theta} _{2,4} ) \\
  \end{smallmatrix}
  \right]}_{\ddot p:=}
\end{equation}

\end{widetext}

\ifthenelse{\boolean{pt-br}}
  {\ptbr
  {que, resumidamente, ser\'a reescrita como}
  }

\noindent
which will be briefly rewritten as

\begin{equation}\nonumber
  \ddot{\Sigma}\cdot\ddot{P}=\ddot{p}
\end{equation}

\ifthenelse{\boolean{pt-br}}
  {\ptbr
  {Para descobrirmos se o sistema \� resolv\'ivel, temos que calcular o rank da matriz de coeficientes
  e o da matriz estendida, se ambos os rank forem iguais, ent�o o sistema \'e poss\'ivel de resolver,
  sen\~ao ser\'a um sistema imposs\'ivel de se resolver.}
  }

To find out if the system is solvable, we have to calculate the rank of the coefficient matrix and the
matrix of the extended matrix. If both rank are equal, then the system is possible to solve. If it's
not, the system is simply not solvable.

\ifthenelse{\boolean{pt-br}}
  {\ptbr
  {Se o rank for igual ao n\'umero de vari\'aveis, ent\~ao o sistema \'e poss\'ivel e determinado (ou
  seja, existira uma \'unica solu\c{c}\~ao), mas se o rank for menor que o n\'umero de vari\'aveis,
  ent\~ao o sistema \'e poss\'ivel e indeterminado (ou seja, algumas das vari\'aveis ficar\'a em fun\c{c}\~ao
  de outras, possibilitando uma infinidade de solu\c{c}\~oes).}
  }

If the rank is equal to the number of variables, then the system is possible and determined (that is,
there would be a single solution), but if the rank is less than the number of variables, then the system
is possible and undetermined (that is, Some of the variables will be in function of others, allowing
an infinity of solutions).

\ifthenelse{\boolean{pt-br}}
  {\ptbr
  {Nesse caso, as nossas vari\'aveis s\~ao as fun\c{c}\~oes de probabilidade $\mathcal{P}_{\dot{Z}_{1,2,3,4}}(z_1,z_2,z_3,z_4)$,
  como para cada $z$ existem dois valores ($-1$ e $+1$), ent\~ao, com quatro vari\'aveis temos 16 fun\c{c}\~oes
  de probabilidade. Poder\'iamos pensar que s\'o temos 15 vari\'aveis, pois se 15 dessas fun\c{c}\~oes
  estiverem definidas, o valor da \'ultima ser\'a definida subtraindo de 1 a soma dos valores das outras
  quinze, j\'a que a soma de todas deve ser igual a 1, conforme um dos axiomas de Kolmogorov, mas estamos
  deixando aberto a possibilidade do axioma ser violado. Assim todas as 16 fun\c{c}\~oes estar\~ao livres
  para assumir os valores que resulte da solu\c{c}\~ao do sistema, se o sistema for poss\'ivel de se
  resolver.}
  }

In this case, our variables are the probability functions$\mathcal{P}_{\dot{Z}_{1,2,3,4}}(z_1,z_2,z_3,z_4)$,
as for every $z$ there are two values ($-1$ and $+1$), so with four variables we have 16 probability
functions. We could think that we have only 15 variables, because if 15 of these functions are defined,
the value of the last one will be defined subtracting from 1 the sum of the values of the other fifteen,
since the sum of all the variables must be equal to 1, according to one of the axioms Kolmogorov, but
we are open-ended to the possibility of the axiom being violated. Thus all 16 functions will be free
to assume the values that result from the solution of the system, if the system is possible to solve.

\ifthenelse{\boolean{pt-br}}
  {\ptbr
  {Calculamos o rank fazendo o escalonamento das matrizes, e a matriz que escalona a matriz de coeficientes
  \'e a matriz}
  }

We calculate the rank by doing the scheduling of the matrices, and the matrix that scales the matrix
of coefficients is the matrix:

\begin{gather}\nonumber
  \eta:=
  \left[
  \begin{smallmatrix}
    0 & 0 & 0 &  1 & 0 & 0 & -1 &  0 & 0 & -1 & 0 &  0 &  1 &  0 &  0 &  0 \\
    0 & 0 & 0 &  0 & 0 & 0 &  0 & -1 & 0 &  0 & 0 &  0 &  0 &  1 &  0 &  0 \\
    0 & 0 & 0 & -1 & 0 & 0 &  0 &  0 & 0 &  1 & 0 &  0 &  0 &  0 &  0 &  0 \\
    0 & 0 & 0 &  0 & 0 & 0 &  0 &  0 & 0 &  0 & 0 & -1 &  0 &  0 &  1 &  0 \\
    0 & 0 & 0 &  0 & 0 & 0 &  0 &  0 & 0 &  0 & 0 &  0 &  0 &  0 &  0 &  1 \\
    0 & 0 & 0 &  0 & 0 & 0 &  0 &  0 & 0 &  0 & 0 &  1 &  0 &  0 &  0 &  0 \\
    0 & 0 & 0 & -1 & 0 & 0 &  1 &  0 & 0 &  0 & 0 &  0 &  0 &  0 &  0 &  0 \\
    0 & 0 & 0 &  0 & 0 & 0 &  0 &  1 & 0 &  0 & 0 &  0 &  0 &  0 &  0 &  0 \\
    0 & 0 & 0 &  1 & 0 & 0 &  0 &  0 & 0 &  0 & 0 &  0 &  0 &  0 &  0 &  0 \\
    1 & 0 & 0 & -1 & 0 & 0 &  1 &  1 & 0 &  1 & 0 &  1 & -1 & -1 & -1 & -1 \\
    0 & 1 & 0 &  1 & 0 & 0 &  0 &  0 & 0 & -1 & 0 & -1 &  0 &  0 &  0 &  0 \\
    0 & 0 & 1 &  1 & 0 & 0 & -1 & -1 & 0 &  0 & 0 &  0 &  0 &  0 &  0 &  0 \\
    0 & 0 & 0 &  0 & 1 & 0 &  1 &  0 & 0 &  0 & 0 &  0 & -1 &  0 & -1 &  0 \\
    0 & 0 & 0 &  0 & 0 & 1 &  0 &  1 & 0 &  0 & 0 &  0 &  0 & -1 &  0 & -1 \\
    0 & 0 & 0 &  0 & 0 & 0 &  0 &  0 & 1 &  1 & 0 &  0 & -1 & -1 &  0 &  0 \\
    0 & 0 & 0 &  0 & 0 & 0 &  0 &  0 & 0 &  0 & 1 &  1 &  0 &  0 & -1 & -1 \\
  \end{smallmatrix}
  \right]
\end{gather}

\ifthenelse{\boolean{pt-br}}
  {\ptbr
  {que, quando multiplicada, pela esquerda, com a matriz $\ddot{\Sigma}$ resulta no seguinte escalonamento}
  }

\noindent
which, when multiplied from the left, with the $\ddot{\Sigma}$ matrix results in the following scaling

\begin{gather}\nonumber
  \eta\cdot\ddot\Sigma
  =
  \left[
  \begin{smallmatrix}
    1 & 0 & 0 & -1 & 0 & 0 & 0 & 0 & 0 & 0 & 0 & 0 & -1 & 0 & 0 & 1 \\
    0 & 1 & 0 & 1 & 0 & 0 & 0 & 0 & 0 & 0 & 0 & 0 & 0 & -1 & 0 & -1 \\
    0 & 0 & 1 & 1 & 0 & 0 & 0 & 0 & 0 & 0 & 0 & 0 & 0 & 0 & -1 & -1 \\
    0 & 0 & 0 & 0 & 1 & 0 & 0 & -1 & 0 & 0 & 0 & 0 & 1 & 0 & 0 & -1 \\
    0 & 0 & 0 & 0 & 0 & 1 & 0 & 1 & 0 & 0 & 0 & 0 & 0 & 1 & 0 & 1 \\
    0 & 0 & 0 & 0 & 0 & 0 & 1 & 1 & 0 & 0 & 0 & 0 & 0 & 0 & 1 & 1 \\
    0 & 0 & 0 & 0 & 0 & 0 & 0 & 0 & 1 & 0 & 0 & -1 & 1 & 0 & 0 & -1 \\
    0 & 0 & 0 & 0 & 0 & 0 & 0 & 0 & 0 & 1 & 0 & 1 & 0 & 1 & 0 & 1 \\
    0 & 0 & 0 & 0 & 0 & 0 & 0 & 0 & 0 & 0 & 1 & 1 & 0 & 0 & 1 & 1 \\
    0 & 0 & 0 & 0 & 0 & 0 & 0 & 0 & 0 & 0 & 0 & 0 & 0 & 0 & 0 & 0 \\
    0 & 0 & 0 & 0 & 0 & 0 & 0 & 0 & 0 & 0 & 0 & 0 & 0 & 0 & 0 & 0 \\
    0 & 0 & 0 & 0 & 0 & 0 & 0 & 0 & 0 & 0 & 0 & 0 & 0 & 0 & 0 & 0 \\
    0 & 0 & 0 & 0 & 0 & 0 & 0 & 0 & 0 & 0 & 0 & 0 & 0 & 0 & 0 & 0 \\
    0 & 0 & 0 & 0 & 0 & 0 & 0 & 0 & 0 & 0 & 0 & 0 & 0 & 0 & 0 & 0 \\
    0 & 0 & 0 & 0 & 0 & 0 & 0 & 0 & 0 & 0 & 0 & 0 & 0 & 0 & 0 & 0 \\
    0 & 0 & 0 & 0 & 0 & 0 & 0 & 0 & 0 & 0 & 0 & 0 & 0 & 0 & 0 & 0 \\
  \end{smallmatrix}
  \right]
\end{gather}

\ifthenelse{\boolean{pt-br}}
  {\ptbr
  {como h\'a 9 linhas n\~ao zeradas, temos que o rank \'e 9.}
  }

\noindent
as there are 9 unzero lines, we have that the rank is 9.

\ifthenelse{\boolean{pt-br}}
  {\ptbr
  {Usando novamente a matriz $\eta$, calculamos o rank da matriz aumentada 
  $\begin{bmatrix}
  \ddot\Sigma & \ddot p \\
  \end{bmatrix}$
  (matriz obtida ao se acrescentar \`a matriz de coeficientes $\ddot{\Sigma}$ a matriz $\ddot{p}$) obtemos}
  }

Using the matrix   eta again, we calculate the rank of the increased matrix
$ \begin{bmatrix}
\ddot\Sigma & \ddot p \\
\end {bmatrix}$
(That is, the matrix obtained by adding to the matrix of coefficients  $ \ddot{\Sigma} $ the matrix
$ \ddot{p}$), obtaining

\begin{widetext}

\begin{equation}\nonumber
  \eta\cdot
  \begin{bmatrix}
  \ddot\Sigma & \ddot p \\
  \end{bmatrix}
  =
  \left[
  \begin{smallmatrix}%{ccccccccccccccccc}
    1 & 0 & 0 & -1 & 0 & 0 & 0 & 0 & 0 & 0 & 0 & 0 & -1 & 0 & 0 & 1 
    & \frac{\sin^2 (\bar{\theta}_{1,3})-\cos ^2(\bar{\theta}_{1,4})-\cos ^2(\bar{\theta}_{2,3})+\sin^2(\bar{\theta}_{2,4})}{2}
    \\
    0 & 1 & 0 & 1 & 0 & 0 & 0 & 0 & 0 & 0 & 0 & 0 & 0 & -1 & 0 & -1 
    & \frac{\cos^2(\bar{\theta}_{2,4})-\sin^2(\bar{\theta}_{1,4})}{2}
    \\
    0 & 0 & 1 & 1 & 0 & 0 & 0 & 0 & 0 & 0 & 0 & 0 & 0 & 0 & -1 & -1 
    & \frac{\cos^2(\bar{\theta}_{2,3})-\sin^2(\bar{\theta}_{1,3})}{2} 
    \\
    0 & 0 & 0 & 0 & 1 & 0 & 0 & -1 & 0 & 0 & 0 & 0 & 1 & 0 & 0 & -1 
    & \frac{\cos^2(\bar{\theta}_{2,4})-\sin^2(\bar{\theta}_{2,3})}{2} 
    \\
    0 & 0 & 0 & 0 & 0 & 1 & 0 & 1 & 0 & 0 & 0 & 0 & 0 & 1 & 0 & 1 
    & \frac{\sin^2(\bar{\theta}_{2,4})}{2} 
    \\
    0 & 0 & 0 & 0 & 0 & 0 & 1 & 1 & 0 & 0 & 0 & 0 & 0 & 0 & 1 & 1 
    & \frac{\sin^2(\bar{\theta}_{2,3})}{2} 
    \\
    0 & 0 & 0 & 0 & 0 & 0 & 0 & 0 & 1 & 0 & 0 & -1 & 1 & 0 & 0 & -1 
    & \frac{\cos^2(\bar{\theta}_{1,4})-\sin^2 (\bar{\theta}_{1,3})}{2} 
    \\
    0 & 0 & 0 & 0 & 0 & 0 & 0 & 0 & 0 & 1 & 0 & 1 & 0 & 1 & 0 & 1 
    & \frac{\sin^2(\bar{\theta}_{1,4})}{2} 
    \\
    0 & 0 & 0 & 0 & 0 & 0 & 0 & 0 & 0 & 0 & 1 & 1 & 0 & 0 & 1 & 1 
    & \frac{\sin^2(\bar{\theta}_{1,3})}{2} 
    \\
    0 & 0 & 0 & 0 & 0 & 0 & 0 & 0 & 0 & 0 & 0 & 0 & 0 & 0 & 0 & 0 & 0 \\
    0 & 0 & 0 & 0 & 0 & 0 & 0 & 0 & 0 & 0 & 0 & 0 & 0 & 0 & 0 & 0 & 0 \\
    0 & 0 & 0 & 0 & 0 & 0 & 0 & 0 & 0 & 0 & 0 & 0 & 0 & 0 & 0 & 0 & 0 \\
    0 & 0 & 0 & 0 & 0 & 0 & 0 & 0 & 0 & 0 & 0 & 0 & 0 & 0 & 0 & 0 & 0 \\
    0 & 0 & 0 & 0 & 0 & 0 & 0 & 0 & 0 & 0 & 0 & 0 & 0 & 0 & 0 & 0 & 0 \\
    0 & 0 & 0 & 0 & 0 & 0 & 0 & 0 & 0 & 0 & 0 & 0 & 0 & 0 & 0 & 0 & 0 \\
    0 & 0 & 0 & 0 & 0 & 0 & 0 & 0 & 0 & 0 & 0 & 0 & 0 & 0 & 0 & 0 & 0 \\
  \end{smallmatrix}
  \right]
\end{equation}

\end{widetext}

\ifthenelse{\boolean{pt-br}}
  {\ptbr
  {onde observamos novamente que 9 linhas n\~ao foram zeradas, portanto o rank tamb\'em \'e 9. Assim,
  temos que o sistema \'e poss\'ivel de se resolver.}
  }

\noindent
where we again note that nine lines have not been zeroed, so rank is also nine. Thus, we have that the
system is possible to solve.

\ifthenelse{\boolean{pt-br}}{
  \ptbr{
  Uma vez que o rank \'e 9, temos que as probabilidades n\~ao s\~ao determinadas de forma un\'ivoca,
  sendo portanto dependentes das probabilidades que ser\~ao atribu�das aos 7 dos 16 eventos relacionados
  \`as vari\'aveis $(Z_1,Z_2,Z_3,Z_4)$. Escolheremos os seguintes eventos}
  }

Since the rank is 9, we have that the probabilities are not determined univocally, and therefore are
dependent on the probabilities that will be attributed to the 7 of the 16 events related to the variables
$(Z_1,Z_2,Z_3,Z_4)$. We will choose the following events

\begin{eqnarray}\nonumber
  (z_1,z_2,z_3,z_4)\in\{&&(-1,-1,+1,+1),(-1,+1,+1,+1),
  \\\nonumber
  &&(+1,-1,+1,+1),(+1,+1,-1,-1),
  \\\nonumber
  &&(+1,+1,-1,+1),(+1,+1,+1,-1),
  \\\nonumber
  &&(+1,+1,+1,+1)\}
\end{eqnarray}

\ifthenelse{\boolean{pt-br}}{
  \ptbr{
  Para esses eventos, se definirmos as probabilidades $\mathcal{P}_{\dot Z_{1,2,3,4}}(z_1,z_2,z_3,z_4)$
  e as probabilidades $\mathcal{P}_{\dot Z_{j,k}}(z_j,z_k)$ e $\mathcal{P}_{\dot Z_{j,k}}(z_j,-z_k)$
  (sendo que $(j,k)\in\{(1,3),(1,4),(2,3),(2,4)\}$), os demais eventos poder\~ao ser encontrados.}
  }

For these events, if we define the probabilities $\mathcal{P}_{\dot Z_{1,2,3,4}}(z_1,z_2,z_3,z_4)$
and the probabilities $\mathcal{P}_{\dot Z_{j,k}}(z_j,z_k)$ e $\mathcal{P}_{\dot Z_{j,k}}(z_j,-z_k)$
(and $(j,k)\in\{(1,3),(1,4),(2,3),(2,4)\}$), other events can be found.

\ifthenelse{\boolean{pt-br}}{
  \ptbr{
  Para facilitar, vamos colocar um \textperiodcentered\ nas probabilidades em que estamos atribuindo
  os valores (esses pontos \textperiodcentered\ s\~ao apenas para marca\c{c}\~ao, servir\~ao para indicar
  de qual conjunto de equa\c{c}\~oes estamos determinando os valores de $P$ e, portanto, colocamos um
  ponto para os valores determinados a partir do primeiro conjunto de f\'ormulas, depois colocamos um
  segundo ponto para os vlores determinados a partir do segundo conjunto de f\'ormulas, e assim por
  diante). Dessa forma temos as seguintes probabilidades com seus valores j\'a determinados}
  }

To make it simplier, let's put a \textperiodcentered\ in the probabilities in which we are assigning
the values (these points \textperiodcentered\ are only for marking they, will serve to indicate from
which set of equations we are determining the values of $ P $ and, therefore, we put a point for the
values determined from the first set of formulas. Then we put a second point for the values determined
a from the second set of formulas, and so on). In this way we have the following probabilities with
their already determined values

\begin{eqnarray}\nonumber
  \dot P_{1,2,3,4}^{--++}
  ,\ \dot P_{1,2,3,4}^{-+++}
  ,\ \dot P_{1,2,3,4}^{+-++}
  ,\ \dot P_{1,2,3,4}^{++--}
  ,
  \\\nonumber
  \ \dot P_{1,2,3,4}^{++-+}
  ,\ \dot P_{1,2,3,4}^{+++-}
  ,\ \dot P_{1,2,3,4}^{++++}
  ,\\
  \ \dot P_{1,3}^{++}
  ,\ \dot P_{1,4}^{++}
  ,\ \dot P_{2,3}^{++}
  ,\ \dot P_{2,4}^{++}
\label{eq:probCHSHcte}
\end{eqnarray}

\ifthenelse{\boolean{pt-br}}{
  \ptbr{
  Lembrando que $P_{j,k}^{++}=P_{j,k}^{--}=\tfrac{1}{2}-P_{j,k}^{+-}=\tfrac{1}{2}-P_{j,k}^{-+}$, dessa
  forma conseguimos determinar todas as outras probabilidades de $(Z_j,Z_k)$.}
  }

Remembering that $P_{j,k}^{++}=P_{j,k}^{--}=\tfrac{1}{2}-P_{j,k}^{+-}=\tfrac{1}{2}-P_{j,k}^{-+}$, in
this way we can determine all the other probabilities of $(Z_j,Z_k)$.

\ifthenelse{\boolean{pt-br}}{
  \ptbr{
  Uma vez que
  }}

Once

\begin{eqnarray}
  \dot P_{1,3}^{++}
  =P_{1,2,3,4}^{+-+-}+\dot P_{1,2,3,4}^{+-++}+\dot P_{1,2,3,4}^{+++-}+\dot P_{1,2,3,4}^{++++}
\label{eq:probCHSHcte2.1}
  \\
  \dot P_{2,3}^{++}
  =P_{1,2,3,4}^{-++-}+\dot P_{1,2,3,4}^{+++-}+\dot P_{1,2,3,4}^{-+++}+\dot P_{1,2,3,4}^{++++}
\label{eq:probCHSHcte2.2}
  \\
  \dot P_{1,4}^{++}
  =P_{1,2,3,4}^{+--+}+\dot P_{1,2,3,4}^{++-+}+\dot P_{1,2,3,4}^{+-++}+\dot P_{1,2,3,4}^{++++}
\label{eq:probCHSHcte2.3}
  \\
  \dot P_{2,4}^{++}
  =P_{1,2,3,4}^{-+-+}+\dot P_{1,2,3,4}^{-+++}+\dot P_{1,2,3,4}^{++-+}+\dot P_{1,2,3,4}^{++++}
\label{eq:probCHSHcte2.4}
\end{eqnarray}

\ifthenelse{\boolean{pt-br}}{
  \ptbr{
  podemos determinar $P_{1,2,3,4}^{+-+-}$, $P_{1,2,3,4}^{-++-}$, $P_{1,2,3,4}^{+--+}$ e $P_{1,2,3,4}^{-+-+}$
  a partir de (\ref{eq:probCHSHcte}). Vamos colocar dois \textperiodcentered\ (pontos) em cima dessas
  probabilidades.}
  }

\noindent
we can determine $P_{1,2,3,4}^{+-+-}$, $P_{1,2,3,4}^{-++-}$, $P_{1,2,3,4}^{+--+}$ e $P_{1,2,3,4}^{-+-+}$
from (\ref{eq:probCHSHcte}). Let's put two \textperiodcentered\ (points) over these probabilities.

\ifthenelse{\boolean{pt-br}}{
  \ptbr{
  Agora, temos que
  }}

Now we have to

\begin{eqnarray}
  \dot P_{1,3}^{+-}
  =P_{1,2,3,4}^{+---}+\ddot P_{1,2,3,4}^{+--+}+\dot P_{1,2,3,4}^{++--}+\dot P_{1,2,3,4}^{++-+}
\label{eq:probCHSHcte3.1}
  \\
  \dot P_{2,3}^{+-}
  =P_{1,2,3,4}^{-+--}+\ddot P_{1,2,3,4}^{-+-+}+\dot P_{1,2,3,4}^{++--}+\dot P_{1,2,3,4}^{++-+}
\label{eq:probCHSHcte3.2}
  \\
  \dot P_{2,4}^{-+}
  =P_{1,2,3,4}^{---+}+\dot P_{1,2,3,4}^{--++}+\ddot P_{1,2,3,4}^{+--+}+\dot P_{1,2,3,4}^{+-++}
\label{eq:probCHSHcte3.3}
\end{eqnarray}

\ifthenelse{\boolean{pt-br}}{
  \ptbr{
  dessa forma determinamos $P_{1,2,3,4}^{+---}$, $P_{1,2,3,4}^{-+--}$ e $P_{1,2,3,4}^{---+}$ a partir
  das equa\c{c}\~oes (\ref{eq:probCHSHcte}-\ref{eq:probCHSHcte2.4}). Vamos colocar
  tr\^es \textperiodcentered\ (pontos).
  }}

\noindent
in this way we determine $P_{1,2,3,4}^{+---}$, $P_{1,2,3,4}^{-+--}$ and $P_{1,2,3,4}^{---+}$ starting
from the equations (\ref{eq:probCHSHcte}-\ref{eq:probCHSHcte2.4}). Let's put
three \textperiodcentered\ (points).

\ifthenelse{\boolean{pt-br}}{
  \ptbr{
  Finalmente, temos
  }}

Finally, we have

\begin{eqnarray}\nonumber
  \dot P_{1,3}^{-+}
  =P_{1,2,3,4}^{--+-}+\dot P_{1,2,3,4}^{--++}+\ddot P_{1,2,3,4}^{-++-}+\dot P_{1,2,3,4}^{-+++}
  \\\nonumber
  \dot P_{1,3}^{--}
  =P_{1,2,3,4}^{----}+\dddot P_{1,2,3,4}^{---+}+\dddot P_{1,2,3,4}^{-+--}+\ddot P_{1,2,3,4}^{-+-+}
\end{eqnarray}

\ifthenelse{\boolean{pt-br}}{
  \ptbr{
  onde determinamos os valores de $P_{1,2,3,4}^{--+-}$ e $P_{1,2,3,4}^{----}$ a partir das equa\c{c}\~oes
  (\ref{eq:probCHSHcte}-\ref{eq:probCHSHcte3.3}). Valor colocar quatro \textperiodcentered\ (pontos)
  em cima dessas probabilidades.
  }}

\noindent
where we determine the values of $P_{1,2,3,4}^{--+-}$ and $P_{1,2,3,4}^{----}$ from the equations
(\ref{eq:probCHSHcte}-\ref{eq:probCHSHcte3.3}). Let's put four \textperiodcentered\ (points) over these
probabilities.

\ifthenelse{\boolean{pt-br}}{
  \ptbr{
  Dessa forma, concluimos a determina\c{c}\~ao dos valores das probabilidades de todos os 16 eventos
  poss\'iveis (que colocaremos na forma matricial apenas por comodidade) 
  }}

In this way, we conclude that the determination of the probability values of all 16 possible events
(which we will put in the matrix form only for convenience)

\begin{equation}
\label{eq:probCHSHcteMatriz}
  \left[\begin{matrix}
    \ddddot P_{1,2,3,4}^{----} &\dddot P_{1,2,3,4}^{---+} &\ddddot P_{1,2,3,4}^{--+-} &\dot P_{1,2,3,4}^{--++}
    \\
    \dddot P_{1,2,3,4}^{-+--}  &\ddot P_{1,2,3,4}^{-+-+}  &\ddot P_{1,2,3,4}^{-++-}   &\dot P_{1,2,3,4}^{-+++}
    \\
    \dddot P_{1,2,3,4}^{+---}  &\ddot P_{1,2,3,4}^{+--+}  &\ddot P_{1,2,3,4}^{+-+-}   &\dot P_{1,2,3,4}^{+-++}
    \\
    \dot P_{1,2,3,4}^{++--}    &\dot P_{1,2,3,4}^{++-+}   &\dot P_{1,2,3,4}^{+++-}    &\dot P_{1,2,3,4}^{++++}
  \end{matrix}\right]
\end{equation}

\ifthenelse{\boolean{pt-br}}{
  \ptbr{
  a partir das probabilidades (\ref{eq:probCHSHcte}). Relembrando que os pontos colocados acima de $P$
  s\~ao apenas para marca\c{c}\~ao.
  }}

\noindent
from these probabilities (\ref{eq:probCHSHcte}). Recalling that the points placed above $ P $ are just
for tagging.

\ifthenelse{\boolean{pt-br}}{
  \ptbr{
  Para o exemplo calculado em (\ref{eq:CHSHex}), onde temos
  }}

For the example calculated in (\ref{eq:CHSHex}), where we have

\begin{equation}\nonumber
  \dot P_{j,k}^{++}=\tfrac{\cos^2(\bar\theta_{j,k})}{2}
  ,\ ({\bar\theta}_{1,3},{\bar\theta}_{1,4},{\bar\theta}_{2,3},{\bar\theta}_{2,4})
  =(\tfrac{2\cdot\pi}{3},\pi,\tfrac{\pi}{3},\tfrac{2\cdot\pi}{3})
\end{equation}

\ifthenelse{\boolean{pt-br}}{
  \ptbr{
  a partir da\'i encontramos
  }}

\noindent
from there we find

\begin{equation}\nonumber
  \begin{array}{llll}
    \dot P_{1,3}^{--}=\dot P_{1,3}^{++}
    &=\tfrac{1}{8}
    ,&
    \dot P_{1,3}^{-+}=\dot P_{1,3}^{+-}
    &=\tfrac{3}{8}
    ,\\
    \dot P_{1,4}^{--}=\dot P_{1,4}^{++}
    &=\tfrac{1}{2}
    ,&
    \dot P_{1,4}^{-+}=\dot P_{1,4}^{+-}
    &=0
    ,\\
    \dot P_{2,3}^{--}=\dot P_{2,3}^{++}
    &=\tfrac{1}{8}
    ,&
    \dot P_{2,3}^{-+}=\dot P_{2,3}^{+-}
    &=\tfrac{3}{8}
    ,\\
    \dot P_{2,4}^{--}=\dot P_{2,4}^{++}
    &=\tfrac{1}{8}
    ,&
    \dot P_{2,4}^{-+}=\dot P_{2,4}^{+-}
    &=\tfrac{3}{8}
  \end{array}
\end{equation}

\ifthenelse{\boolean{pt-br}}{
  \ptbr{
  ou seja
  }}

\noindent
that is

\begin{equation}\noindent
  \ddot p=
  \left[
  \begin{smallmatrix}
    \frac{1}{8} & \frac{3}{8} & \frac{3}{8} & \frac{1}{8} & \frac{1}{2} & 0 & 0 & \frac{1}{2} & \frac{1}{8}
    & \frac{3}{8} & \frac{3}{8} & \frac{1}{8} & \frac{1}{8} & \frac{3}{8} & \frac{3}{8} & \frac{1}{8}
    \\
  \end{smallmatrix}
  \right]
\end{equation}

\ifthenelse{\boolean{pt-br}}{
  \ptbr{
  agora, atribu\'imos valores \`as probabilidades $P_{1,2,3,4}$ de (\ref{eq:probCHSHcte}). Vamos atribuir
  valor zero a todas as probabilidades
  }}

\noindent
we now assign values to the probabilities $P_{1,2,3,4}$ of (\ref{eq:probCHSHcte}). Let's assign the
value zero to all probabilities

\begin{equation}
\label{eq:probCHSHcte0}
  \begin{array}{ll}
    P_{1,2,3,4}^{--++}
    &=\dot P_{1,2,3,4}^{-+++}
    =\dot P_{1,2,3,4}^{+-++}
    =\dot P_{1,2,3,4}^{++--}
    =\\
    &=\dot P_{1,2,3,4}^{++-+}
    =\dot P_{1,2,3,4}^{+++-}
    =\dot P_{1,2,3,4}^{++++}
    =0
  \end{array}
\end{equation}

\ifthenelse{\boolean{pt-br}}{
  \ptbr{
  a partir das equa\c{c}\~oes (\ref{eq:probCHSHcte0}) e (\ref{eq:probCHSHcte2.1}-\ref{eq:probCHSHcte2.4}),
  temos
  }}

\noindent
from the equations (\ref{eq:probCHSHcte0}) and (\ref{eq:probCHSHcte2.1}-\ref{eq:probCHSHcte2.4}), we
have

\begin{equation}\nonumber
  \tfrac{1}{8}
  =P_{1,2,3,4}^{+-+-}
  ,\ 
  \tfrac{1}{8}
  =P_{1,2,3,4}^{-++-}
  ,\ 
  \tfrac{1}{2}
  =P_{1,2,3,4}^{+--+}
  ,\ 
  \tfrac{1}{8}
  =P_{1,2,3,4}^{-+-+}
\end{equation}

\ifthenelse{\boolean{pt-br}}{
  \ptbr{
  A partir das equa\c{c}\~oes (\ref{eq:probCHSHcte3.1}-\ref{eq:probCHSHcte3.3}) e dos valores encontrados
  em (\ref{eq:probCHSHcte2.1}-\ref{eq:probCHSHcte2.4}) e dos valores atribu\'idos em (\ref{eq:probCHSHcte0}),
  temos
  }}

From the equations (\ref{eq:probCHSHcte3.1}-\ref{eq:probCHSHcte3.3}) and values found in 
(\ref{eq:probCHSHcte2.1}-\ref{eq:probCHSHcte2.4}) and of the values attributed (\ref{eq:probCHSHcte0}),
we have

\begin{eqnarray}
\label{eq:probCHSHcte0.1-}
  \tfrac{3}{8}
  =P_{1,2,3,4}^{+---}+\tfrac{1}{2}+0+0\Rightarrow P_{1,2,3,4}^{+---}=-\tfrac{1}{8}
  \\
  \tfrac{3}{8}
  =P_{1,2,3,4}^{-+--}+\tfrac{1}{8}+0+0\Rightarrow P_{1,2,3,4}^{-+--}=+\tfrac{1}{4}
  \\
\label{eq:probCHSHcte0.2-}
  \tfrac{3}{8}
  =P_{1,2,3,4}^{---+}+0+\tfrac{1}{2}+0\Rightarrow P_{1,2,3,4}^{---+}=-\tfrac{1}{8}
\end{eqnarray}

\ifthenelse{\boolean{pt-br}}{
  \ptbr{
  observamos que em (\ref{eq:probCHSHcte0.1-}) e em (\ref{eq:probCHSHcte0.2-}), as parcelas 0 s\~ao
  referentes aos valores qua atribu\'imos \`as probabilidades $P_{1,2,3,4}$. Se tiv\'essemos atribu\'ido
  um outro valor compat\'ivel com os axiomas de Kolmogorov (valor maior que zero), as probabilidades
  $P_{1,2,3,4}^{+---}$ e $P_{1,2,3,4}^{---+}$ seriam ainda mais negativas. Portanto, fica evidente que
  para o exemplo (\ref{eq:CHSHex}), essas duas probabilidades ter\~ao valores negativos, violando os
  axiomas de Kolmogorov.
  }}

we observed that in(\ref{eq:probCHSHcte0.1-}) and in (\ref{eq:probCHSHcte0.2-}), plots 0 are the values
we assign to the probabilities $ P_ {1,2,3,4}$. If we had assigned another value compatible with Kolmogorov's
axioms (a value greater than zero), the probabilities $P_{1,2,3,4}$ and $P_{1,2,3,4}$ would be even
more negative. Therefore, it is evident that on example (\ref{eq:CHSHex}), these two probabilities will
have negative values, violating Kolmogorov's axioms.

\ifthenelse{\boolean{pt-br}}{
  \ptbr{
  Determinando as probabilidades que restarem, temos
  }}

Determining the odds remaining, we have

\begin{eqnarray}\nonumber
  \tfrac{3}{8}
  =P_{1,2,3,4}^{--+-}+0+\tfrac{1}{8}+0\Rightarrow P_{1,2,3,4}^{--+-}=+\tfrac{1}{4}
  \\\nonumber
  \tfrac{1}{8}
  =P_{1,2,3,4}^{----}-\tfrac{1}{8}+\tfrac{1}{4}+\tfrac{1}{8}\Rightarrow P_{1,2,3,4}^{----}=-\tfrac{1}{8}
\end{eqnarray}

\ifthenelse{\boolean{pt-br}}{
  \ptbr{
  portanto, temos os seguintes valores para as probabilidades (a matriz (\ref{eq:probCHSHcteMatriz})
  ser\'a igual a)
  }}

\noindent
therefore, we have the following values for the probabilities (the matrix (\ref{eq:probCHSHcteMatriz})
will be equal to)

\begin{equation}\nonumber
  \left[\begin{matrix}
    -\tfrac{1}{8} &-\tfrac{1}{8} &+\tfrac{1}{4} &0
    \\
    +\tfrac{1}{4} &+\tfrac{1}{8} &+\tfrac{1}{8} &0
    \\
    -\tfrac{1}{8} &+\tfrac{1}{2} &+\tfrac{1}{8} &0
    \\
    0             &0             &0             &0
  \end{matrix}\right]
\end{equation}

\ifthenelse{\boolean{pt-br}}{
  \ptbr{
  portanto a matriz $\ddot P^{\top}$ (onde $\top$ indica que a transposi\c{c}\~ao da matriz) ser\'a
  }}

\noindent
so the matrix $ \ddot P^{\top}$ (where $ \top $ indicates that the transpose of the matrix) will be

\begin{equation}
\label{eq:probCHSHcteMatrizEx}
  \ddot P^{\top}=
  \left[\begin{smallmatrix}
    -\tfrac{1}{8}  &-\tfrac{1}{8} &+\tfrac{1}{4} &0
    &+\tfrac{1}{4} &+\tfrac{1}{8} &+\tfrac{1}{8} &0
    &-\tfrac{1}{8} &+\tfrac{1}{2} &+\tfrac{1}{8} &0
    &0             &0             &0             &0
  \end{smallmatrix}\right]
\end{equation}

\ifthenelse{\boolean{pt-br}}{
  \ptbr{
  com isso obtemos que $\ddot\Sigma\cdot\ddot P=\ddot p$ exatamente como em (\ref{eq:CHSHmatriz}).
  }}

\noindent
with this we obtain that $\ddot\Sigma\cdot\ddot P=\ddot p$ exactly as in (\ref{eq:CHSHmatriz}).

\ifthenelse{\boolean{pt-br}}{
  \ptbr{
  Podemos observar que a soma de todos os valores em (\ref{eq:probCHSHcteMatrizEx}) \'e igual a 1 (como
  ocorre quando somamos as probabilidades de todos os eventos, por\'em, probabilidades n\~ao podem ter
  valores negativos). 
  }}

We can see that the sum of all values in (\ref{eq:probCHSHcteMatrizEx}) is equal to 1 (as occurs when
we sum the probabilities of all events, but probabilities can not have negative values).

\ifthenelse{\boolean{pt-br}}{
  \ptbr{
  De (\ref{eq:Correlacao}), temos que 
  }}

From (\ref{eq:Correlacao}), We have to

\begin{equation}\nonumber
  \mathcal{E}_{\dot Z_{j,k}}(Z_j\cdot Z_k)=P_{j,k}^{--}-P_{j,k}^{-+}-P_{j,k}^{+-}+P_{j,k}^{++}
\end{equation}

\ifthenelse{\boolean{pt-br}}{
  \ptbr{
  podemos encontrar os quatro valores esperados (com $(j,k)\in\{(1,3),(1,4),(2,3),(2,4)\}$) realizando
  a multiplica\c{c}\~ao matricial pela seguinte matriz
  }}

\noindent
we can find the four expected values (with $(j,k)\in\{(1,3),(1,4),(2,3),(2,4)\}$) performing the matrix
multiplication by the following matrix

\begin{equation}\nonumber
  \ddot E:=
  \left[\begin{smallmatrix}
    1 & -1 & -1 & 1 & 0 & 0 & 0 & 0 & 0 & 0 & 0 & 0 & 0 & 0 & 0 & 0 \\
    0 & 0 & 0 & 0 & 1 & -1 & -1 & 1 & 0 & 0 & 0 & 0 & 0 & 0 & 0 & 0 \\
    0 & 0 & 0 & 0 & 0 & 0 & 0 & 0 & 1 & -1 & -1 & 1 & 0 & 0 & 0 & 0 \\
    0 & 0 & 0 & 0 & 0 & 0 & 0 & 0 & 0 & 0 & 0 & 0 & 1 & -1 & -1 & 1 \\
  \end{smallmatrix}\right]
\end{equation}

\ifthenelse{\boolean{pt-br}}{
  \ptbr{
  assim, temos
  }}

\noindent
so we have

\begin{equation}
\label{eq:CorrelacaoEx}
  \left[\begin{smallmatrix}
    \mathcal{E}_{\dot Z_{1,3}}(Z_1\cdot Z_3)\\
    \mathcal{E}_{\dot Z_{1,4}}(Z_1\cdot Z_4)\\
    \mathcal{E}_{\dot Z_{2,3}}(Z_2\cdot Z_3)\\
    \mathcal{E}_{\dot Z_{2,4}}(Z_2\cdot Z_4)\\
  \end{smallmatrix}\right]
  =\ddot E\cdot\ddot p
  =\ddot E\cdot\ddot\Sigma\cdot\ddot P
  =
  \left[\begin{smallmatrix}
    -\frac{1}{2} \\
    1 \\
    -\frac{1}{2} \\
    -\frac{1}{2} \\
  \end{smallmatrix}\right]
\end{equation}

\ifthenelse{\boolean{pt-br}}{
  \ptbr{
  de (\ref{eq:CHSH-Abs}), podemos obter a desigualdade de CHSH, multiplicando a seguinte matriz
  $\ddot\eta:=\left[\begin{smallmatrix}
      +1 &-1 &+1 &+1
    \end{smallmatrix}\right]$
  }}

\noindent
from (\ref{eq:CHSH-Abs}), we can obtain the CHSH inequality by multiplying the following matrix
$\ddot\eta:=\left[\begin{smallmatrix}
      +1 &-1 &+1 &+1
    \end{smallmatrix}\right]$

\begin{equation}\nonumber
  \begin{array}{l}
    \left[\begin{smallmatrix}
      +1 &-1 &+1 &+1
    \end{smallmatrix}\right]
    \cdot
    \left[\begin{smallmatrix}
      \mathcal{E}_{\dot Z_{1,3}}(Z_1\cdot Z_3)\\
      \mathcal{E}_{\dot Z_{1,4}}(Z_1\cdot Z_4)\\
      \mathcal{E}_{\dot Z_{2,3}}(Z_2\cdot Z_3)\\
      \mathcal{E}_{\dot Z_{2,4}}(Z_2\cdot Z_4)\\
    \end{smallmatrix}\right]
    =\ddot\eta\cdot\ddot E\cdot\ddot\Sigma\cdot\ddot P=
    \\
    =
    \mathcal{E}_{\ddot Z}(Z_1\!\cdot\! Z_3)-\mathcal{E}_{\ddot Z}(Z_1\!\cdot\! Z_4)
    +\mathcal{E}_{\ddot Z}(Z_2\!\cdot\! Z_3)+\mathcal{E}_{\ddot Z}(Z_1\!\cdot Z_3)
  \end{array}
\end{equation}

\ifthenelse{\boolean{pt-br}}{
  \ptbr{
  assim temos
  }}

\noindent
so we have

\begin{equation}\nonumber
  \ddot\eta\cdot\ddot E\cdot\ddot p
  =-\tfrac{5}{2}
\end{equation}

\ifthenelse{\boolean{pt-br}}{
  \ptbr{
  encontramos o mesmo resultado encontrado em (\ref{eq:CHSHex}),
  uma viola\c{c}\~ao do tipo ($-2\leq-2.5\leq2$).
  }}

\noindent
we found the same result of (\ref{eq:CHSHex}),
a breach of the type ($-2\leq-2.5\leq2$).

\ifthenelse{\boolean{pt-br}}{
  \ptbr{
  Lembrando que em (\ref{eq:CHSHdem}) n\'os supomos que $\mathcal{P}_{\ddot Z}(Z_1,Z_2,Z_3,Z_4)=|\mathcal{P}_{\ddot
  Z}(Z_1,Z_2,Z_3,Z_4)|$, que n\~ao \'e satisfeito pelos valores encontrados em (\ref{eq:probCHSHcteMatrizEx}),
  ent\~ao n\~ao podemos trocar um pelo outro, j\'a que n\~ao s\~ao iguais para alguns casos. Se us\'assemos
  $|\mathcal{P}_{\ddot Z}(Z_1,Z_2,Z_3,Z_4)|$, ter\'iamos
  }}

Recalling that in (\ref{eq:CHSHdem}) we assume that $\mathcal{P}_{\ddot Z}(Z_1,Z_2,Z_3,Z_4)=|\mathcal{P}_{\ddot
Z}(Z_1,Z_2,Z_3,Z_4)|$, which is not satisfied by the values found in (\ref{eq:probCHSHcteMatrizEx}),
so we can not trade for each other, since they are not the same for some cases. If we used
$|\mathcal{P}_{\ddot Z}(Z_1,Z_2,Z_3,Z_4)|$, we would have

\begin{equation}\nonumber
  \ddot Q^{\top}:=
  \left[\begin{smallmatrix}
    +\tfrac{1}{8}  &+\tfrac{1}{8} &+\tfrac{1}{4} &0
    &+\tfrac{1}{4} &+\tfrac{1}{8} &+\tfrac{1}{8} &0
    &+\tfrac{1}{8} &+\tfrac{1}{2} &+\tfrac{1}{8} &0
    &0             &0             &0             &0
  \end{smallmatrix}\right]
\end{equation}

\ifthenelse{\boolean{pt-br}}{
  \ptbr{
  em vez de (\ref{eq:probCHSHcteMatrizEx}), o que resultaria em
  }}

\noindent
rather than (\ref{eq:probCHSHcteMatrizEx}), which would result in

\begin{equation}\nonumber
  \sum\limits_{(Z_1,Z_2,Z_3,Z_4)\in\{-1,+1\}^4}(|\mathcal{P}_{\ddot Z}(Z_1,Z_2,Z_3,Z_4)|)
  =\tfrac{7}{4}
\end{equation}

\ifthenelse{\boolean{pt-br}}{
  \ptbr{
  que viola um outro axioma de Kolmogorov. 
  }}

\noindent
which violates another axiom of Kolmogorov.

\ifthenelse{\boolean{pt-br}}{
  \ptbr{
  Contudo, tal troca evita a viola\c{c}\~ao da desigualdade de CHSH, pois ao calcularmos (\ref{eq:CHSH}),
  encontramos 
  }}

However, such exchange avoids the violation of the CHSH inequality, because when we calculate
(\ref{eq:CHSH}), we found

\begin{equation}\nonumber
  \ddot\eta\cdot\ddot E\cdot\ddot\Sigma\cdot\ddot Q
  =-1
\end{equation}

\ifthenelse{\boolean{pt-br}}{
  \ptbr{
  portanto, temos que a desigualdade n\~a \'e violada, pois encontramos ($-2\leq-1\leq2$).
  }}

\noindent
therefore, we have that the inequality is not violated, since we find ($-2\leq-1\leq2$).

\ifthenelse{\boolean{pt-br}}{
  \ptbr{
  Portanto, tal troca \'e v\'alida para o c\'alculo da desigualdade de CHSH (pois atende \`as condi\c{c}\~oes
  em (\ref{eq:CHSHdem})), mas se estendermos essa troca para o c\'alculo das probabilidades 
  $\mathcal{P}_{\dot Z_{j,k}}(Z_j,Z_k)$ resultaria em discrep\^ancias, como por exemplo
  }}

Hence, such exchange is valid for the calculation of CHSH inequality (since it meets the conditions in (\ref{eq:CHSHdem})), but if we extend this exchange to the calculation of the probabilities
$\mathcal{P}_{\dot Z_{j,k}}(Z_j,Z_k)$, it would result in discrepancies, such as

\begin{equation}\nonumber
  \mathcal{P}_{\dot Z_{1,4}}(-1,-1)
  =\tfrac{3}{4}
  \neq\mathcal{P}_{\dot Z_{1,4}}(+1,+1)
  =\tfrac{1}{2}
\end{equation}

\ifthenelse{\boolean{pt-br}}{
  \ptbr{
  Podemos observar o que ocorreria com as outras probabilidades realizando a multiplica\c{c}\~ao matricial
  }}

We can observe what would happen with the other probabilities by performing the matrix multiplication

\begin{equation}\nonumber
  \ddot\Sigma\cdot\ddot Q
  =\left[
  \begin{smallmatrix}
    \frac{5}{8} & \frac{3}{8} & \frac{5}{8} & \frac{1}{8} & \frac{3}{4} & \frac{1}{4} & \frac{1}{4}
    & \frac{1}{2} & \frac{7}{8} & \frac{3}{8} & \frac{3}{8} & \frac{1}{8} & \frac{5}{8} & \frac{5}{8}
    & \frac{3}{8} & \frac{1}{8} \\
  \end{smallmatrix}
  \right]^{\top}
  \neq\ddot p
\end{equation}

\ifthenelse{\boolean{pt-br}}{
  \ptbr{
  evidenciando a discrep\^ancia em 8 das 16 probabilidades.
  }}

\noindent
evidencing the discrepancy in 8 of the 16 probabilities.

\ifthenelse{\boolean{pt-br}}{
  \ptbr{
  Tamb\'em haveriam discrep\^ancias no valores atribu�dos aos valores esperados em que tal troca ocorrese,
  resultando em
  }}

There would also be discrepancies in the values attributed to the expected values in which such exchange
occurs, resulting in

\begin{equation}\nonumber
  \ddot E\cdot\ddot\Sigma\cdot\ddot Q
  =\left[
  \begin{smallmatrix}
    -\frac{1}{4} \\
    \frac{3}{4} \\
    \frac{1}{4} \\
    -\frac{1}{4} \\
  \end{smallmatrix}
  \right]
\end{equation}

\ifthenelse{\boolean{pt-br}}{
  \ptbr{
  que s\~ao diferentes aos valores encontrados em (\ref{eq:CorrelacaoEx})
  }}

\noindent
which are different from the values found, resulting in (\ref{eq:CorrelacaoEx})

\ifthenelse{\boolean{pt-br}}
  {\ptbr
  {Concluimos nesta se\c{c}\~ao que o sistema \'e poss\'ivel de resolver, mas \'e indeterminado, pois
  o rank \'e menor que o n\'umero de vari\'aveis. Assim, n\~ao h\'a inconsist\^encia no sistema, a \'unica
  incosist\^encia \'e em rela\c{c}\~ao aos axiomas de Kolmogorov, visto na se\c{c}\~ao anterior.}
  }

We conclude in this section that the system is possible to solve, but it is indeterminate, because the
rank is smaller than the number of variables. Thus, there is no inconsistency in the system, the only
inconsistency is in relation to Kolmogorov's axioms, as seen in the previous section.

\ifthenelse{\boolean{pt-br}}
  {\ptbr
  {Desigualdade b\'asica}
  }

\section{Basic inequality}
\label{sec:Desig_Basica}

\ifthenelse{\boolean{pt-br}}
  {\ptbr
  {Nesta se\c{c}\~ao iremos propor uma desigualdade, da qual as desigualdade de Bell e de CHSH s\~ao
  casos espec�ficos.}
  }

In this section, we shall propose an inequality, in which the inequalities of Bell and CHSH are specific
cases.

\ifthenelse{\boolean{pt-br}}
  {\ptbr
  {Come\c{c}amos por definir os valores assumidos pelas vari\'aveis aleat\'orias $Z_j$}
  }

We will start by defining  the values assumed by the random variables $Z_j$

\begin{equation}\nonumber
  Z_j\in\{-1,+1\}
\end{equation}

\ifthenelse{\boolean{pt-br}}
  {\ptbr
  {sob essa condi\c{c}\~ao temos que}
  }

\noindent
Under this condition we have that

\begin{equation}\nonumber
  Z_j^2=1,
  \qquad |Z_j|=1
\end{equation}

\ifthenelse{\boolean{pt-br}}
  {\ptbr
  {Definida a vari\'avel aleat\'oria, passamos a examinar as express\~oes $|Z_j-Z_k|$ e $|Z_j+Z_k|$,
  resultnado em}
  }

Defined the random variable, we start to examine the expressions $|Z_j-Z_k|$ e $|Z_j+Z_k|$, coming to
the solution

\begin{eqnarray}\nonumber
  |Z_j\pm Z_k|
  &=&|Z_j\pm \underbrace{Z_j^2}_{=1}\cdot Z_k|
  =|Z_j\cdot(1\pm Z_j\cdot Z_k)|=
  \\\nonumber
  &=&\underbrace{|Z_j|}_{=1}\cdot|\underbrace{1\pm Z_j\cdot Z_k}_{\in\{0,+2\}}|
  =1\pm Z_j \cdot Z_k
\end{eqnarray}

\ifthenelse{\boolean{pt-br}}
  {\ptbr
  {Agora, calculando o valor esperado dessa express\~ao, temos que}
  }

Now, calculating the expected value on this equation, we shall have

\begin{eqnarray}\nonumber
  \mathcal{E}_{\dot{Z}_{j,k,l}}(1\pm Z_j\cdot Z_k)
  &=&\mathcal{E}_{\dot Z_{j,k,l}}(|Z_j\pm Z_k|)
  \\\nonumber
  &\geq&|\mathcal{E}_{\dot Z_{j,k,l}}(Z_j)\pm\mathcal{E}_{\dot Z_{j,k,l}}(Z_k)|
\end{eqnarray}

\ifthenelse{\boolean{pt-br}}
  {\ptbr
  {resultando nas seguintes f\'ormulas}
  }

\noindent
resulting in the following formulas

\begin{equation}
\label{eq:basic}
  \left\{
  \begin{array}{c}
    1-\mathcal{E}_{\dot{Z}_{j,k,l}}(Z_j\cdot Z_k)
    \geq|\mathcal{E}_{\dot Z_{j,k,l}}(Z_j)-\mathcal{E}_{\dot Z_{j,k,l}}(Z_k)| 
    \\
    1+\mathcal{E}_{\dot{Z}_{j,k,l}}(Z_j\cdot Z_k)
    \geq|\mathcal{E}_{\dot Z_{j,k,l}}(Z_j)+\mathcal{E}_{\dot Z_{j,k,l}}(Z_k)|
  \end{array}\right.
\end{equation}

\ifthenelse{\boolean{pt-br}}
  {\ptbr
  {que s\~ao v\'alidas na condi\c{c}\~ao de que as vari\'aveis aleat\'orias assumam valores em $\{-1,+1\}$}
  }

The expressions are valide under the condition that these random variables are between $\{-1,+1\}$.

\ifthenelse{\boolean{pt-br}}
  {\ptbr
  {Podemos verificar essa desigualdade usando (\ref{eq:ProbZjZk}), come��ndo por calcular}
  }

We can verify this inequality by using (\ref{eq:ProbZjZk}), starting by calculating

\[
  \mathcal{P}_{Z_{j}}(z_j)=\sum\limits_{z_k\in\{-1,+1\}}(\mathcal{P}_{\dot Z_{j,k}}(z_j,z_k))=\frac{1}{2}
\]

\ifthenelse{\boolean{pt-br}}
  {\ptbr
  {para que possamos usar no c\'alculo do seguinte valor esperado}
  }

So we can use it to calculate the following value

\[
  \mathcal{E}_{\dot Z_{j,k,l}}(Z_j)=\mathcal{E}_{Z_j}(Z_j)=-1\cdot\frac{1}{2}+1\cdot\frac{1}{2}=0
\]

\ifthenelse{\boolean{pt-br}}{
  \ptbr{
  subsituindo nas desigualdades, encontramos
  }}

\noindent
substituting the inequalities, we shall find

\[
  \left\{
  \begin{array}{l}
    1-\cos(2\bar\theta_{j,k})\geq|0-0|
    \\
    1+\cos(2\bar\theta_{j,k})\geq|0+0|
  \end{array}\right.
\]

\ifthenelse{\boolean{pt-br}}
  {\ptbr
  {que resulta em}
  }

That results in

\begin{equation}\nonumber
  1\geq\cos(2\bar\theta_{j,k})\geq-1
\end{equation}

\ifthenelse{\boolean{pt-br}}
  {\ptbr
  {que claramente \'e satisfeito, qualquer que sejam os \^angulos envolvidos. Portanto, a plica\c{c}\~ao
  direta nessa desigualdade n\~ao \'e violada.}
  }

That is clearly satisfactory, whatever the angle envolved. So, the direct application in this inequality isn't violated.

\ifthenelse{\boolean{pt-br}}
  {\ptbr
  {A partir dessa desigualdade, fazendo apenas algumas substitui\c{c}\~oes, podemos demonstrar a desiguadlade
  de Bell. Comecemos por definir a vari\'avel $M_{j,k}$ como sendo o produto das vari\'aveis $Z_j$ e
  $Z_k$}
  }

Starting from this inequality making only some substitutions, we can demonstrate the inequality of Bell. We start by defining the variable $M_{j,k}$ as being the product of the variables $Z_j$ and $Z_k$.

\begin{equation}\nonumber
  M_{j,k}
  =Z_j\cdot Z_k\in \{-1,+1\}
\end{equation}

\ifthenelse{\boolean{pt-br}}
  {\ptbr
  {que obviamente assumir\'a os valores $\{-1,+1\}$, portanto cumpre a condi\c{c}\~ao para a validade
  da desigualade b\'asica. Aplicando nas f\'ormulas (\ref{eq:basic}) as substitui\c{c}\~oes de $Z_j$
  por $M_{j,k}$ e de $Z_k$ por $M_{k,l}$ temos que}
  }

\noindent
that will obviously assume  the values $\{-1,+1\}$, so it fullfils the condition for the basic inequality
validity. Appling in these formulas (\ref{eq:basic}) the substitutions from $Z_j$ to $M_{j,k}$ and from
$Z_k$ to $M_{k,l}$ we have

\begin{equation}\nonumber
  1\pm\mathcal{E}_{\dot{Z}_{j,k,l}}(M_{j,k} \cdot M_{k,l})
  \geq|\mathcal{E}_{\dot Z_{j,k,l}}(M_{j,k})\pm\mathcal{E}_{\dot Z_{j,k,l}}(M_{k,l})|
\end{equation}

\ifthenelse{\boolean{pt-br}}
  {\ptbr
  {portanto}
  }

\noindent
so

\begin{equation}\nonumber
  1\pm\mathcal{E}_{\dot{Z}_{j,k,l}}(Z_j\cdot \underbrace{Z_k\cdot Z_k}_{=1}\cdot Z_l)
  \geq|\mathcal{E}_{\dot Z_{j,k,l}}(Z_j\cdot Z_k)\pm\mathcal{E}_{\dot Z_{j,k,l}}(Z_k\cdot Z_l)|
\end{equation}

\ifthenelse{\boolean{pt-br}}
  {\ptbr
  {resultando nas f\'ormulas}
  }

\noindent
resulting in the formulas

\begin{equation}
\label{eq:basicBell}
  \left\{
  \begin{array}{c}
    1-\mathcal{E}_{\dot{Z}_{j,k,l}}(Z_j\cdot Z_l)
    \geq|\mathcal{E}_{\dot Z_{j,k,l}}(Z_j\cdot Z_k)-\mathcal{E}_{\dot Z_{j,k,l}}(Z_k\cdot Z_l)| 
    \\
    1+\mathcal{E}_{\dot{Z}_{j,k,l}}(Z_j\cdot Z_l)
    \geq|\mathcal{E}_{\dot Z_{j,k,l}}(Z_j\cdot Z_k)+\mathcal{E}_{\dot Z_{j,k,l}}(Z_k\cdot Z_l)|
  \end{array}\right.
\end{equation}

\ifthenelse{\boolean{pt-br}}
  {\ptbr
  {sendo que a primeira desigualdade \'e a desigualdade de Bell. Se somarmos, membro a membro, as desigualdades
  (\ref{eq:basicBell}), realizando a seguinte substitui\c{c}\~ao $k=1$ na primeira desigualdade, $k=2$
  na segunda desigualdade e $(j,l)=(3,4)$ em ambas, teremos}
  }

\noindent
while the first inequality is the inequality of Bell. If we add, term by term, the inequalities (\ref{eq:basicBell}),
doing the following substitution $k=1$ in the first inequality, $k=2$
in the second and $(j,l)=(3,4)$ in both, we have

\begin{eqnarray}\nonumber
  2\geq&&|\mathcal{E}_{\dot Z_{3,1,4}}(Z_3\cdot Z_1)-\mathcal{E}_{\dot Z_{3,1,4}}(Z_1\cdot Z_4)|
    \\
    &&+|\mathcal{E}_{\dot Z_{3,2,4}}(Z_3\cdot Z_2)+\mathcal{E}_{\dot Z_{3,2,4}}(Z_2\cdot Z_4)|
\label{eq:CHSH_Bell}
\end{eqnarray}

\ifthenelse{\boolean{pt-br}}
  {\ptbr
  {denominaremos essa f\'ormula por desigualdade de Bell-CHSH, uma vez que se originou da soma de uma
  desigualdade de Bell original e uma outra modificada e resultou numa desigualdade semelhante \`a de
  CHSH.}
  }

Naming this formula as the inequality of Bell-CHSH, once originated, the addition of one original inequality of Bell and one modified and resulted in one inequality similar to the CHSH.

\ifthenelse{\boolean{pt-br}}
  {\ptbr
  {Rela\c{c}\~ao entre a desigualdade de Clauser-Horne-Shimony-Holt e a desigualdade de Bell}
  }

\section{Relationship between the inequality of Clauser-Horne-Shimony-Holt to the inequality of Bell}
\label{sec:Bell_CHSH}

\ifthenelse{\boolean{pt-br}}
  {\ptbr
  {Nesta se\c{c}\~ao iremos verfificar como a viola\c{c}\~ao da desigualdade de Bell-CHSH est\'a relacionada
  com a viola\c{c}\~ao da desigualdade de CHSH.}
  }

In this section we shall verify how the viotation of the inequality of Bell- CHSH is related  with the violation of the inequality of CHSH.

\ifthenelse{\boolean{pt-br}}
  {\ptbr
  {A desigualdade de CHSH \'e dada por}
  }

The inequality of CHSH is given by

\begin{equation}\nonumber
  2\geq|\mathcal{E}_{\ddot Z}(Z_3\cdot Z_1)-\mathcal{E}_{\ddot Z}(Z_1\cdot Z_4)
    +\mathcal{E}_{\ddot Z}(Z_3\cdot Z_2)+\mathcal{E}_{\ddot Z}(Z_2\cdot Z_4)|
\end{equation}

\ifthenelse{\boolean{pt-br}}
  {\ptbr
  {com $\ddot Z=\dot Z_{1,2,3,4}$, significando que assumimos a exist\^encia da fun\c{c}\~ao de probabilidade
  para $(Z_1,Z_2,Z_3,Z_4)$, uma vez que o valor esperado \'e dependente de $(Z_1,Z_2,Z_3,Z_4)$. Estenderemos
  tal suposi\c{c}\~ao para a f\'ormula (\ref{eq:CHSH_Bell}), passando a escrever $\mathcal{E}_{\ddot
  Z}$ nela.}
  }

With $\ddot Z=\dot Z_{1,2,3,4}$, meaning that it assumed the existence of the probability function  for $(Z_1,Z_2,Z_3,Z_4)$, once this expected value is dependent of $(Z_1,Z_2,Z_3,Z_4)$. We shall understant such suposition from the formula (\ref{eq:CHSH_Bell}), now writing $\mathcal{E}_{\ddot Z}$
 in it.

\ifthenelse{\boolean{pt-br}}
  {\ptbr
  {Pela desigualdade triangular, temos que }
  }

By the triangular inequality, we have that

\begin{widetext}

\begin{equation}\nonumber
  |\mathcal{E}_{\ddot Z}(Z_3\cdot Z_1)-\mathcal{E}_{\ddot Z}(Z_1\cdot Z_4)|
    +|\mathcal{E}_{\ddot Z}(Z_3\cdot Z_2)+\mathcal{E}_{\ddot Z}(Z_2\cdot Z_4)|
    \geq
    |\mathcal{E}_{\ddot Z}(Z_3\cdot Z_1)-\mathcal{E}_{\ddot Z}(Z_1\cdot Z_4)
    +\mathcal{E}_{\ddot Z}(Z_3\cdot Z_2)+\mathcal{E}_{\ddot Z}(Z_2\cdot Z_4)|
\end{equation}

\end{widetext}

\ifthenelse{\boolean{pt-br}}
  {\ptbr
  {portanto, temos que se a desigualdade de CHSH \'e violada, ent\~ao a desigualdade (\ref{eq:CHSH_Bell})
  tamb\'em ser\'a.}
  }

So we have that the inequality of CHSH is violated, so the inequality (\ref{eq:CHSH_Bell}) will be as well.

\begin{widetext}

  \begin{table}
    
    \ifthenelse{\boolean{pt-br}}{\ptbr
      {Gr\'aficos da regi\~ao de viola\c{c}\~ao da desigualdade de CHSH (gr\'aficos \`a esquerda)
      e da viola\c{c}\~ao da desigualdade de Bell-CHSH (gr\'aficos \`a direita). Na primeira linha est\~ao
      os gr\'aficos de superf\'icies de n\'ivel que violam a desigualdade. Na segunda linha foram feitos
      cortes para diferentes valores de $\theta_4$, para cada valor de $\theta_4$ tem-se as regi\~oes
      com os valores de $(\theta_2,\theta_3)$ em que s\~ao violadas as desigualdades}}
    
    \caption{
    Graphics of the region of the violation of the inequality of CHSH (graphics
    on the left) and the violation of the inequality of Bell-CHSH (graphics on the right). On the first
    line are the graphics of the level surface that violates the inequality. On the second line were
    made cuts for different values of $\theta_4$,  for each value of $\theta_4$ there are regions with
    the values of $ (\theta_2,\theta_3)$  where the inequalities were violated
  }
  
  \begin{tabular}{c|c}
    \includegraphics
      [scale=0.12]
      {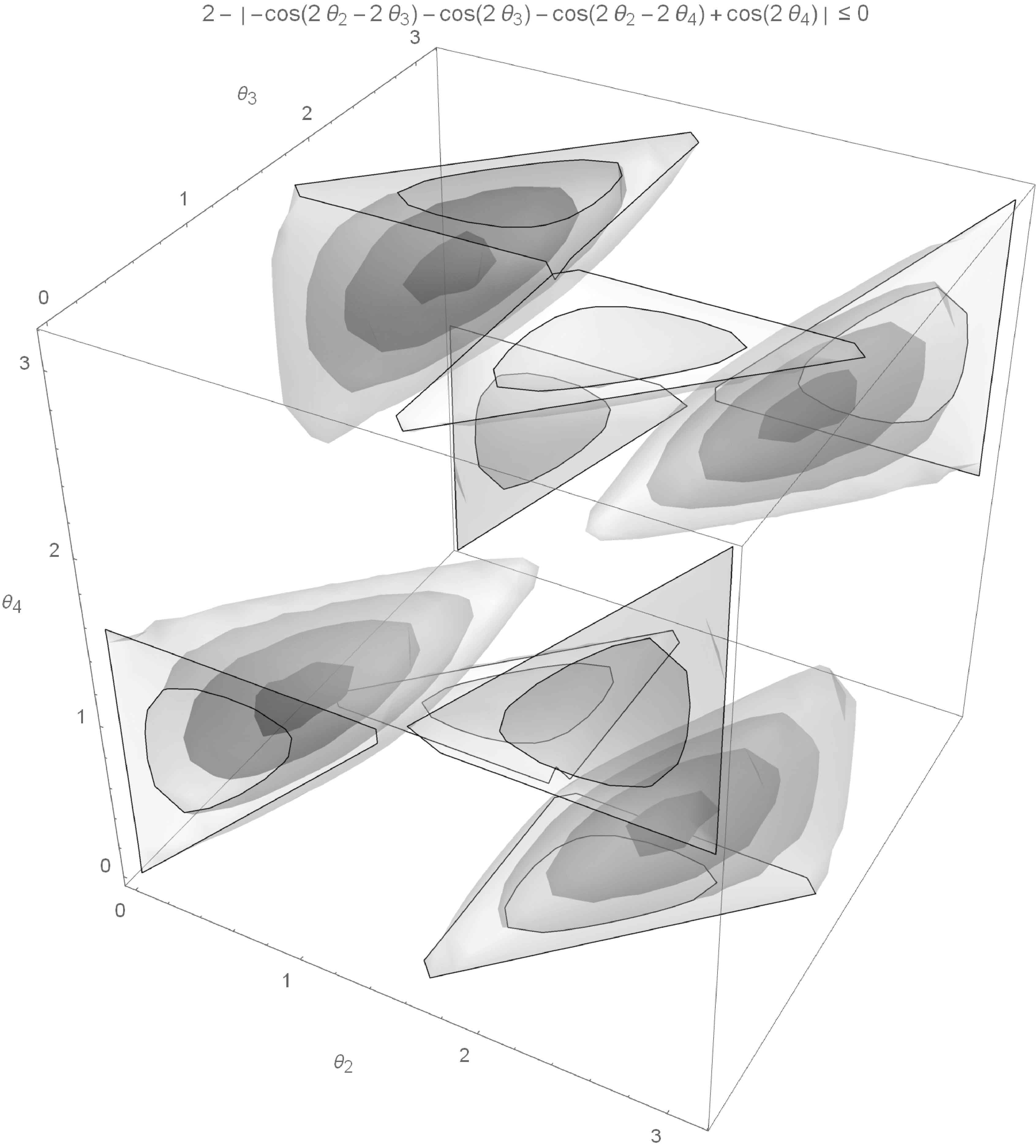} 
    & 
    \includegraphics
      [scale=0.12]
      {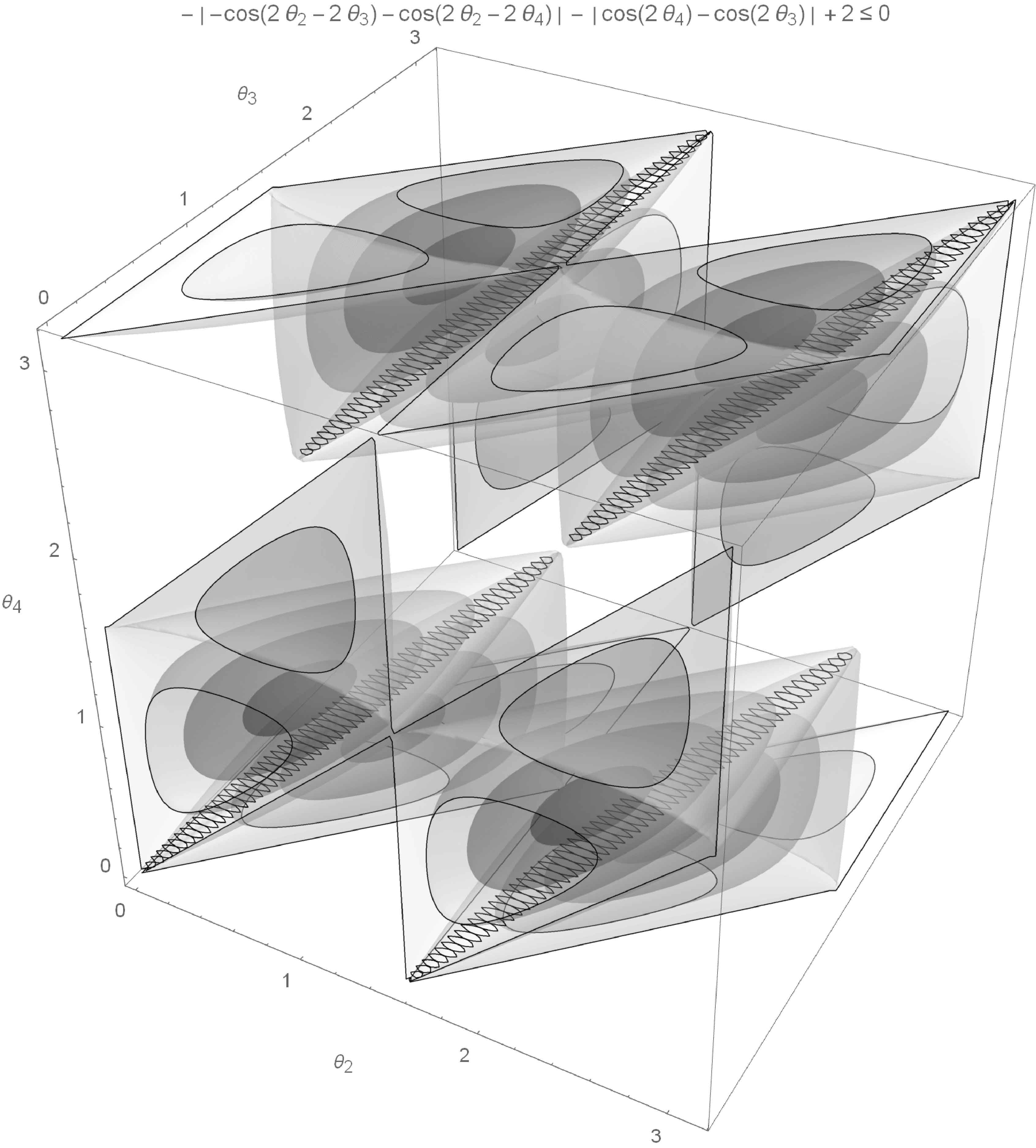}
    \\\hline
    \includegraphics
      [scale=0.09]
      {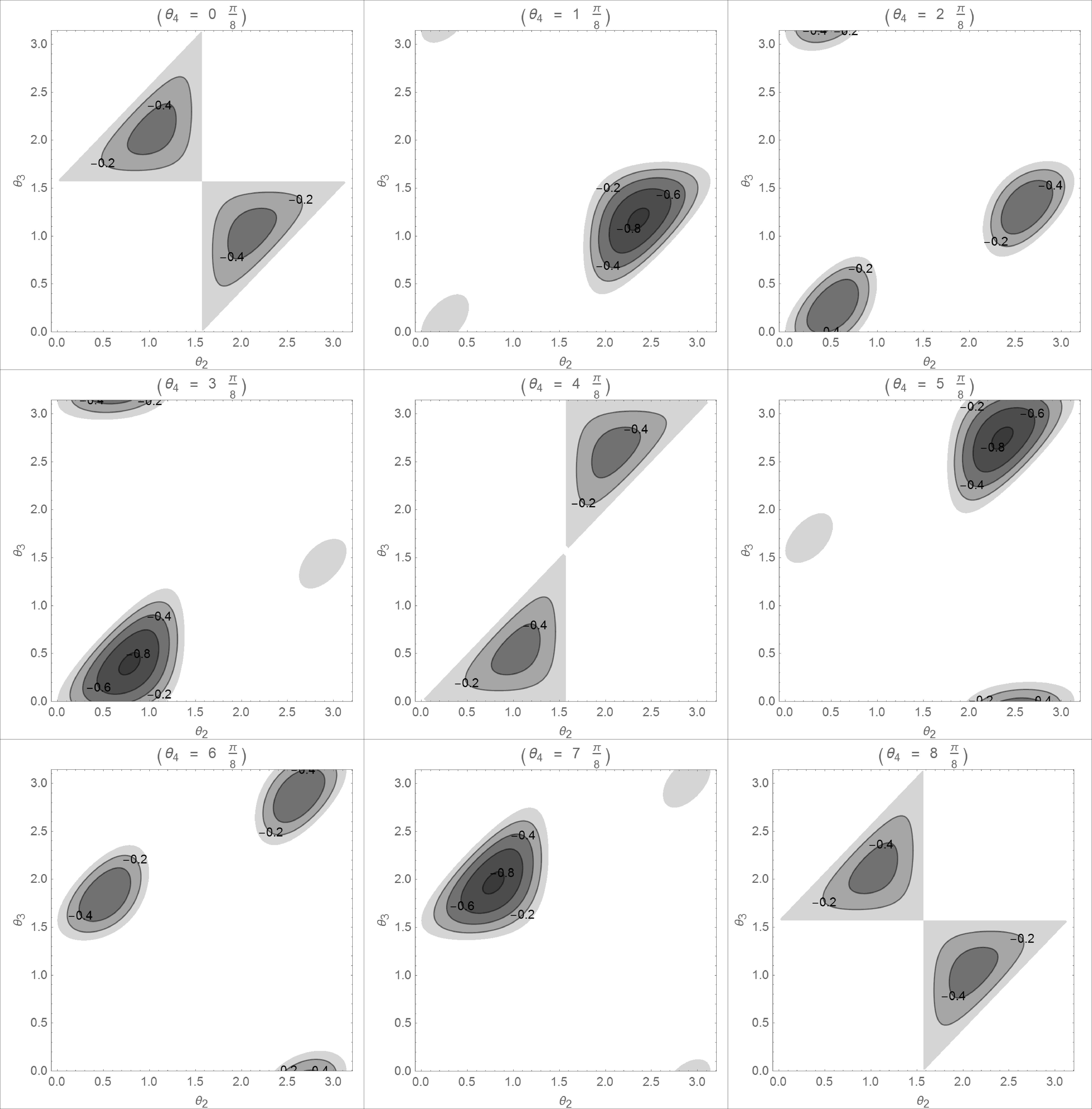}
    & 
    \includegraphics
      [scale=0.09]
      {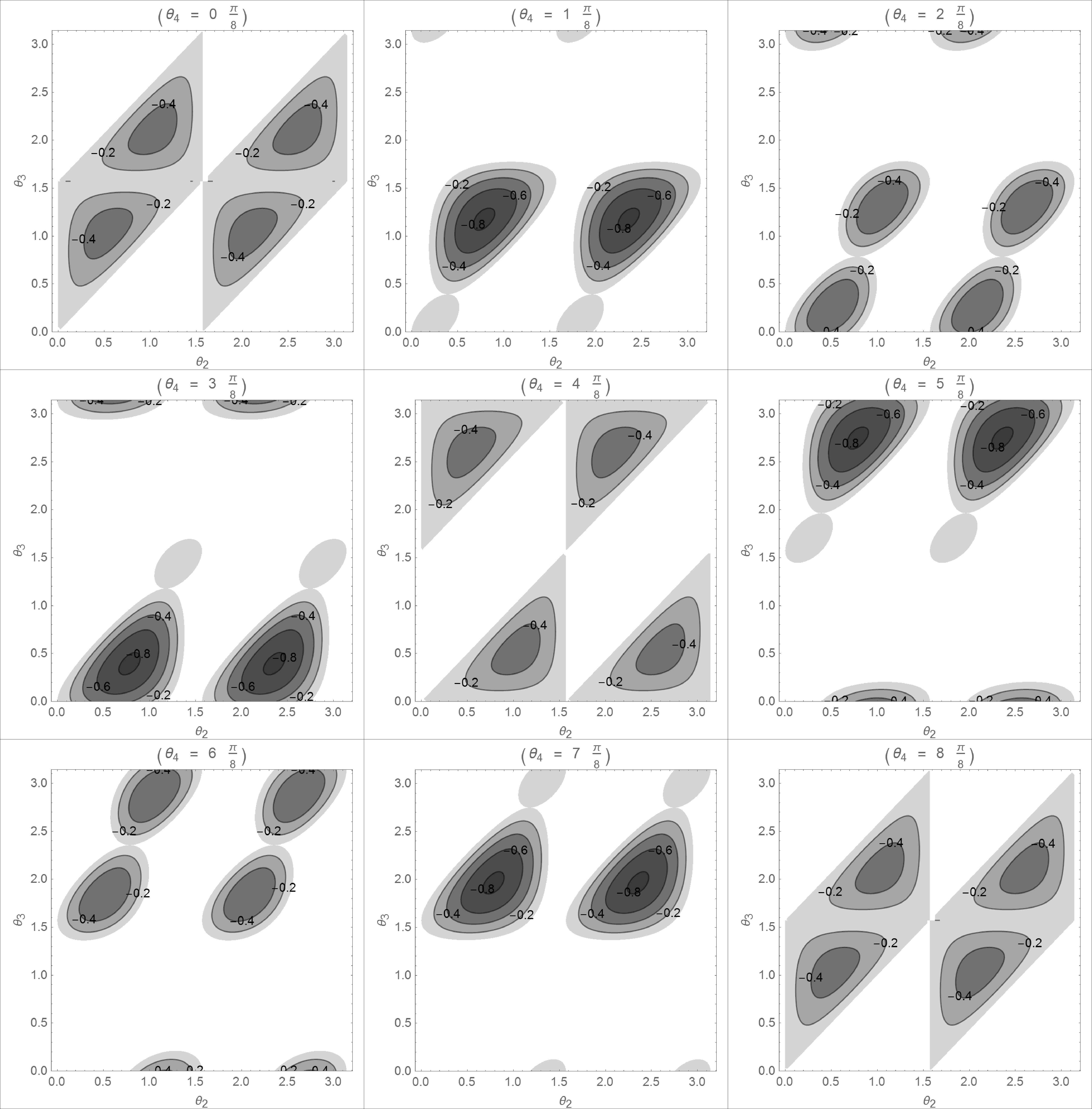}
  \end{tabular}
  \end{table}

\begin{table}

  \ifthenelse{\boolean{pt-br}}
  {\ptbr{Gr\'aficos da desigualdade de CHSH (gr\'aficos \`a esquerda)
  e da desigualdade de Bell-CHSH (gr\'aficos \`a direita). Foram feitos
  cortes para diferentes valores de $\theta_2$, para cada valor de $\theta_2$ tem-se as regi\~oes com
  os valores de $(\theta_3,\theta_4)$}}
  
  \caption{
  Graphics of the inequality of CHSH (on the left) and the inequality of Bell-CHSH (graphics on the right). On this graphics were made cuts for different values of $\theta_2$,  for each value of $\theta_2$ there are region with the values of $(\theta_3,\theta_4)$
}
  
  \begin{tabular}{cc}
    \includegraphics
      [scale=0.28]
      {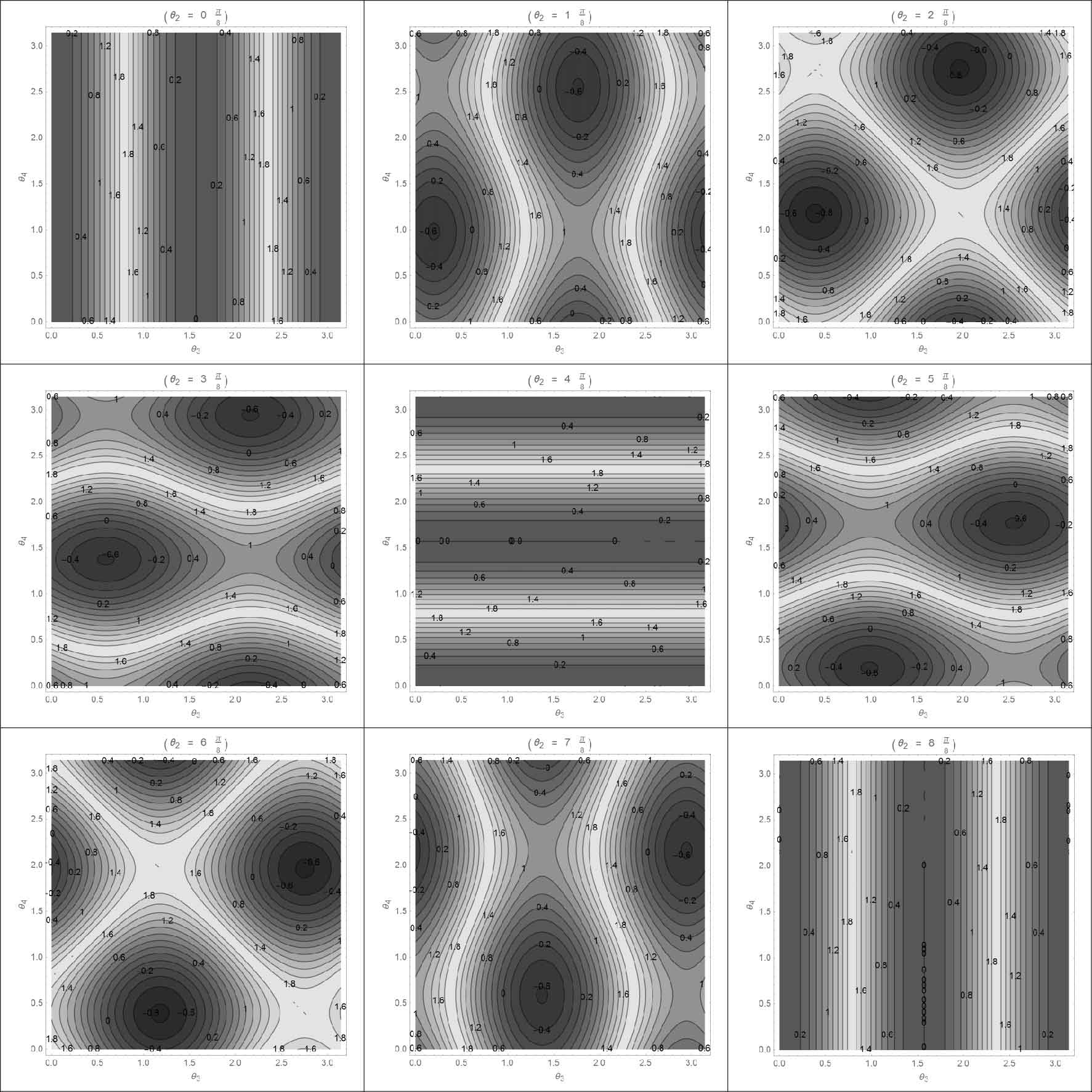} 
    & 
    \includegraphics
      [scale=0.28]
      {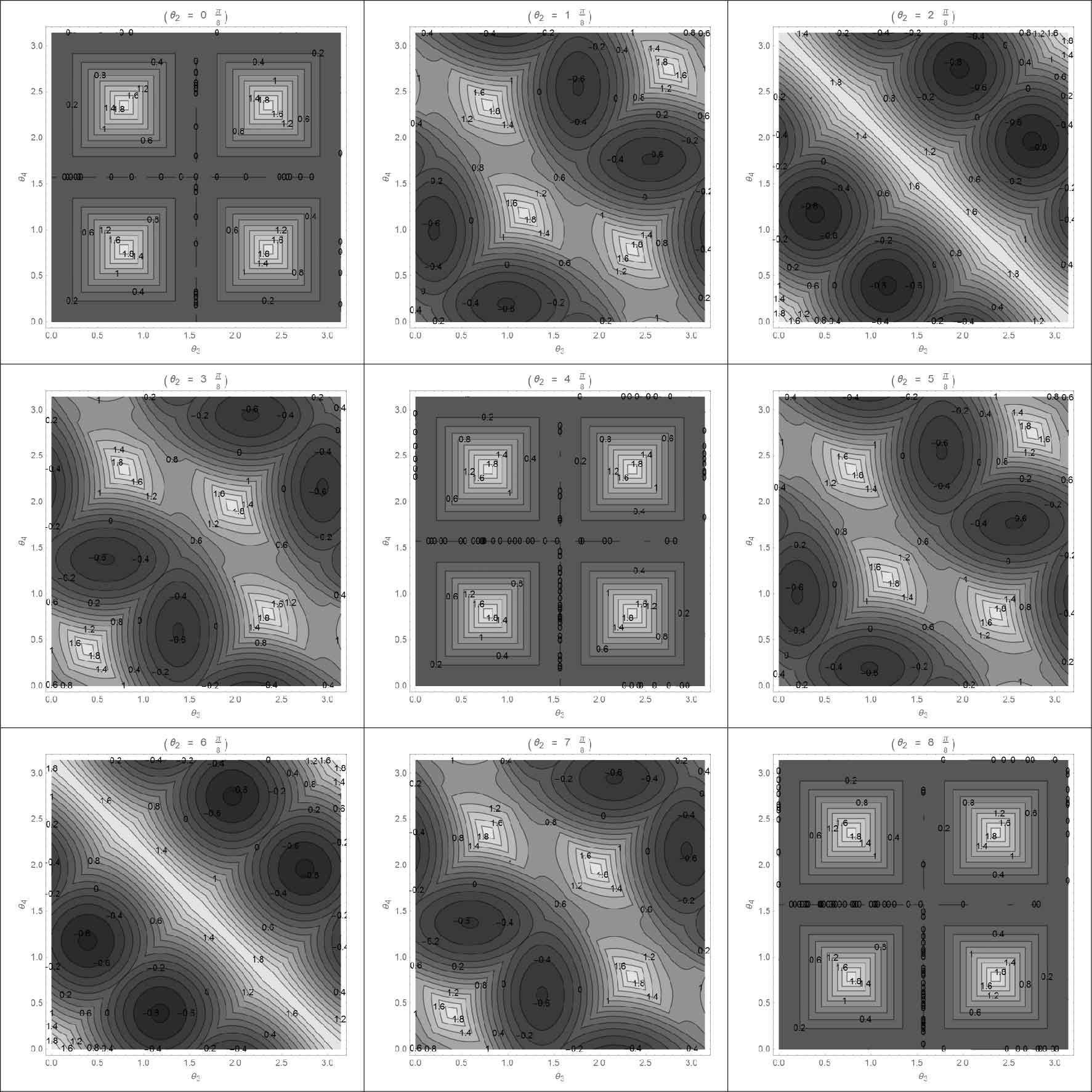}
  \end{tabular}
\end{table}

\end{widetext}

\ifthenelse{\boolean{pt-br}}
  {\ptbr
  {Nova perspectiva da modelagem probabil\'istica do experimento}
  }

\section{New perspective of the probabilistic modeling of the experiment}
\label{sec:Prob_Cond}

By analyzing the experiment, it is observed that there is an associated probability for each possible route, supposing that they are among themselves regardless the paths. Therefore, $p_{1}$ is given for the photon 1 when it deviates and $(1-p_{1})$ for not deviating. Similary, for the photon 2, $p_{2}$ is given for the probability of deviation and $(1-p_{2})$ otherwise. Therefore, a random variable $W_k$ can be associated to each route, with $k\in\{1,2\}$ (for each photon). The photon $k$ responsible for the deviation corresponds to $W_k=2$, and $W_k=1$. In that way, the random variables regarding the passage of the photons to the polarizers is given by $(Z,Z^{'})$ and regarding the route traveled is $(W_{1},W_{2})$ until arriving at the polarizers 

\begin{widetext}

  \begin{figure}[ht]

    \includegraphics[scale=0.2]{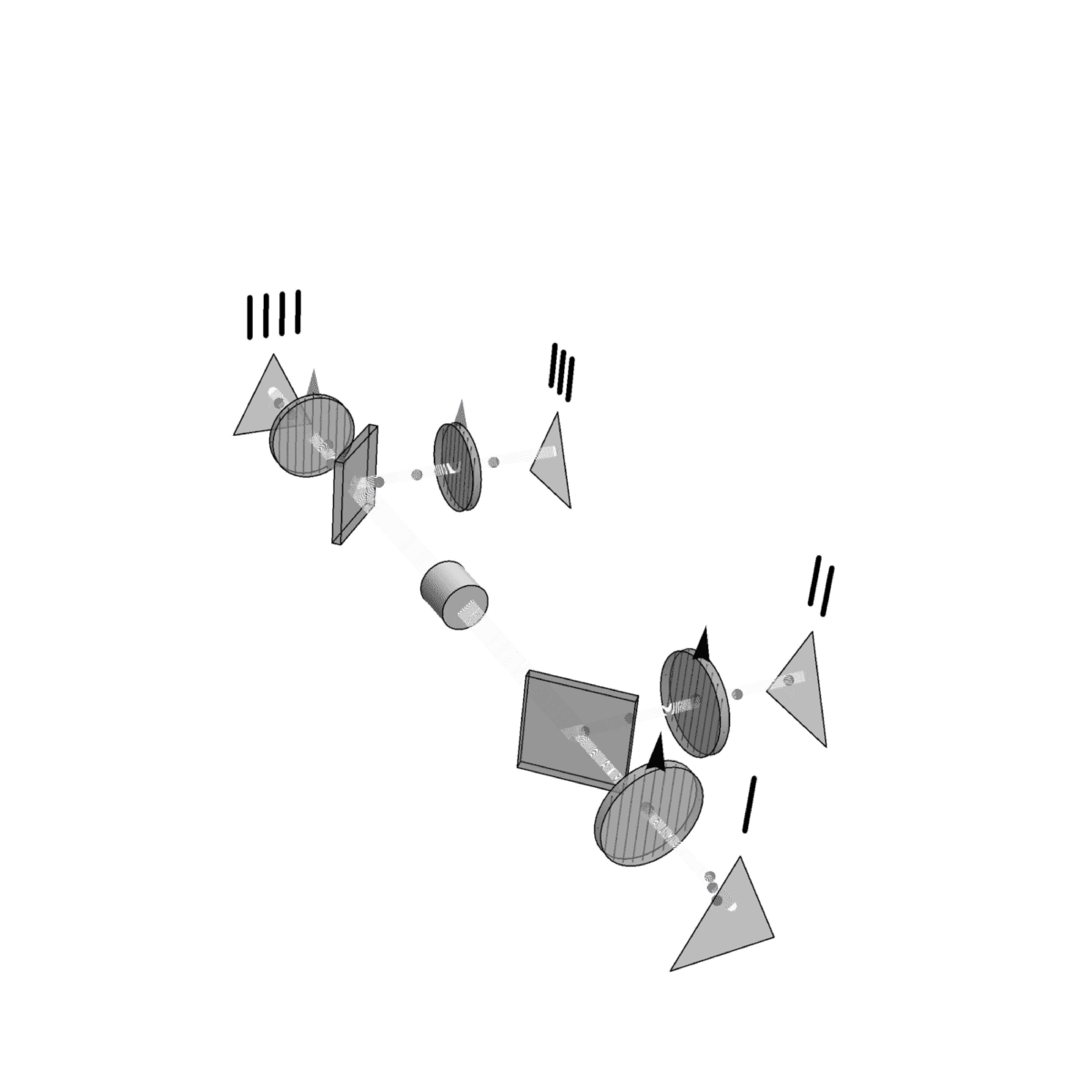}
    
    \ifthenelse{\boolean{pt-br}}{
    \ptbr{
    Esquema experimental do experimento de Alain Aspect: o cilindro no centro
    \'e a fonte de f\'otons, os ret\^angulos s\~ao os espelhos semitransparentes, os c\'irculos s\~ao
    polarizadores e os tri\^angulos s\~ao os detectores
    }}
    
    \caption{ 
    Experimental scheme of the experiment of Alain Aspect: the cylinder in the center is the source of photons, the rectangles are the semitransparent mirrors, the circles are polarizers and the triangles are the detectors}
  \end{figure}

  \begin{eqnarray}\nonumber
    &&{{\mathcal P}}_{\ddot Z,\ddot W}\left(z_{1},z_{2},z^{'}_1,z^{'}_2,w_{1},w_{2}\right)
    =\left\{
    \begin{array}{c}
      {{\mathcal P}}_{W_{1}}(w_{1})\cdot {{\mathcal P}}_{W_{2}}(w_{2})\cdot {{\mathcal P}}_{Z_{1},Z^{'}_1}(z_{1},z^{'}_1,{\vartheta}_{1,1})
      \Leftarrow (w_{1},w_{2},z_{2},z^{'}_2)=(1,1,0,0)
      \\ 
      {{\mathcal P}}_{W_{1}}(w_{1})\cdot {{\mathcal P}}_{W_{2}}(w_{2})\cdot {{\mathcal P}}_{Z_{1},Z^{'}_2}(z_{1},z^{'}_2,{\vartheta}_{1,2})
      \Leftarrow (w_{1},w_{2},z_{2},z^{'}_1)=(1,2,0,0)
      \\ 
      {{\mathcal P}}_{W_{1}}({w}_{1})\cdot {{\mathcal P}}_{W_{2}}(w_{2})\cdot {{\mathcal P}}_{Z_{2},Z^{'}_1}(z_{2},z^{'}_1,{\vartheta}_{2,1})
      \Leftarrow (w_{1},w_{2},z_{1},z^{'}_{2})=(2,1,0,0)
      \\ 
      {{\mathcal P}}_{W_{1}}(w_{1})\cdot {{\mathcal P}}_{W_{2}}(w_{2})\cdot {{\mathcal P}}_{Z_{2},Z^{'}_2}(z_{2},z^{'}_2,{\vartheta}_{2,2})
      \Leftarrow (w_{1},w_{2},z_{1},z^{'}_1)=(2,2,0,0)
    \end{array}
    \right.,
    \\\nonumber
    &&Z_1^{'}:=Z_3,\quad Z_2^{'}:=Z_4,\quad \ddot Z:=(Z_1,Z_2,Z_1^{'},Z_2^{'}),\quad \ddot W:=(W_1,W_2),
    \quad \vartheta_{j,k}:=\theta_k^{'}-\theta_j,\quad \theta_1^{'}:=\theta_3,\quad \theta_2^{'}:=\theta_4
  \end{eqnarray}
  
\end{widetext}

\[
  {\mathcal P}_{W_k}\left(w_k\right)
  =\left\{
  \begin{array}{ll}
    p_k&\Leftarrow w_k=2
    \\ 
    1-p_k&\Leftarrow w_k=1
  \end{array}
  \right.,\quad
  0\le p_k\le 1,\quad
  k\in\left\{1,2\right\}
\]

Where $Z_j=0$ and $Z_k^{'}=0$ mean the absence of photons, in other words, the path was not correspondent to $(W_{1},W_{2})=(j,k)$ (There is another path).

Therefore, the conditional probability, given that the photons traveled for the route related with the random variables $(W_{1},W_{2})$ with same values, $(w_{1},w_{2})$ will be 

\begin{eqnarray}\nonumber
  &&{{\mathcal P}}_{\ddot Z;\ddot W}\left(z_{1},z_{2},z^{'}_1,z^{'}_2;w_{1},w_{2}\right)=
  \\*
  &&=\left\{
  \begin{array}{c}
    {{\mathcal P}}_{Z_{1},Z^{'}_1}(z_{1},z^{'}_1,{\vartheta}_{1,1})\Leftarrow (w_{1},w_{2},z_{2},z^{'}_2)=(1,1,0,0)
    \\ 
    {{\mathcal P}}_{Z_{1},Z^{'}_2}(z_{1},z^{'}_2,{\vartheta}_{1,2})\Leftarrow (w_{1},w_{2},z_{2},z^{'}_1)=(1,2,0,0)
    \\ 
    {{\mathcal P}}_{Z_{2},Z^{'}_1}(z_{2},z^{'}_1,{\vartheta}_{2,1})\Leftarrow (w_{1},w_{2},z_{1},z^{'}_2)=(2,1,0,0)
    \\ 
    {{\mathcal P}}_{Z_{2},Z^{'}_2}(z_{2},z^{'}_2,{\vartheta}_{2,2})\Leftarrow (w_{1},w_{2},z_{1},z^{'}_1)=(2,2,0,0)
  \end{array}
  \right.
  \nonumber
\end{eqnarray}

The conditional expectation (the point sign and comma separates the random variables $(Z,Z^{'})$ of those observed $(W_{1},W_{2})$) of the product $Z\cdot Z^{'}$ will be 

\begin{eqnarray}\nonumber
  &&{{\mathcal E}}_{\ddot Z;\ddot W}(Z_{w_{1}}\cdot Z^{'}_{w_2};w_{1},w_{2})=
  \\\nonumber
  &&=\sum_{(z_{w_{1}},z_{w_{2}}^{'})\in\{-1,+1\}^2}
    (z_{w_{1}}\cdot z^{'}_{w_2}\cdot{{\mathcal P}}_{Z_{w_1},Z_{w_2};\ddot W}(z_{w_1},z_{w_2}^{'};w_{1},w_{2}))
\end{eqnarray}

\noindent
therefore 

\begin{equation}\nonumber
  {{\mathcal E}}_{\ddot Z;\ddot W}(Z_{w_{1}}\cdot Z^{'}_{w_2};w_{1},w_{2})
  =\rm{cos}(2{\vartheta}_{w_{1},w_{2}})
\end{equation}

\noindent
thereby, with that understanding, the substitution done in the inequality would not be related to the expected value ${{\mathcal E}}_{\ddot Z}$, but to the conditional expectation ${{\mathcal E}}_{\ddot
Z;\ddot W}$, therefore 

\begin{eqnarray}\nonumber
  &&\rm{cos}(2 {\vartheta}_{1,1})-\rm{cos}(2 {\vartheta}_{1,2})+\rm{cos}(2 {\vartheta}_{2,1})+\rm{cos}(2   {\vartheta}_{2,2})=
  \\\nonumber
  &&={{\mathcal E}}_{\ddot Z;\ddot W}(Z_{1}\cdot Z^{'}_1;1,1)
    -{{\mathcal E}}_{\ddot Z;\ddot W}(Z_{1}\cdot Z^{'}_2;1,2)+
  \\\nonumber
  &&\qquad+{{\mathcal E}}_{\ddot Z;\ddot W}(Z_{2}\cdot Z^{'}_1;2,1)
    +{{\mathcal E}}_{\ddot Z;\ddot W}(Z_{2}\cdot Z^{'}_2;2,2)
\end{eqnarray}

This result makes no sense from the standpoint of statistic because it is mixing conditional expectation from different events. The conditional probability of a certain event possesses the same properties of the probability theory, but the event that has happened is maintained fixed.

The expectation of $Z_j\cdot Z^{'}_k$ with such function of probability would be given by 

\begin{equation}
  {{\mathcal E}}_{\ddot Z,\ddot W}\left(Z_j\cdot Z^{'}_k\right)
  ={{\mathcal E}}_{\ddot W}({{\mathcal E}}_{\ddot Z;\ddot W}(Z_j\cdot Z^{'}_k;W_{1},W_{2}))
  \nonumber
\end{equation}

\begin{widetext}

  \begin{equation}
    {{\mathcal E}}_{\ddot Z;\ddot W}\left(Z_j\cdot Z^{'}_k;W_{1},W_{2}\right)=\left\{\begin{array}{ll}
    {{\mathcal P}}_{W_{1}}(W_{1})\cdot {{\mathcal P}}_{W_{2}}(W_{2})\cdot \mathrm{cos}(2 {\vartheta}_{W_{1},W_{2}})
    &\Leftarrow ((j=W_{1})\wedge (k=W_{2}))  
    \\ 
    0
    &\Leftarrow \neg((j=W_{1})\wedge (k=W_{2})) \end{array}
    \right.
  \nonumber
  \end{equation}
\end{widetext}

\begin{eqnarray}\nonumber
  &&{{\mathcal E}}_{\ddot Z,\ddot W}\left(Z_j\cdot Z^{'}_k\right)
  ={{\mathcal P}}_{W_{1}}(j)\cdot {{\mathcal P}}_{W_{2}}(k)\cdot \rm{cos}(2 {\vartheta}_{j,k})
\end{eqnarray}

Therefore, the inequality will be 

\begin{eqnarray}\nonumber
  &&|(1-p_{1})\cdot(1-p_{2})\cdot\mathrm{cos}(2{\vartheta}_{1,1})
    \\
    \nonumber
    &&\qquad-(1-p_{1})\cdot p_{2}\cdot\mathrm{cos}(2{\vartheta}_{1,2})+ 
    \\
    \nonumber
    &&\qquad+p_{1}\cdot(1-p_{2})\cdot\mathrm{cos}(2 {\vartheta}_{2,1})+p_{1}\cdot p_{2}\cdot \mathrm{cos}(2 {\vartheta}_{2,2})|\le
    \\
    \nonumber
    &&\le|(1-p_{1})\cdot (1-p_{2})+(1-p_{1})\cdot p_{2}
    \\
    \nonumber
    &&\qquad+p_{1}\cdot(1-p_{2})+p_{1}\cdot p_{2}|=
    \\
    \nonumber
  &&=\left|\left(1-p_{1}\right)\cdot\left(1-p_{2}+p_{2}\right)+p_{1}\cdot\left(1-p_{2}+p_{2}\right)\right|
  \\
  \nonumber
  &&=\left|\left(1-p_{1}\right)+p_{1}\right|=1\le 2
\end{eqnarray}

Hence, besides being reasonable with the theory of the probability, the inequality is mathematically impossible to be violated. Nevertheless, if values are attributed in the original inequality,

\begin{eqnarray}\nonumber
  |\rm{cos}(2{\vartheta}_{1,1})-\rm{cos}(2{\vartheta}_{1,2})+\rm{cos}(2{\vartheta}_{2,1})+\rm{cos}(2{\vartheta}_{2,2})|=
  \\
  \nonumber
  = |1+1+1+1|= 4
\end{eqnarray}

\noindent
which shows that it is mathematically possible to violate the original inequality.

\section{Computational Simulations}
\label{sec:Sim}

\ifthenelse{\boolean{pt-br}}
  {\ptbr
  {Nesta se\c{c}\~ao ser\'a demonstrado como os dados amostrais s\~ao utilizados, evidenciando
  a interpreta\c{c}\~ao te\'orica da se\c{c}\~ao anterior, em que utiliza-se a probabilidade
  condicional no c\'alculo dos valores esperados. Portanto, as f\'ormulas te\'oricas para os conjuntos
  de dados amostrais se mostrar\~ao em conformidade com as f\'ormulas te\'oricas populacionais.}
  }

In this section is demonstrated how the sample data is used, evidencing the theoretical comprehention of the previous section, in which is used conditional probability in the calculus of the expected values. So, these theoretical formulas for the set of sample data will be shown in accordance with the theoretical populational formulas.

Computer simulations were conducted, generating up to 100 samples. Each sample was composed of 100 numbers, which estimated the probabilities that are used in the inequality of Clauser-Horne-Shimony-Horne in order to verify if there is a violation or not regarding the samples. The numbers ranged from 1 to 1000; thereby, there was a total of 100 estimates (an estimate of each sample) of probabilities.

The estimations were based on the theory in order to generate samples of a distribution through a sample of uniform distribution. Each number was associated with an event. The events, similar to the Aspect experiment scheme (performed to test the inequality of Clauser-Horne-Shimony-Holt), are as follows:

\begin{itemize}
  \item 
  When two photons pass through the semitransparent mirrors. The probability of this event is $\frac{1}{4}$. The numbers from 1 to 250 correspond to this event. In this event $j$ is equal to 1 
  \item
  When the first photon pass through the semitransparent mirrors and the second turns away. The probability of this event is $\frac{1}{4}$. The figures corresponding to this event range from 251 to 500. In this event $j$ is equal to $2$ 
  \item
  When the first photon is deflected and the second photon passes. The probability is $\frac{1}{4}$. The references range from 501 to 750. In this event $j$ is equal to 3 
  \item
  When two photons fall away. The probability is $\frac{1}{4}$. The correspondent numbers range from 751 to 1000. In the event $j$ is equal to 4. 
\end{itemize}

It is noted that in each of these events the number of intervals associated with them have 250 numbers, so they are equiprobable. To make it simplier, after knowing that the event held at the generated number, n defined as $D$, the value is calculated by the following expression: $D(j-1)\cdot 250$. Thus, it follows that:

\begin{itemize}
  \item
  If a number $D$ is generated from 1 to 250, $j$ is 1. Therefore, the following operation is performed: attribute to $D$ the value of $D(j-1)\cdot 250$. So if $j=1$ the value obtained will own $D$, which will be 1 to 250. 
  \item
  If $j=2$, therefore $251\le D\le 500$, the value obtained is $D-250$, which will be 1 and 250 
  \item
  If $j=3$, therefore $501\le D\le 750$, the value obtained is $D-500$, which will be 1 and 250 
  \item
  If $j=4$, therefore $751\le D\le 1000$, the value obtained is $D-750$, which will be 1 and 250 
\end{itemize}

For each of these events, each photon encounters in their paths (any path) a polarizer. Each polarizer is oriented in a particular direction and these orientations are maintained fixed for all 1000 values generated within the sample; changes happen only from sample to sample. By passing through the polarizer, there are two possibilities: either the photon passes or not. Therefore, there are four events for each of the four possible trajectories obtained by passing the mirrors. These four events are:  

\begin{itemize}
  \item
  Both photons do not pass through the respective polarizers. The event probability is $p_j$ (in which $j$ refers to one of four events related to the mirrors). For this event, the new value of $D$ must satisfy the following inequality $0<D\le p_j\cdot 250$ 
  \item
  The first photon does not pass and the second pass. The probability of this event is $(0.5-p_j)$. For this event, the new value of $D$ must satisfy the following inequality: $p_j\cdot 250<D\le 0.5\cdot 250$ 
  \item
  The first photon passes and the second not. The probability of this event is $(0.5-p_j)$. For this event, the new value of $D$ must satisfy the following inequality: $0.5<D\le (1-p_j)\cdot $ 250 
  \item
  Both photons pass. The probability of this event is $p_j$. For this event, the new value of $D$ must satisfy the following inequality: $(1-p_j)\cdot 250<D\le 250$ 
\end{itemize}

\section{Results of simulations}
\label{sec:Sim_Resultados}

The following tables show the values obtained by the theory as well as the statistics obtained through the simulation.

Each table refers to a set of values that were obtained by fixing the value of parameter ${\theta}_{2}$ (the value of ${\theta}_{2}$ is in the cell of black background, in the upper left corner). At the top (on the cells gray background), are the values of ${\theta}_{3}$ and on left (on gray background of cells) are the values of ${\theta}_{4}$.

The ${\theta}_{1}$ is set to 0 without loss of generality.

The values of ${\theta}_j$ (with $j\in \{2,3,4\}$) are submultiples of $\pi $: $\{\frac{\pi}{16},\frac{2\pi}{16},\frac{3\pi}{16},\frac{4\pi}{16},\frac{5\pi}{16},\frac{6\pi}{16},\frac{7\pi}{16}\}$.

There were 7 values for each of the 3 parameters (${\theta}_{2}$  ${\theta}_{3}$ and ${\theta}_{4}$); for example, a total of ${7}^{3}=343$ combinations of values.

The cells are the values and the color bars are related these values. The bars in white represent positive values, and the wider the bars, the higher the values. The gray bars represent negative values and the wider the bars, the smaller the values. Negative values represent a violation of inequalities. 

The following will be presented in seven tables (one for each value of ${\theta}_{2}$). They were obtained theoretically from CHSH inequality that is given by:

\begin{eqnarray}\nonumber
  2-|-{{\mathcal E}}_{Z_{1},Z_{3}}(Z_{1}\cdot Z_{3})+{{\mathcal E}}_{Z_{1},Z_{4}}(Z_{1}\cdot Z_{4})
    \\
    \nonumber
    +{{\mathcal E}}_{Z_{2},Z_{3}}(Z_{2}\cdot Z_{3})+{{\mathcal E}}_{Z_{2},Z_{4}}(Z_{2}\cdot Z_{4})|
  \ge 0
\end{eqnarray}

\noindent
performing the following substitution 

\begin{eqnarray}
\label{eq:subEspErrado}
  &&{{\mathcal E}}_{Z_j,Z_k}(Z_j\cdot Z_k)
  =\mathrm{cos}(2 {\bar\theta}_{j,k}),
\end{eqnarray}

\noindent
results in the following inequality

\begin{eqnarray}\nonumber
  2-|-\mathrm{cos}(2{\bar\theta}_{1,3})+\mathrm{cos}(2{\bar\theta}_{1,4})
  +\mathrm{cos}(2{\bar\theta}_{2,3})+\mathrm{cos}(2{\bar\theta}_{2,4})|
  \\\nonumber
  \ge 0
\end{eqnarray}

\noindent
and, if the first member of the inequality is greater than or equal to 0, the inequality is obeyed; otherwise there is a violation of inequality.

The values correspond to the value obtained from the first member of inequality (expression that comes before the inequality sign $\ge $). 

\begin{widetext}

  \begin{table}[ht]
    \caption{Theoretical values of the first member of the CHSH inequality for ${\theta}_{1}=0$ e ${\theta}_{2}=\frac{\pi}{16}$ }
    \includegraphics[scale=0.4]{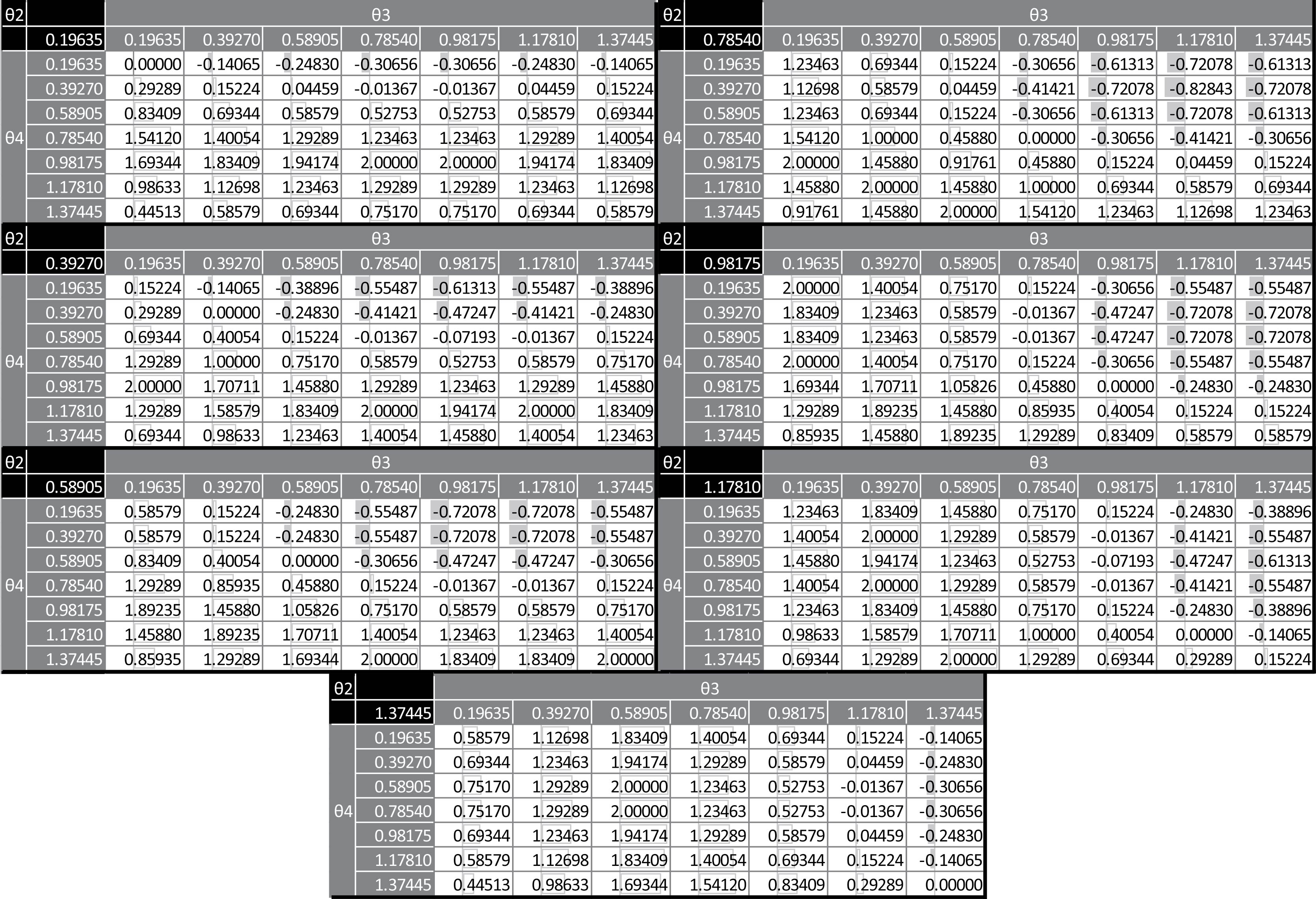}
  \end{table}

\end{widetext}

In the tables above, it is observed that there were violations (negative values) of the CHSH inequality.

The following tables are obtained through the average of 100 estimates of the CHSH inequality. Each sample contained 1000 elements, which were used to calculate the inequality of CHSH that is given by:

\begin{equation}
\label{eq:subEstEspErrado}
  \textstyle
  {\widehat{{\mathcal E}}}_{Z_j,Z_k}(Z_j\cdot Z_k)
  =\frac{n^{--}_{j,k}-{n}^{-+}_{j,k}-n^{+-}_{j,k}+n^{++}_{j,k}}{n_{j,k}}
\end{equation}

\noindent
where  

\begin{itemize}
  \item
  $n^{--}_{j,k}$ is the number of occurrences of the event when none of the photons passes through the polarizers $j$ and $k$ 
  \item
  $n^{-+}_{j,k}$ is the number of occurrences of the event when the first photon does not pass and the second passes through the polarizers $j$ e $k$, respectively 
  \item
  $n^{+-}_{j,k}$ is the number of occurrences of the event when the first photon passes and the second not pass the polarizers $j$ e $k$, respectively 
  \item
  $n^{++}_{j,k}$ is the number of occurrences of the event when both photons pass through the polarizers $j$ e $k$ 
  \item
  $n_{j,k}=(n^{--}_{j,k}+n^{-+}_{j,k}+n^{+-}_{j,k}+n^{++}_{j,k})$ is the number of photons that have taken the path to the polarizers $j$ e $k$ regardless of whether or not it passes through the polarizers.
  \item
  $j\in \{1,2\}$ e $k\in \{3,4\}$ 
\end{itemize}

The inequality based in each sample is given by

\begin{eqnarray}\nonumber
  2-|-{\widehat{{\mathcal E}}}_{Z_{1},Z_{3}}(Z_{1}\cdot Z_{3})
    +{\widehat{{\mathcal E}}}_{Z_{1},Z_{4}}(Z_{1}\cdot Z_{4})
    \\
    \nonumber
    +{\widehat{{\mathcal E}}}_{Z_{2},Z_{3}}(Z_{2}\cdot Z_{3})
    +{\widehat{{\mathcal E}}}_{Z_{2},Z_{4}}(Z_{2}\cdot Z_{4})|
  \ge 0
\end{eqnarray}

If the first member of inequality is greater than or equal to 0, the inequality is obeyed; otherwise there is a violation of the inequality. The values correspond to the sample average of 100 values (one value for each of the 100 samples) that were obtained from the first member of the inequality. It was calculated for each sample. 

\begin{widetext}

  \begin{table}[ht]
    \caption{Sampled values for ${\theta}_{1}=0$. On the left are the average values of 100 sample values of the first member of CHSH's inequality. On the right are the standard deviations of the values obtained from 100 samples.}
    \includegraphics
      [scale=0.4]
      {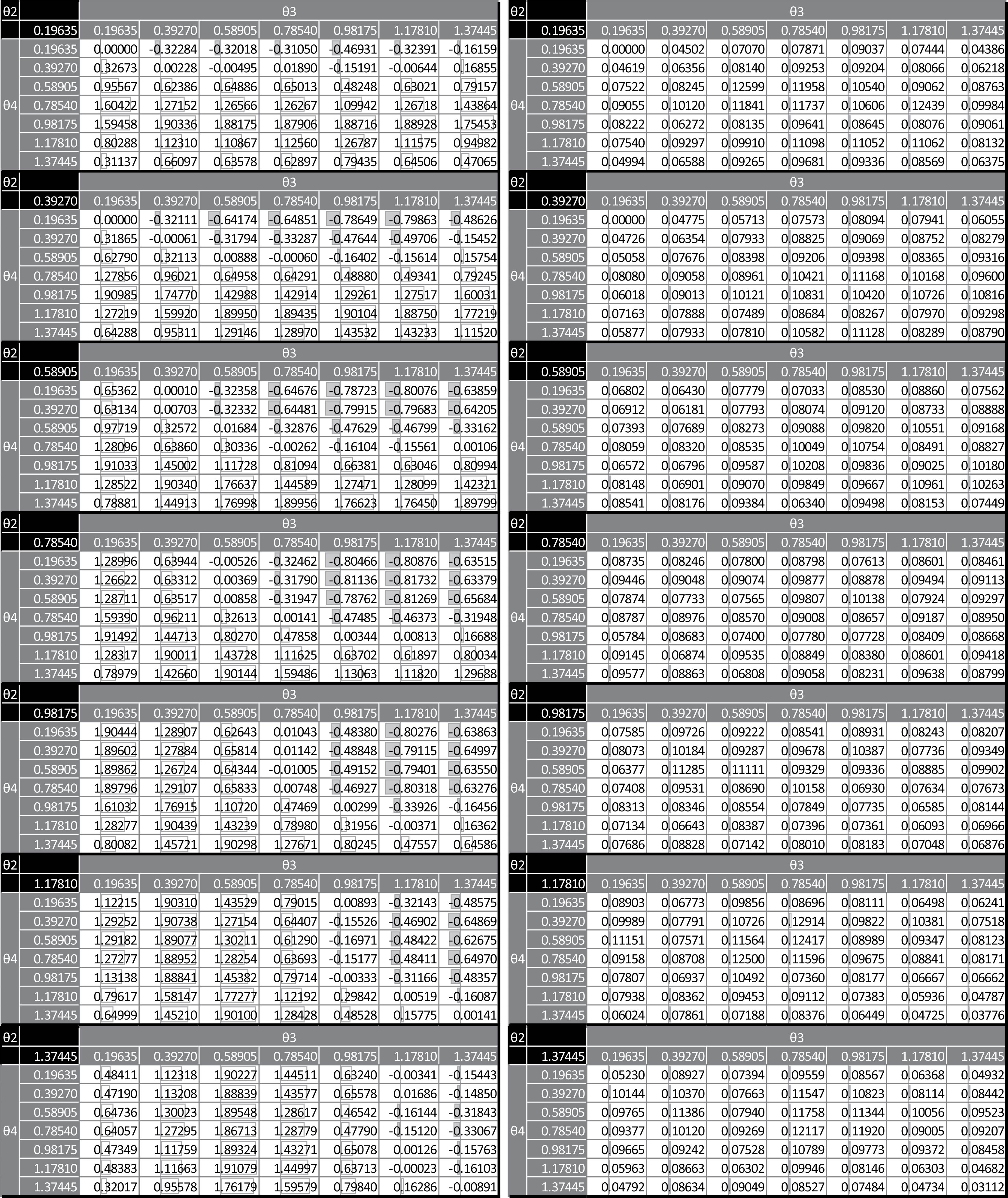}
  \end{table}

\end{widetext}

\ifthenelse{\boolean{pt-br}}
  {\ptbr
  {Devemos lembrar que as substitui\c{c}\~oes (\ref{eq:subEspErrado}) e (\ref{eq:subEstEspErrado}) s\~ao
  as f\'ormulas encontradas na literatura. Contudo, de acordo com o que apresentamos neste artigo, tais
  valores esperados s\~ao valores esperados condicionais.}
  }

We ought remember that the substitutions (\ref{eq:subEspErrado}) and (\ref{eq:subEstEspErrado}) are formulas found in literature. However, according to what was presented in this article,  such expected values are expected conditional values.

\ifthenelse{\boolean{pt-br}}
  {\ptbr
  {A f\'ormula (\ref{eq:subEstEspErrado}) tem em seu denominador a quantidade $n_{j,k}$, que representa
  apenas a quantidade de observa\c{c}\~oes relacionadas aos polarizadores $j$ e $k$, o que evidencia
  que a estimativa do valor esperado \'e condicional, ou seja, \'e a estimativa quando se considera
  apenas as observa\c{c}\~oes em que as part\'iculas tomaram o caminho especificamente para os polarizadores
  $j$ e $k$, desconsiderando as outras observa\c{c}\~oes. Se tal estimativa fosse para o valor esperado
  do experimento, ent\~ao dever\'iamos encontrar no denominado a quantidade $n$ (em vez de $n_{j,k}$),
  pois assim todas as observa\c{c}\~oes do experimento estariam no c\'alculo do valor esperado.}
  }

The formula (\ref{eq:subEstEspErrado}) has its denominador the quantity $n_{j,k}$, that represents only the quantity of observations related to the polarizers $j$ and $k$, evidencing that the estimate of the expected vallue is conditional, that is, it is the estimation when it is considered only the observations in which the particles were taking place  specificly for those polarizers $j$ and $k$, disregarding other observations. If such estimation was made for the expected value for the experiment, so we should find in the denominator the quantity $n$ (rather than $n_{j,k}$) because all these observations would be in the calculus of the expected value.

\ifthenelse{\boolean{pt-br}}
  {\ptbr
  {Assim, na f\'ormula (\ref{eq:subEspErrado}), o correto seria escrever que}
  }

So, in the formula (\ref{eq:subEspErrado}), the correct way would be to write

\begin{eqnarray}\nonumber
  \mathcal{E}_{Z_j,Z_K^{'};\ddot W}(Z_j\cdot Z_K^{'};j,K)=\cos(2\bar\theta_{j,k}),
\end{eqnarray}

\ifthenelse{\boolean{pt-br}}
  {\ptbr
  {e em (\ref{eq:subEstEspErrado}), seria}
  }

\noindent
in which (\ref{eq:subEstEspErrado}) would be

\begin{eqnarray}\nonumber
  \hat{\mathcal{E}}_{Z_j,Z_k^{'};\ddot W}(Z_j\cdot Z_k^{'};j,K)
  =\tfrac{n^{--}_{j,k}-{n}^{-+}_{j,k}-n^{+-}_{j,k}+n^{++}_{j,k}}{n_{j,k}}
\end{eqnarray}

\ifthenelse{\boolean{pt-br}}
  {\ptbr
  {com k=3 quando K=1 e k=4 quando K=2.}
  }

\noindent
with $k=3$ when $K=1$ and $K=4$ when $K=2$.

As for the CHSH inequality, where the expected values they are estimated from the conditioned probabilities of the event of photons that have or not been deflected to pass through the semitransparent mirrors, the CHSH inequality is kept constant. However, the expected values are as follows: 

\begin{eqnarray}\nonumber
  &&{{\mathcal E}}_{Z_j,Z_k^{'}}(Z_j\cdot Z_k^{'})
  ={{\mathcal P}}_{W_{1},W_{2}}(j,k)\cdot \mathrm{cos}(2{\vartheta}_{j,k})
  \\\nonumber
  &&Z_1^{'}=Z_3,\ Z_2^{'}=Z_4,\ \vartheta_{j,k}=\theta_j-\theta_k^{'},\ \theta_1^{'}=\theta_3,\ \theta_2^{'}=\theta_4
\end{eqnarray}

\noindent
since ${{\mathcal P}}_{W_{1},W_{2}}$ is the function of probability related with the pass or reflection of photons by transparent mirrors, in the case it was adopted that all possible trajectories are equally likely, therefore 

\[\textstyle
  {{\mathcal P}}_{W_{1},W_{2}}(j,k)
  =\frac{1}{4}
\]

\noindent
with $j$ and $k$ belonging to the set $\{1,2,3,4\}$, the expected value that is replaced in the inequality is 

\[\textstyle
  {{\mathcal E}}_{Z_j,Z_k}(Z_j\cdot Z_k)
  =\frac{1}{4}\cdot \rm{cos}(2{\bar\theta}_{j,k})
\]

\noindent
resulting in the following inequality 

\begin{eqnarray}\nonumber\textstyle
  2-\left|\frac{-\rm{cos}(2{\bar\theta}_{1,3})
  +\rm{cos}(2{\bar\theta}_{1,4})
  +\rm{cos}(2{\bar\theta}_{2,3})
  +\rm{cos}(2{\bar\theta}_{2,4})}{4}\right|
  \ge 0
\end{eqnarray}

\noindent
if the first member of the inequality is greater than or equal to 0, the inequality is obeyed; otherwise there is a violation of the inequality. The values correspond to values that were obtained from the first member of inequality.

\begin{widetext}

  \begin{table}[ht]
    \caption{Theoretical values for ${\theta}_{1}=0$ from first member of the CHSH inequality, modeled via conditional probability }
    \includegraphics[scale=0.4]{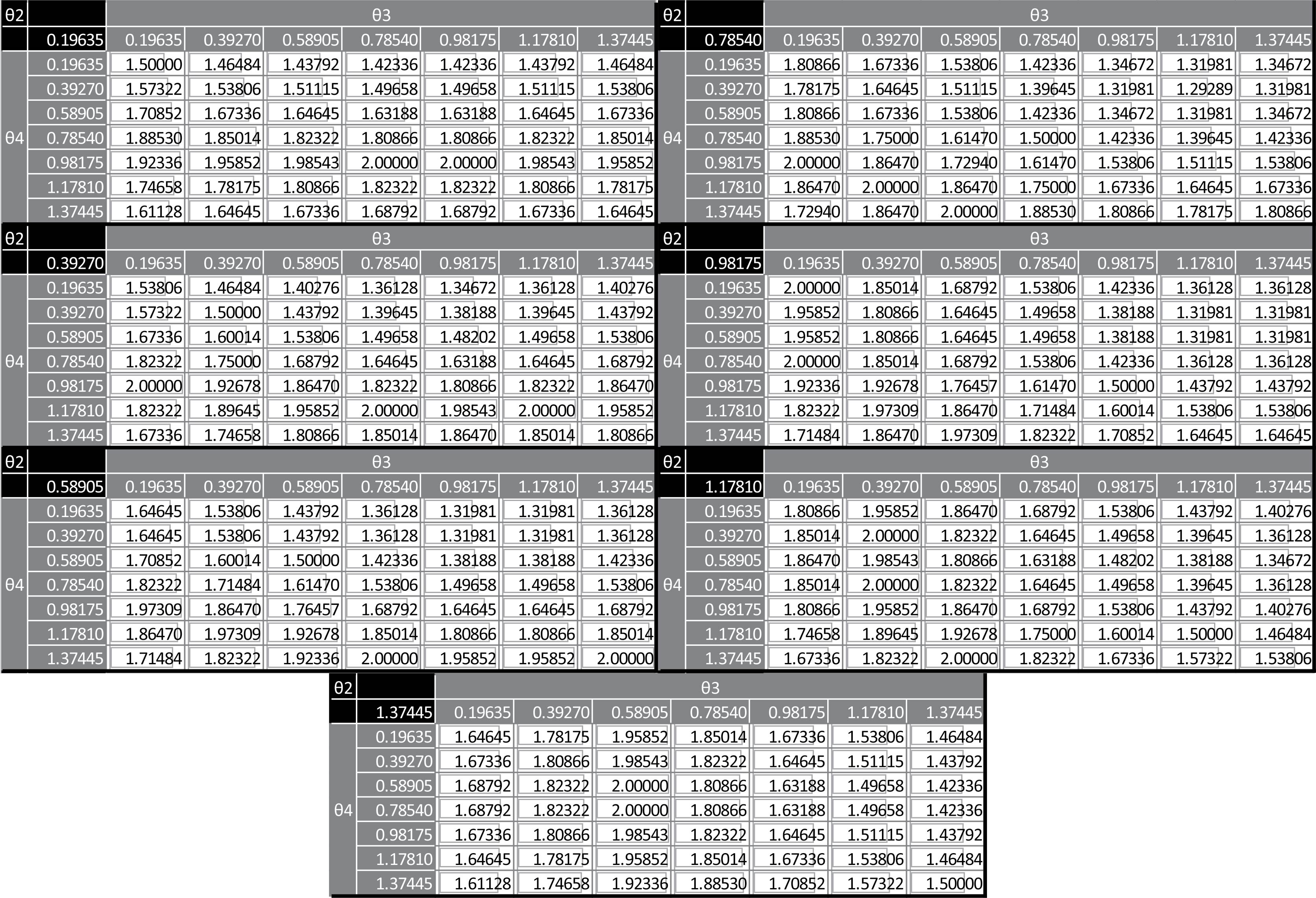}
  \end{table}

\end{widetext}

On the tables above, it is observed that there was no violation (negative values) of the CHSH's inequality (odds conditional modeled).

The following tables show estimates that were obtained by the average of 100 estimates of CHSH's inequality, which represent the modeled conditional probability. The same samples were previously used, but the calculation of expected values were performed by using the expected sample values (i.e.: the total number of events $n$ were used as the denominator) 

\[\textstyle
  {\widehat{{\mathcal E}}}_{Z_j,Z_k}(Z_j\cdot Z_k)=\frac{n^{--}_{j,k}-n^{-+}_{j,k}-n^{+-}_{j,k}+n^{++}_{j,k}}{n}
\]

\noindent
wherein:  
\begin{itemize}
  \item
  $n^{--}_{j,k}$, $n^{-+}_{j,k}$, $n^{+-}_{j,k}$ e $n^{++}_{j,k}$ have the same previously attributed meaning 
  \item
  $n=(n_{1,3}+n_{1,4}+n_{2,3}+n_{2,4})$ is the total number of events when there are 1000 occurrences for each sample 
\end{itemize}

Since the CHSH's inequality can be rewritten as 

\[
  2-|{{\mathcal E}}_{Z_{1},Z_{2},Z_{3},Z_{4}}(Z_{1}\cdot Z_{3}+Z_{1}\cdot Z_{4}+Z_{2}\cdot Z_{3}+Z_{2}\cdot Z_{4})|
  \ge 0
\]

\noindent
then, the inequality (through the sample values) should be calculated by 

\begin{widetext}
  \[\textstyle
    2-\left|\frac{-(n^{--}_{1,3}-n^{-+}_{1,3}-n^{+-}_{1,3}+n^{++}_{1,3})+(n^{--}_{1,4}-n^{-+}_{1,4}-n^{+-}_{1,4}+n^{++}_{1,4})+(n^{--}_{2,3}-n^{-+}_{2,3}-n^{+-}_{2,3}+n^{++}_{2,3})+(n^{--}_{2,4}-n^{-+}_{2,4}-n^{+-}_{2,4}+n^{++}_{2,4})}{n}\right|\ge 0
  \]
  that is 
  \[\textstyle
    2-\left|\frac{-(n^{--}_{1,3}-n^{-+}_{1,3}-n^{+-}_{1,3}+n^{++}_{1,3})+(n^{--}_{1,4}-n^{-+}_{1,4}-n^{+-}_{1,4}+n^{++}_{1,4})+(n^{--}_{2,3}-n^{-+}_{2,3}-n^{+-}_{2,3}+n^{++}_{2,3})+(n^{--}_{2,4}-n^{-+}_{2,4}-n^{+-}_{2,4}+n^{++}_{2,4})}{+(n^{--}_{1,3}+n^{-+}_{1,3}+n^{+-}_{1,3}+n^{++}_{1,3})+(n^{--}_{1,4}+n^{-+}_{1,4}+n^{+-}_{1,4}+n^{++}_{1,4})+(n^{--}_{2,3}+n^{-+}_{2,3}+n^{+-}_{2,3}+n^{++}_{2,3})+(n^{--}_{2,4}+n^{-+}_{2,4}+n^{+-}_{2,4}+n^{++}_{2,4})}\right|\ge 0
  \]
\end{widetext}

\noindent
which is evident that the numerator will always be less or equal to the denominator; thereby, the fraction (as it can be seen) remained between $-1$ and $+$ 1. Therefore, the inequality becomes

\[
  2-|\pm 1|=2-1=1\ge 0
\]

\noindent
thus, the CHSH's inequality will never be experimentally violated if the data are modeled by this formula.

Since the expected value is being calculated based on $(Z_{1},Z_{2},Z_{3},Z_{4})$, it makes sense to divide the total number of events related to any of the variables $Z$. Therefore, it is plausible that the denominator is given by $n$.

The inequality based in each sample is given by 

\begin{eqnarray}\nonumber
  2-|-{\widehat{{\mathcal E}}}_{Z_{1},Z_{3}}(Z_{1}\cdot Z_{3})
  +{\widehat{{\mathcal E}}}_{Z_{1},Z_{4}}(Z_{1}\cdot Z_{4})
  \\*
  +{\widehat{{\mathcal E}}}_{Z_{2},Z_{3}}(Z_{2}\cdot Z_{3})
  +{\widehat{{\mathcal E}}}_{Z_{2},Z_{4}}(Z_{2}\cdot Z_{4})|
  \ge 0 
  \nonumber
\end{eqnarray}

\noindent
if the first member of the inequality is greater than or equal to 0, the inequality is obeyed; otherwise there is a violation of the inequality. The values correspond to the average sample of the 100 values that were obtained from the calculation of the first member of the inequality for each sample. 

\begin{widetext}

  \begin{table}[ht]
    \caption{Sampled values for ${\theta}_{1}=0$. On the left are the average values of 100 sample values of the first member of CHSH's inequality. On the right are the standard deviations of the values obtained from 100 samples.}
    \includegraphics[scale=0.4]{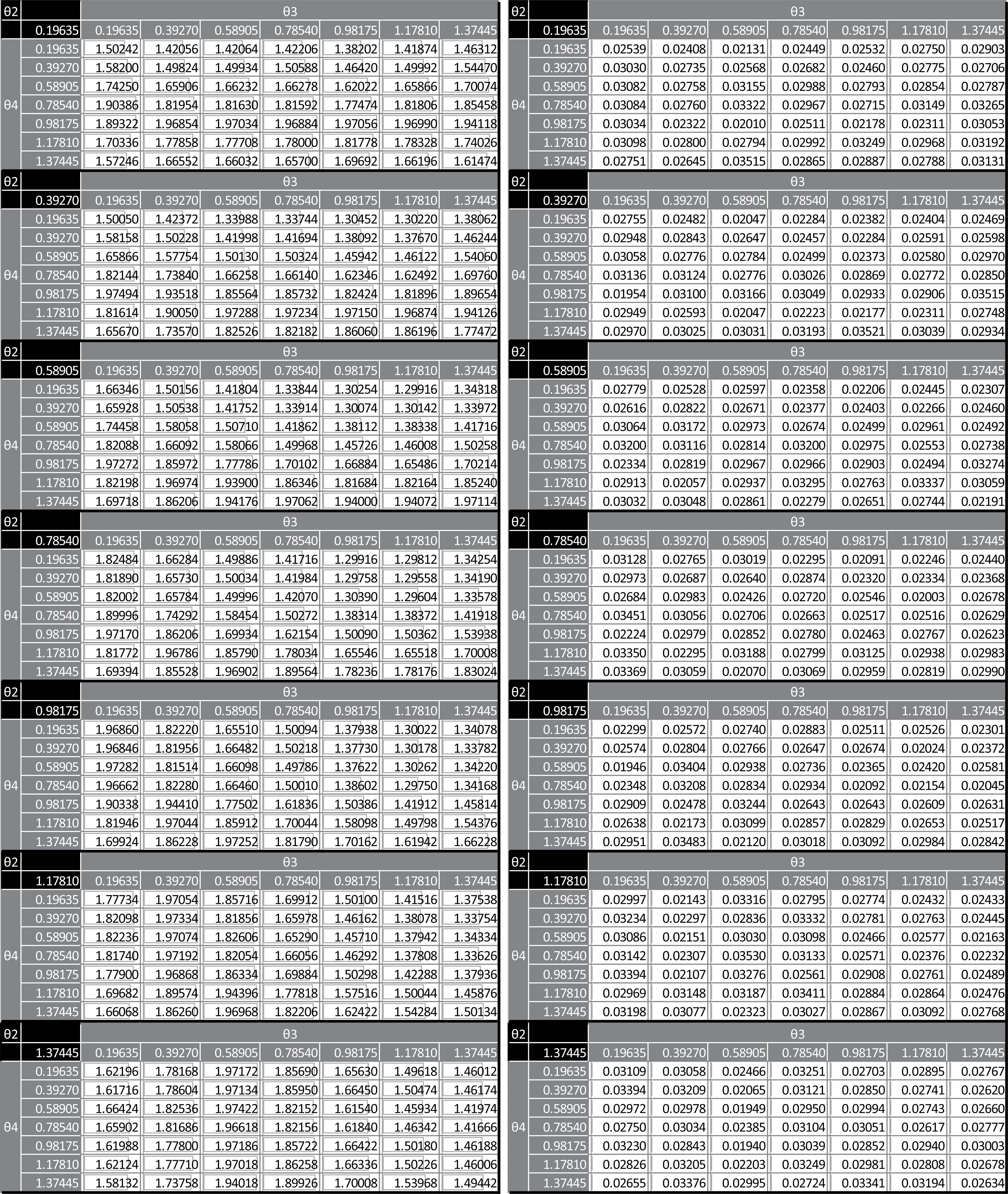}
  \end{table}

\end{widetext}

\section{Analysis and conclusions}
\label{sec:Conclusao}

\ifthenelse{\boolean{pt-br}}
  {\ptbr
  {Neste trabalho, demonstramos (na se\c{c}\~ao \ref{sec:variavel_oculta}) que qualquer depend\^encia
  entre a vari\'avel aleat\'oria $Z_j$ e a vari\'avel oculta $\lambda$ n\~ao altera em nada os valores
  esperados calculados. Portanto qualquer considera\c{c}\~ao feita acerca da vari\'avel oculta $\lambda$
  n\~ao tem relev\^ancia para os valores esperados calculados, isso n\~ao significa que tal vari\'avel
  oculta n\~ao possa ter algum papel importante na cria\c{c}\~ao de um modelo que forne\c{c}a as fun\c{c}\~oes
  de probabilidade usadas.}
  }

In this work, we showed (on the section \ref{sec:variavel_oculta}) that any dependence between the random variable $Z_j$ and the hidden variable $\lambda$ does not change the calculated expected values at all. Therefore any consideration made of the hidden variable $\lambda$ is not relevant to the calculated expected values, this does not mean that such hidden variable can not play any important role in the creation of a model that provides the probability functions used.

\ifthenelse{\boolean{pt-br}}
  {\ptbr{
  De fato, a vari\'avel oculta $\lambda$ tem seu papel reduzido, essencialmente por causa do resultado
  do experimento se reduzir \`a apenas dois valores ($-1$ e $+1$). Nesse caso, quem assume o papel relevante
  \'e a vari\'avel $Z_j$. Uma vez que a desigualdade de CHSH \'e expressa em termos de valores esperados
  de $Z_j\cdot Z_k$ (tal produto tamb\'em resulta em apenas dois valores: $-1$ e $+1$), o foco na vari\'avel
  $Z_j$ fica compat\'ivel com a desigualdade de CHSH, ou seja, tratar de $Z_j$, em vez de $\lambda$,
  n\~ao influ\^encia nas conclus\~oes obtidas atrav\'es de c\'alculos envolvendo os valores esperados
  de $Z_j\cdot Z_k$.}}

In fact, the hidden variable $ \lambda $ has its role reduced, essentially because the result of the experiment reduces to only two values ($ -1 $ and $ + 1 $). In this case, the variable $ Z_j $ plays the relevant role. Since the CHSH inequality is expressed in terms of expected values of $ Z_j \cdot Z_k $ (this product also results in only two values: $ -1 $ and $ + 1 $), the focus on the variable $ Z_j $ compatible with the CHSH inequality, that is, treating $ Z_j $, instead of $ \lambda $, does not influence the conclusions obtained by calculations involving the expected values of $ Z_j \cdot Z_k $.

\ifthenelse{\boolean{pt-br}}
  {\ptbr
  {Mostramos (na se\c{c}\~ao \ref{sec:Sistema}) que o conunto de fun\c{c}\~oes de probabilidade usados
  na literatura podem determinar uma fun\c{c}\~ao $\mathcal{P}_{\ddot Z}$ de tal forma que se marginalizarmos
  algumas da vari\'aveis, n\'os recuperamos o conjunto de fun\c{c}\~oes usadas. Tamb\'em demonstramos
  (na se\c{c}\~ao \ref{sec:CHSH-Kolmogorov}), contudo, que s\'o \'e violada a desigualdade de CHSH quando
  os axiomas de Kolmogorov forem violados. Assim, concluimos que, embora possamos encontrar a fun\c{c}\~ao
  que gere o conjunto de fun\c{c}\~oes usadas, essa fun\c{c}\~ao nem sempre respeitar\'a os axiomas
  de Kolmogorov.}
  }

We presented (on the section \ref{sec:Sistema}) that the set of probability functions used in the literature can determine a function $\mathcal{P}_{\ddot Z}$   in such a way that if we marginalize some of the variables, we retrieve the set of functions used. We also demonstrated (on the section \ref{sec:CHSH-Kolmogorov}), however, that only the CHSH inequality is violated when Kolmogorov's axioms are violated. Thus, we concluded that although we can find the function that generates the set of functions used, this function does not always respect Kolmogorov's axioms.

\ifthenelse{\boolean{pt-br}}
  {\ptbr
  {Analisamos o esquema experimental (na se\c{c}\~ao \ref{sec:Prob_Cond}), e observamos que no conjunto
  de fun\c{c}\~oes n\~ao fica especificado as probabilidades relacionadas aos caminhos tomados pelas
  part\'iculas. Atrav\'es do esquema experimental, n\'os propomos considerar as probabilidades relacionadas
  aos caminhos, da\'i n\'os geramos um novo conjunto de probabilidades, das quais, atrav\'es das probabilidades
  condicionais, podemos recuperar o conjunto encontrado na literatura. Al\'em disso demonstramos que
  desse novo conjunto de probabilidades, podemos obter os valores esprados de $Z_j\cdot Z_k$, e que
  ao substituirmos na desigualdade de CHSH, nenhuma viola\c{c}\~ao \'e encontrada. Portanto, al\'em
  dessa fun\c{c}\~ao gerar o conjunto de fun\c{c}\~oes de probabilidades, ele respeita a desigualdade
  de CHSH, o que significa que respeita os axiomas de Komogorov, portanto \'e justific\'avel afirmar
  que \'e uma fun\c{c}\~ao de probabilidade e que descreve o experimento estudado.}
  }

We analyzed the experimental scheme (on the section \ref{sec:Prob_Cond}), and we observed that in the set of functions the probabilities related to the paths taken by the particles are not specified. Through the experimental scheme, we propose to consider the probabilities related to the paths, hence we generated a new set of probabilities, from which, through the conditional probabilities, we could retrieve the set found in the literature. Furthermore we showed that from this new set of probabilities, we can obtain the scaled values of $Z_j\cdot Z_k$, and that when we substituted in the inequality of CHSH, no violation was found. Therefore, besides this function generate the set of functions of probabilities, it respects the CHSH inequality, which means that it respects the axioms of Komogorov, therefore it is justifiable to say that it is a function of probability and that describes the experiment studied.

\ifthenelse{\boolean{pt-br}}
  {\ptbr
  {N\'os simulamos o experimento (nas se\c{c}\~oes \ref{sec:Sim} e \ref{sec:Sim_Resultados}), para demonstrar
  como foi o uso de dados amostrais encontrado na literatura e como seria o uso desses mesmos dados
  com base no que foi apresentado neste trabalho. Esses mesmos dados que apresentaram a viola\c{c}\~ao
  da desigualdade de CHSH (na forma de uso encontrada na literatura), no novo modo de us\'a-los (que
  n\'os apresentamos) n\~ao apresentou qualquer viola\c{c}\~ao na desigualdade de CHSH (cujos valores
  esperados foram calculados com base nas fun\c{c}\~oes de probabilidade que cont\'em a informa\c{c}\~ao
  da probabilidade do percurso das part\'iculas).}
  }

We simulated the experiment (on the sections \ref{sec:Sim} and \ref{sec:Sim_Resultados}) to demonstrate how the use of sample data was found in the literature and how the use of the same data would be based on what was presented in this study. These same data that presented the CHSH inequality violation (in the form of use found in the literature), in the new way of using them (which we presented) did not show any violation in the CHSH inequality (whose expected values were calculated based on the Probability functions that contain information on the probability of the particle travel).

\ifthenelse{\boolean{pt-br}}
  {\ptbr
  {Nas pr\'oprias f\'ormulas das estimativas (usadas na literatura), mostramos (na se\c{c}\~ao \ref{sec:Sim_Resultados})
  que o denominador $n_{j,k}$ demonstra que o valor esperado era relacionado a parte do experimento
  (ou seja, \`a parte relacionada ao caminho percorrido), e que a nossa f\'ormula (cujo denominador
  era $n$) al\'em de considerar o experimento inteiro (todas as observa\c{c}\~oes s\~ao consideradas,
  mesmos as relacionadas a outros caminhos) ela tamb\'em n\~ao apresenta qualquer viola\c{c}\~ao da
  desigualdade de CHSH.}
  }

In the same formulas of the estimates (used in the literature), we showed (on the section \ref{sec:Sim_Resultados}) that the denominator $n_{j,k}$, demonstrated that the expected value was related to part of the experiment (i.e. related to the path traveled), and that our formula (whose denominator was $n$) besides considering the whole experiment (all observations are considered, same as those related to other paths) it also does not show any violation of the CHSH inequality.

\ifthenelse{\boolean{pt-br}}
  {\ptbr
  {Um outro resultado deste trabalho, foi a proposta (na se\c{c}\~ao \ref{sec:Desig_Basica}) de uma
  desigualdade b\'asica, da qual podemos gerar a desigualdade de Bell e a desigualdade de CHSH. Verificamos
  que essa desigualdade, quando diretamente usada, n�o tem nenhuma viola\c{c}\~ao. Depois mostramos
  que com substitui\c{c}\~oes simples, podemos encontrar as desigualdade de Bell e uma desigualdade
  do tipo de CHSH, em que viola\c{c}\~oes da desigualdade original de CHSH implicariam em viola\c{c}\~oes
  desta desigualdade tamb\'em. Assim, demonstramos as rela\c{c}\~oes entre a desigualdade de Bell e
  a de CHSH (na se\c{c}\~ao \ref{sec:Bell_CHSH}) e a rela\c{c}\~ao entre a desigualdade de Wigner
  e a de CHSH (na se\c{c}\~ao \ref{sec:CHSH-Kolmogorov}).}
  }

Another result of this work was the proposal (on the section \ref{sec:Desig_Basica}) of a basic inequality, from which we can generate the Bell inequality and CHSH inequality. We found that this inequality, when directly used, has no violation. Then we showed that with simple substitutions, we can find Bell inequality and CHSH type inequality, in which violations of the original CHSH inequality would imply in violations of this inequality as well. Thus, we showed the relations between the Bell inequality and the CHSH inequality
(in the section \ref{sec:Bell_CHSH}) and the relationship between the Wigner inequality and the CHSH inequality (on the section \ref{sec:CHSH-Kolmogorov}).

\ifthenelse{\boolean{pt-br}}
{\ptbr
{De uma forma geral, concluimos que o esquema experimental justifica considerarmos que as fun\c{c}\~oes
usadas s\~ao na verdade probabilidades condicionais e que, a partir dessa considera\c{c}\~ao , encontramos
novas fun\c{c}\~oes de probabilidades que nos fornecem valores esperados (com um fator de corre\c{c}\~ao
dado pelas probabilidades relacionadas aos percursos) que ao substituirmos na desigualdade de CHSH, nenhuma viola\c{c}\~ao ocorre, qualquer que sejam os par\^ametros. }
}

In general, we conclude that the experimental scheme justifies considering that the functions used are in fact conditional probabilities and that, from this consideration, we find new functions of probabilities that give us expected values (with a correction factor) that when we replace in CHSH inequality, no violation occurs, regardless of the parameters.

\ifthenelse{\boolean{pt-br}}
{\ptbr
{Concluimos assim que o experimento de Alain Aspect pode ser modelado pela Estat\'istica cl\'assica de forma a satisfazer plenamente a desigualdade de CHSH e que a viola\c{c}\~ao de CHSH s\'o seria poss\'ivel
se houver uma viola\c{c}\~ao dos axiomas de Kolmogorov.}
}

We conclude that Alain Aspect's experiment can be modeled by Classical Statistics in order to fully satisfy the CHSH inequality and that CHSH violation would only be possible if there is a violation of Kolmogorov's axioms.

\section*{Acknowledgements}

We are grateful to Marcelo Silva de Oliveira, Lucas Monteiro Chavez, Devanil Jaques de Souza for valuable discussions.

%\newpage

%\bibliographystyle{apsrev4-1}
%\bibliography{CHSH_FormatPhysRevA}

\begin{thebibliography}{9}
\section*{References}

\bibitem{citeulike_679960}
Aspect, Alain and Dalibard, Jean and Roger, G\'erard,
\textit{Experimental Test of Bell's Inequalities Using Time- Varying Analyzers},
Phys. Rev. Lett. \textbf{49} (December 1982), no. 25, 1804-1807.

\bibitem{Bell1964a}
John S. Bell,
\textit{On the Einstein-Podolsky-Rosen paradox},
Physics \textbf{1} (1964), 195-200.

\bibitem{PhysRevLett_23_880}
Clauser, John F. and Horne, Michael A. and Shimony, Abner and Holt, Richard A.,
\textit{Proposed experiment to test local hidden-variable theories},
Phys. Rev. Lett. \textbf{23} (October 1969), 880-884

\bibitem{einstein1935b}
A. Einstein, B. Podolsky, and N. Rosen,
\textit{Can Quantum-Mechanical Description of Physical Reality Be Considered Complete?},
Physical Review \textbf{47} (May 1935), no. 10, 777-780.

\bibitem{citeulike_3633707}
A. Khrennikov,
\textit{Epr-bohm experiment and Bell's inequality: Quantum physics meets probability theory},
Theoretical and Mathematical Physics \textbf{157} (October 2008), no. 1, 1448-1460.

\bibitem{citeulike_8683605}
Andrei Y. Khrennikov,
\textit{Contextual Approach to Quantum Formalism},
1st ed., Springer, 2009.

\bibitem{magalhaes2006probabilidade}
M. N. Magalh\~aes,
\textit{Probabilidade e Vari{\'a}veis Aleat{\'o}rias},
Edusp, 2006.

\bibitem{Mood_IntroTheoryStats}
Alexsander M. Mood, Franklin A. Graybill, and Duane C. Boes,
\textit{Introduction to the Theory of Statistics (McGraw-Hill Series in Probability and Statistics)},
3rd ed., McGraw-Hill Companies, 1974.

\bibitem{citeulike_9323624}
Itamar Pitowsky,
\textit{Quantum Probability - Quantum Logic},
Springer-Verlag Berlin and Heidelberg GmbH \& Co. K, 1989.

\end{thebibliography}

%================================================================================================================
%\begin{comment}

%\end{comment}

\end{document}